\documentclass[aps,prd,reprint,nofootinbib,floatfix,longbibliography]{revtex4-2}

\usepackage{amsmath,amssymb}
\usepackage{booktabs}
\usepackage{graphicx}
\usepackage{float}
\usepackage{flafter}
\usepackage{hyperref}
\hypersetup{colorlinks=true,allcolors=blue,
  pdftitle={Exact time-correlated coincidence modeling with reset
    boundaries},
  pdfauthor={Jinjing Li}}
\usepackage{siunitx}
\usepackage{tikz}
\usetikzlibrary{arrows.meta,decorations.pathreplacing,patterns}

\begin{document}

\title{Exact time-correlated coincidence modeling with reset boundaries}

\author{Jinjing Li}
\email{lijinjing@hnu.edu.cn}
\affiliation{School of Physics and Electronics, Hunan University, 410082 Changsha, China}
\affiliation{Hunan Provincial Key Laboratory of High-Energy Scale Physics and Applications, 410082 Changsha, China}
\date{\today}

\begin{abstract}
Delayed-coincidence searches identify a rare prompt--delayed signal, but
their accidental background is not a simple product of marginal rates:
muon vetoes, event dead time, and delayed events created before the
current window condition which event sequences can be recorded.  We
derive exact ordered coincidence rates, within a stated stochastic model,
for a recorded stream of uncorrelated prompt-like singles and correlated
prompt--delayed sources subject to Poisson reset boundaries and event dead
time.  The calculation separates two tasks: a Markov history chain carries
the delayed events still pending at reset boundaries across windows, and a
current-window propagator evaluates the ordered within-window integrals in
closed form with block-matrix exponentials, the matrix-analytic toolkit of
applied probability.  The construction
extends to any prescribed finite multiplicity by increasing the block-chain
depth.  Here we report every ordered one-, two-, and three-fold rate formed
from uncorrelated singles, correlated prompts, and recorded delayed events,
together with the genuine/accidental split of prompt--delayed pairs,
multiplicity efficiencies, and the aggregate rate for multiplicity four or
more.  For the default window-close convention, we derive a
three-term a posteriori error bound for the finite pending-population
truncation.  An independent
streaming toy Monte Carlo validates the ordered rates, two-fold time
densities, matched dead-time conventions, and aggregate high multiplicity.
Within the stated assumptions, the construction is exact on the retained
finite state spaces and provides explicit, computable truncation-error bounds.
\end{abstract}

\maketitle

\section{Introduction}
\label{sec:intro}

Time-correlated coincidence selection isolates a rare prompt--delayed signal
in a high-rate detector stream by requiring both events within a window of
length \(T_c\).  The method enabled the first direct detection of the electron
antineutrino through inverse beta decay (IBD),
\(\bar\nu_e+p\to e^++n\)~\cite{Cowan1956}, and underlies modern reactor
oscillation measurements~\cite{KamLAND2003,DayaBay2012,RENO2012,
DoubleChooz2012,JUNOFirst2025}.  Its irreducible background is an accidental
coincidence of unrelated singles.  Products of marginal rates do not retain
the selection, veto, and boundary conditioning needed at current precision.
The missing conditioning is concrete detector physics: a neutron created
before the current coincidence window can capture inside it; a muon veto
can remove the interval that would otherwise define an accepted window;
event dead time hides events that would otherwise open or close a window;
and the analyzer records ordered, non-overlapping sequences, so a pair
counts only when no other event shares its window.  The accidental rate
therefore depends on the delayed population still awaiting capture, on the
location of the window relative to the veto and dead-time boundaries, and
on the ordering logic of the selection itself.

The combined Daya Bay gadolinium- and hydrogen-capture result gives
\(\sin^2 2\theta_{13}=0.0833\pm0.0022\), a relative uncertainty of about
\(2.6\%\)~\cite{DayaBay2023nGd,DayaBay2024nH}.  A schematic propagation from
this oscillation amplitude to the near--far rate ratio gives a total
rate-control scale of about \(0.2\%\), including counting and systematic
effects.  As a separate experimental benchmark, the final
background-subtracted IBD rates reported for individual Daya Bay detectors
have relative uncertainties from \(0.13\%\) to \(0.24\%\) for neutron capture
on gadolinium and from \(0.25\%\) to \(0.61\%\) for the nominal
neutron-capture-on-hydrogen
analysis~\cite{DayaBay2023nGd,DayaBay2024nH}.  Those tabulated rate errors
include counting statistics and background-subtraction uncertainties, but
not other systematic contributions such as relative detection-efficiency
uncertainties; they are therefore a supporting benchmark rather than a
complete rate-uncertainty budget.

The current \(0.2\%\) scale is experimental context, not an accuracy ceiling
for the present calculation.  A dedicated next-generation reactor
measurement has been proposed to bring \(\sin^2\theta_{13}\) to sub-percent
relative precision~\cite{FutureTheta13Subpercent2023}; taking \(1\%\) as a
representative goal would push the corresponding total rate-control scale
below \(0.1\%\).  A more precise external reactor constraint is also relevant
to accelerator measurements of \(\delta_{\rm CP}\).  Leptonic CP violation
has not yet been established independently of the neutrino mass
ordering~\cite{T2KNOvA2025}, while the DUNE and Hyper-Kamiokande
sensitivity studies both rely on an external \(\theta_{13}\)
constraint, which the DUNE study shows to improve \(\delta_{\rm CP}\)
resolution and to reduce parameter degeneracies at finite
exposure~\cite{DUNEPotential2020,HyperKDesignReport2018}.  The present
method therefore does not stop
at validating an approximation at the per-mille level: it derives exact
coincidence rates within the stated stochastic model, with separately
controlled finite-state truncations.  This distinction matters directly for
the accidental background.  At the high-occupancy \(T_0=0\) benchmarks used
here, the earlier leading-order formulas overestimate the pure accidental
\(ss\) rate by \(0.42\%\)--\(1.38\%\) (Supplemental Table~S2), a model bias
that does not shrink with exposure while the statistical error does, so its
relative weight in the error budget grows.  Exact accidental-coincidence rates under the actual
veto and reset-boundary conventions are therefore needed before that bias
becomes limiting in a future sub-per-mille rate budget.

The challenge is sharpest for neutron capture on hydrogen (nH).
Gadolinium capture releases an \(\sim8\,\)MeV gamma cascade above the
natural-radioactivity continuum, so Daya Bay, RENO, and Double Chooz obtain
strong accidental rejection from the delayed-energy
cut~\cite{DayaBay2023nGd,RENO2025,DoubleChooz2012}.  Hydrogen capture instead
releases \(2.2\,\)MeV, in the same band as natural gamma backgrounds.  It has
supported dedicated oscillation analyses at Daya Bay and
RENO~\cite{DayaBay2024nH,RENOnH2020}.  Double Chooz instead used a total
neutron-capture selection that integrates captures on H, C, and Gd in a
single oscillation analysis~\cite{DoubleChooz2020}.  These analyses control
the larger accidental component with dedicated background models and
coincidence selections.  JUNO controls its hydrogen-capture channel through
material radiopurity and event selection.  Its design study predicted a
\(7.2\,\)Hz natural-radioactivity singles rate, below the \(10\,\)Hz design
target, for reconstructed energy \(E_{\rm rec}>0.7\,\mathrm{MeV}\) and the
default fiducial volume, a liquid-scintillator radius
\(r_{\rm LS}<17.2\,\mathrm{m}\); at the same threshold the prediction was
\(59\,\)Hz in the full \(r_{\rm LS}<17.7\,\mathrm{m}\) detector
volume~\cite{JUNORadiopurity2021}.  The first-data analysis used an
even tighter prompt fiducial selection, \(R<16.5\,\mathrm{m}\) and
\(\lvert z\rvert<15.5\,\mathrm{m}\), and measured a residual accidental rate
of \((4.9\pm0.3)\times10^{-2}\) per day with off-time
windows~\cite{JUNOFirst2025}.  These controls suppress the background but do
not remove the need to evaluate its residual rate under the actual detector
selection.

The same ordered-time accounting appears in geoneutrino and
reactor-antineutrino measurements in liquid-scintillator
detectors~\cite{KamLANDGeo2011,BorexinoGeo2020,SNOplus2025},
accidental-coincidence backgrounds from unrelated S1 and S2 signals in
dual-phase xenon detectors~\cite{PandaXIIAccidental2022,LZ2023,
XENONnT2023}, pulse-pile-up rejection and time-coincidence vetoes in
double-beta-decay searches~\cite{GERDA2020,KamLANDZen2023,CUORE2020}, and
random-coincidence estimation in positron-emission
tomography~\cite{PET1997,PETrandoms2016}.  Reactor IBD is the concrete
realization here;
the calculation applies whenever the delayed-time distribution can be
represented by a finite mixture of exponential components and the relevant
boundaries are specified.

We fix the event notation before using it.  The symbol \(s\) denotes an
uncorrelated prompt-like single, an isolated event with no physical partner;
\(e\) denotes the prompt member of a correlated prompt-delayed source (the
positron signal in the reactor realization), and \(n\)
a recorded delayed member (the neutron capture).  The source rates are
\(R_s\) for the uncorrelated singles and \(R_{\rm corr}\) for the
correlated prompt-delayed pairs.  The delayed-arm detection efficiency
\(\epsilon_n\) is the probability that the delayed member of an \(e\) is detected
and can therefore appear as an \(n\); whether it is recorded is decided
downstream by the veto and dead-time state (Sec.~\ref{sec:model}).  Ordered strings such as \(en\) and \(ss\) name accepted coincidence
windows; for \(en\), ``True'' means the recorded \(n\) is the delayed member of
the triggering \(e\), while ``False'' means it is an old delayed member left
pending by an earlier correlated source.  The detector model that fixes the
delayed-time mixture, the reset-boundary rate, and the dead time follows in
Sec.~\ref{sec:model}.

The relevant earlier approximate calculation is that of
Yu \emph{et al.}~\cite{YuWangChen2015}, developed for low correlated
occupancy and applied to the Daya Bay nH
analysis~\cite{DayaBayNH2014}.  The experimental analysis is application
context, not a separate analytic method.

The central memory variable has a direct detector meaning: it is the
number of delayed events that have been created but have not yet occurred
when a new window may open.  The calculation separates this memory across
windows from the ordering of recorded events inside the current window; a
Markov history chain performs the first task, and finite matrix
propagators perform the second.
The present treatment accordingly carries the pending population at each
reset seam (the stitched-axis boundary point left by a muon veto, defined
in Sec.~\ref{sec:model}) as a Markov state, separates inherited old
neutrons from self neutrons
born from prompts recorded in the current window, propagates unrecorded
births and captures through
event dead time, and evaluates the ordered within-window integrals with
block-matrix exponentials.  These are the matrix-analytic methods of
applied probability~\cite{Neuts1981,LatoucheRamaswami1999,BladtNielsen2017},
standard in that field but comparatively rare in particle-physics
analyses; Sec.~\ref{sec:exact-derivation} therefore opens with the
rationale for this choice and a compact overview of the concepts used
(Sec.~\ref{subsec:toolkit}).  The treatment retains the full coupling of
muons, singles,
prompts, and neutrons without linearizing in \(R_{\rm corr}\).  We provide a
finite-state construction that treats these ingredients jointly for the
stated detector model.  The resulting
one-, two-, and three-fold rates in the alphabet \(s,e,n\) include the
True/False split of the \(en\) sample and remain valid in IBD-rich,
low-threshold, and long-window regimes.
The block-chain construction itself is not limited to three-fold windows:
any prescribed finite multiplicity follows by adding event blocks and, as
needed, increasing the current-state headroom.  The present work reports the
full ordered inventory through three-fold and separately validates the
aggregate rate for multiplicity four or more.
Throughout, \emph{exact} is a stated contract rather than an unqualified
claim:
exact within the stochastic model of Sec.~\ref{sec:model} (independent
homogeneous Poisson source streams with the hyperexponential delay
law~(\ref{eq:delay-density}), under the stated reset and dead-time
conventions) and, numerically, exact on the pending-population state
spaces retained by the calculation: the history space
\(\sum_i h_i\le N_{\max}\) and the doubled current-window space
\(\sum_i(o_i+m_i)\le C=N_{\max}+H\).  The two truncations are
controlled separately in
Sec.~\ref{subsec:truncation-contract}.

Two scope statements are needed at the outset.  First, every source stream
is independent of the reset stream: muon-correlated backgrounds,
cosmogenic \(^{9}\)Li/\(^{8}\)He and muon-induced fast neutrons, the
dominant correlated backgrounds of real reactor analyses, enter only
through the independent-Poisson idealization and are otherwise outside
the model (Sec.~\ref{sec:model}).  Second, all rates are energy
integrated.  Within this scope, the method is intended as the primary evaluation of
the accidental-coincidence rate under a given selection, as a design
instrument for choosing that selection, and as an exact reference for
its simulation chain.

Simple products show the scale of the missing conditioning.  For the IBD-rich
short-window configuration
(\(R_{\rm corr}=\SI{5}{Hz}\), \(T_c=\SI{400}{\micro s}\),
\(R_s=\SI{50}{Hz}\), \(\epsilon_n=0.8\), and \(T_0=0\)), the exact total
\(en\) rate is \(\SI{3.416}{Hz}\), \(14.6\%\) below the simple
prompt-times-efficiency bound
\(R_{\rm corr}\epsilon_n=\SI{4}{Hz}\).  The more informative exact False
component (the \(en^{\rm False}\) rows of
Table~\ref{tab:exact-rates-grid})
is \(\SI{1.2653e-4}{Hz}\) at this point and \(\SI{2.7090e-5}{Hz}\) at
\(T_c=\SI{1500}{\micro s}\), whereas the analogous source-rate product
\(R_{\rm corr}(R_{\rm corr}\epsilon_n)T_c\) gives
\(\SI{8.0e-3}{Hz}\) and \(\SI{3.0e-2}{Hz}\), overestimating the exact rates
by factors of about 63 and \(1.1\times10^3\).  These are deliberately crude baselines;
Sec.~\ref{sec:event-placement} makes the more relevant comparison with the
earlier time-correlated treatment~\cite{YuWangChen2015}.

Section~\ref{sec:model} states the model and validation grid.
Section~\ref{sec:exact-derivation} constructs the exact treatment, the
Markov history chain with its block propagator and truncation-error
contract; Sec.~\ref{sec:event-placement} then defines the secondary
factorized comparison used to interpret the validation overlays.
Sections~\ref{sec:toymc}
and~\ref{sec:results} validate rates, efficiencies, and time distributions;
Sec.~\ref{sec:regimes} gives the physical rate regimes.
Section~\ref{sec:discussion} reports the numerical robustness of
the finite history cap.  The donor placements of the factorized
construction and its factorized-to-exact \(en\) comparison, the
channel-by-channel comparison of the earlier leading-order treatment
with the exact rates, the bulk per-channel rate tables, the per-shard
rate histograms, the full-size two-fold time-density overlays, and the
selection-efficiency scan table are collected in the Supplemental
Material.
Readers interested primarily in the physical mechanism may read
Secs.~\ref{sec:model}, \ref{subsec:toolkit}, \ref{sec:pending-physics},
\ref{sec:regimes}, and \ref{sec:discussion} before returning to the full
operator construction.

\section{Detector model}
\label{sec:model}

We first state the model in the source-agnostic form in which it is derived and
then specialize it to the reactor-antineutrino application used for validation.
Physically, the model has three ingredients: unrelated detector activity
supplies isolated recorded events, a correlated source supplies a prompt
event that may later produce a detected partner, and resets such as muons
interrupt the coincidence selection.  The first two ingredients are defined
now; the reset process and the analysis time axis built from it are
introduced later in this section.
The detector records a single time-ordered stream of \emph{recorded events},
meaning energy deposits that pass the analysis cuts and are written out, fed by two
kinds of true sources.  The uncorrelated singles \(s\) form a Poisson process of
rate \(R_s\), structureless in time.  The time-correlated sources form a Poisson
process of latent firings with rate \(R_{\rm corr}\), a firing being a
physical source occurrence counted before veto and recording losses.  Each firing supplies a
prompt \(e\) and is marked either with one detected delayed daughter, with
probability \(\epsilon_n\), or with no detected daughter, with probability
\(1-\epsilon_n\).  Conditional on the detected-daughter branch, the delay is
drawn from the normalized density
\begin{equation}
  \rho(t)=\sum_{i=1}^{K} f_i\,\tau_i^{-1}\,e^{-t/\tau_i},
  \qquad \sum_{i=1}^{K} f_i=1,
  \label{eq:delay-density}
\end{equation}
a finite mixture of \(K\ge1\) exponential components with weights
\(f_i\ge0\) and lifetimes \(\tau_i>0\).  Each component represents a delayed channel with
constant hazard \(1/\tau_i\).  A detected delayed daughter whose capture
has not yet occurred is called \emph{pending}.  The mixture form is chosen
for a structural reason, not only for fit convenience: each component is
exponential, hence memoryless, so the future of the pending population
depends only on the number of pending daughters in each component, not on
their individual ages, and this count vector is the finite state that
permits the exact treatment of Sec.~\ref{sec:exact-derivation} (the
underlying concepts are reviewed in Sec.~\ref{subsec:toolkit}).
Mathematically, Eq.~(\ref{eq:delay-density}) is a hyperexponential
density, the diagonal special case of the phase-type
family~\cite{Neuts1981,BladtNielsen2017}, and the only structural
assumption the derivation makes about the delayed arm is that it admits
such a representation.
This is in addition to the independent Poisson parent process,
one-daughter Bernoulli marking, and independence from the reset stream.
The rates \(R_s\) and \(R_{\rm corr}\) are intensities of these latent
sources, not of the written-out stream: whether a fired event is recorded
is decided downstream by the veto and dead-time state, and a prompt that
fires inside a blind stretch still produces its delayed daughter, which
the propagator carries as a pending neutron (Sec.~\ref{sec:propagator}).
Five terms keep fixed meanings throughout the paper.  A source
\emph{fires} latent events at its Poisson intensity.  A fired delayed
neutron is \emph{detected} with probability \(\epsilon_n\); an
undetected one never enters any stream.  A fired (and, for the delayed
arm, detected) event is \emph{recorded} if it survives the veto and
the dead-time state at its arrival time.  A detected neutron whose
capture has not yet occurred is \emph{pending}.  A window, with its
recorded members, is \emph{accepted} if no muon crosses it.
A \emph{time-correlated background} is any further source sharing this
prompt-delayed form; such a source is merged into the correlated source
exactly, not as a leading-order convenience.  Readers interested in a
single correlated source may skip
the following proposition; it proves that independent sources with the
same prompt-side recording rule combine without approximation.

\emph{Proposition (correlated-source merging).}  Let sources
\(a=1,\ldots,M\) be mutually independent homogeneous Poisson streams
of latent prompt-delayed firings with intensities \(R_a\), each
independent of the reset stream and subject to the same prompt-side
acceptance and recording rule as \(e\).  A firing of source \(a\)
carries delayed mark \(i\) with probability \(\epsilon_a f_{ai}\), or
the zero-daughter mark with probability \(1-\epsilon_a\), on a common
lifetime set \(\{\tau_i\}\) (extend the set first if a source needs a
new lifetime).  Then every rate computed in this paper is exactly
reproduced by the single merged source
\begin{equation}
  \begin{aligned}
    R_{\rm corr}&=\sum_a R_a,&
    b_i&=\sum_a R_a\epsilon_a f_{ai},\\
    b_\Sigma&=\sum_i b_i,&
    \epsilon_n&=\frac{b_\Sigma}{R_{\rm corr}},\qquad
    f_i=\frac{b_i}{b_\Sigma},
  \end{aligned}
  \label{eq:source-merging}
\end{equation}
when \(R_{\rm corr}>0\) and \(b_\Sigma>0\).  If \(b_\Sigma=0\), set
\(\epsilon_n=0\) and choose any normalized \(f_i\), which then drops
out because every \(b_i\) vanishes; if \(R_{\rm corr}=0\), the merged
source is absent.  In particular, the latent zero-daughter firing
intensity is retained exactly:
\(\sum_a R_a(1-\epsilon_a)=R_{\rm corr}-b_\Sigma
=R_{\rm corr}(1-\epsilon_n)\).
The proof is two superposition facts: the union of independent Poisson
streams is Poisson with the summed rate, carrying independent marks
with the mixed detection-and-capture law; and every operator entry of
the exact construction depends on the correlated sources only through
the three aggregate intensities \(R_{\rm corr}\),
\(R_{\rm corr}(1-\epsilon_n)\), and
\(R_{\rm corr}\epsilon_n f_i = b_i\): the seam
means~(\ref{eq:seam-means}) and the blind
generator~(\ref{eq:ablind}) through \(b_i\), the visible
generator~(\ref{eq:avis}) through \(R_{\rm corr}\), and the event
matrices~(\ref{eq:event-matrices}) through the recorded/unrecorded
split \(R_{\rm corr}(1-\epsilon_n)\) and \(b_i\), so
(\ref{eq:source-merging}) matches each entry term by term.
Sources with distinguishable prompt-side behavior need their own label
instead; a source that merges keeps no separate label in the model.

\paragraph{Coincidence selection and event packing.}
The stream is bundled into coincidence windows of fixed length \(T_c\).  A
recorded event that does not already belong to an open window becomes a
\emph{trigger} and opens one; every recorded event within \(T_c\) of the trigger
is packed into that window, which then closes.  The ordered list of member types
names the window, so the sample is partitioned into one-fold windows
(\(s,e,n\)), two-fold windows (the nine ordered pairs \(ss,se,\dots,nn\)),
three-fold windows (twenty-seven ordered triples), and higher multiplicities,
the multiplicity being the number of members.  In a True \(en\) window the
recorded \(n\) is the triggering \(e\)'s own \emph{self} neutron; in a False
(accidental) one it is an unrelated \emph{old} neutron that merely fell inside
\(T_c\).  Only the True component carries the physical prompt-delayed
correlation; the False component is the irreducible accidental background of the
method.  The packing rule is one-sided: it conditions only on the interval after
the trigger.  An additional isolation requirement \emph{before} the
trigger (a quiet pre-window, as in some multiplicity cuts) would change
the window-opening operator, not the within-window propagator: a
mechanical but genuine kernel extension, not a redefinition of \(T_c\).

\paragraph{Reactor antineutrinos.}
In a reactor experiment the time-correlated source is inverse beta decay,
\(\bar\nu_e+p\to e^+ + n\): the prompt \(e\) is the positron (ionization plus
annihilation) and the delayed \(n\) is the capture of the thermalized neutron,
detected with efficiency \(\epsilon_n\).  The IBD cross section and prompt-positron
spectrum that fix \(R_{\rm corr}\) are reviewed in
Refs.~\cite{VogelBeacom1999,Strumia2003,Vogel2015}.  The uncorrelated singles \(s\) are
then the natural-radioactivity and noise background, and the window length
\(T_c\) is chosen to match the neutron-capture time, so that a genuine
prompt-delayed pair almost always lands in one window while the accidental load
stays tolerable.

\paragraph{The capture-time mixture.}
The delayed density~\eqref{eq:delay-density} is set by how the neutron slows down
and is captured.  After production the neutron thermalizes in a few microseconds
and then captures on one of the available nuclei; the mean capture time is fixed
by the thermal capture cross section and the target density, so a larger cross
section gives a shorter time.  The two validation components,
\((\tau_1,\tau_2)=(200,30)\,\si{\micro s}\), are the two channels of a typical
scintillator: \(\tau_1\simeq\SI{200}{\micro s}\) is capture on free hydrogen
(nH), whose small thermal cross section gives the long time, and
\(\tau_2\simeq\SI{30}{\micro s}\) is capture on dissolved gadolinium (nGd),
whose very large cross section gives the short time even at sub-percent
loading: the scintillators of Daya Bay and RENO (\(0.1\%\) Gd by mass) and
Double Chooz (\SI{1}{g} Gd per litre) have mean nGd capture times of about
30, 26, and \SI{30}{\micro s},
respectively~\cite{DayaBay2023nGd,DayaBay2018,RENO2018,DoubleChooz2012}.
Both validation values include the moderation time.  The weights \((f_1,f_2)=(0.8,0.2)\) are
a validation choice rather than a property of any one detector.  For a
real detector the pair \((f_i,\tau_i)\) is read from its measured
delayed-time spectrum with the same fits that experiments already
perform for their capture-time acceptances~\cite{DayaBay2018,RENO2018}.  The
mixture is treated as run-constant here, so slow drifts of gadolinium
concentration or optics that move \((f_i,\tau_i)\) over a multi-year
run enter as parameter updates between stationary stretches, not as
new model structure.  The two window
lengths straddle this mixture.  \(T_c=\SI{400}{\micro s}\) spans only two nH
lifetimes, so an appreciable fraction of delayed captures leaks past the window
edge and feeds the pool of pending old neutrons, giving the larger False
(accidental) \(en\) rate of the two choices.  \(T_c=\SI{1500}{\micro s}\)
contains nearly the whole mixture and starves that pool; the False rate falls
by a factor of about five, whereas the crude memoryless product
\(R_{\rm corr}(R_{\rm corr}\epsilon_n)T_c\), which grows linearly in \(T_c\),
overestimates it by three orders of magnitude (Sec.~\ref{sec:intro}).  The two windows thus probe cross-window
pending-neutron memory from opposite ends, in the accidental-coincidence
background that precision reactor analyses evaluate and subtract at the percent
level~\cite{DayaBay2018,RENO2018}.

\paragraph{Time-correlated backgrounds.}
The cosmogenic \(\beta\)-delayed neutron emitters \(^{9}\)Li and \(^{8}\)He,
produced by through-going muons, are a standard correlated background of
reactor IBD: their \(\beta\) decay is the prompt and the subsequent neutron
capture is the delayed, so they share the IBD prompt-delayed structure and, with
the neutron captured in the same medium, the same density~\eqref{eq:delay-density}.
Muon-induced fast neutrons provide another important example: a fast neutron
can first scatter in the target and give a prompt-like recoil signal, then
thermalize and capture as the delayed member.  Because that delayed member is
again a neutron capture in the same medium, its capture-time law enters the
same delayed arm, with the fast-neutron rate and any distinct moderation or
capture mixture folded into \(R_{\rm corr}\) and \((f_i,\tau_i)\).  Other
radioactive correlated backgrounds that survive the selection are handled the
same way when they can be represented as one prompt-like recorded event followed
by one delayed-like recorded event.  These sources therefore need no dedicated
state; separating them in a given analysis only means adding their rates and
delay components to the correlated source model, exactly as in the
merging proposition~(\ref{eq:source-merging}).

One condition governs
this folding: the source must itself be an independent homogeneous
Poisson stream whose production is uncorrelated with the reset stream.
Cosmogenic sources violate this at production, since the
\(^{9}\)Li/\(^{8}\)He and fast-neutron parents are the muons that drive
the veto, but the veto and the Poisson reset stream together remove most of
the consequence.  Fast neutrons from tagged muons fall inside the veto and
are excised; those from untagged muons come from an independent thinning of
the muon stream and are therefore Poisson and independent of the resets.
\(^{9}\)Li/\(^{8}\)He decays do cluster behind their parent muon on the
scale of their \SI{0.2}{s} lifetimes, but window survival across the
\emph{next} reset is memoryless and the pending pool at a seam is
stationary, so they enter the recorded rates linearly in their local
intensity and average to the same stationary rate; only their negligible
self-coincidences see the clustering.  The genuine residual is the
population of spallation neutrons still pending when a veto ends,
suppressed by \(e^{-V/\tau_i}\) and hence relevant only for the slow nH
channel, where, for vetoes of a few \(\tau_i\), it can be comparable to
the IBD-fed pending pool.  It then shifts the \(n\) one-fold rate and
every \(n\)-bearing accidental sequence (\(sn\), \(ns\), \(nn\),
False \(en\), and their three-fold extensions) immediately after seams,
by amounts that can reach tens of percent of those channels, while the
total \(en\) rate, dominated by True pairs, is essentially unchanged.
The residual is not part of the model here, but would enter as an additive
component of the Poisson seed at each reset,
Eq.~(\ref{eq:seam-means}), rather than as new structure.  We therefore
treat cosmogenic sources in the independent-Poisson idealization
throughout.
Purely uncorrelated cosmogenic or radioactive activity folds into
\(R_s\).  This uniform treatment
is why the \(en\) sample represents IBD in the validation context yet covers any
prompt-delayed signal of this form in any other application of the framework.

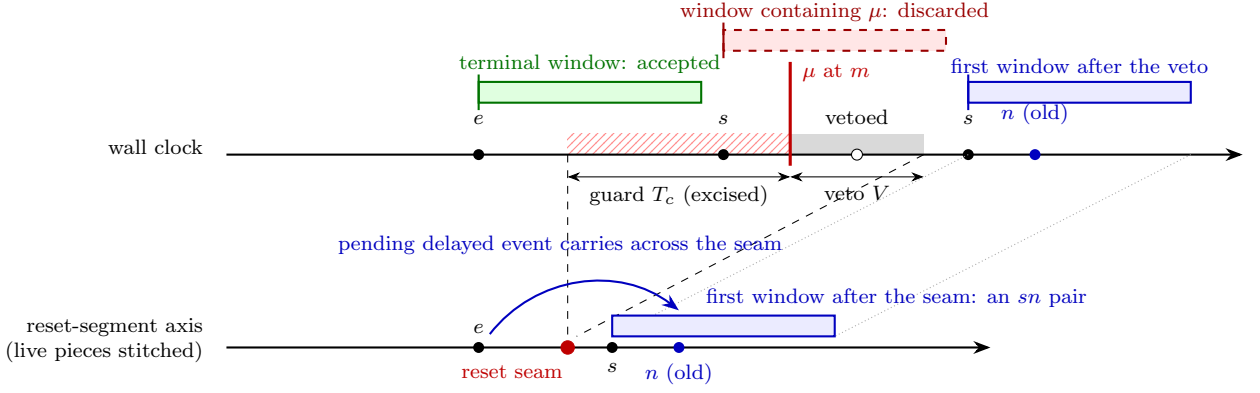
\begin{figure*}[t]
  \centering
  \resizebox{0.92\textwidth}{!}{%
  \begin{tikzpicture}[
    >=Stealth,
    muon/.style={red!75!black,thick,line width=1.2pt},
    win/.style={line width=0.8pt},
    dim/.style={<->,thin},
    label/.style={font=\footnotesize},
    ev/.style={circle,fill=black,inner sep=0pt,minimum size=4.2pt},
    evn/.style={circle,fill=blue!75!black,inner sep=0pt,minimum size=4.2pt},
    vet/.style={circle,draw=black,fill=white,inner sep=0pt,minimum size=4.2pt},
    map/.style={densely dotted,thin,gray!70}
  ]
    \fill[pattern=north east lines, pattern color=red!45]
      (4.5,3.2) rectangle (7.5,3.48);
    \fill[gray!30] (7.5,3.2) rectangle (9.3,3.48);
    \draw[->,thick] (-0.1,3.2) -- (13.6,3.2);
    \node[label,left] at (-0.3,3.3) {wall clock};
    \draw[muon] (7.5,3.05) -- (7.5,4.45);
    \node[label,red!75!black,right] at (7.55,4.3) {$\mu$ at $m$};
    \draw[dim] (4.5,2.9) -- (7.5,2.9)
      node[midway,below,label] {guard $T_c$ (excised)};
    \draw[dim] (7.5,2.9) -- (9.3,2.9)
      node[midway,below,label] {veto $V$};
    \node[ev] (e1) at (3.3,3.2) {};
    \node[label,above] at (3.3,3.5) {$e$};
    \node[ev] (s1) at (6.6,3.2) {};
    \node[label,above] at (6.6,3.5) {$s$};
    \node[vet] at (8.4,3.2) {};
    \node[label,above] at (8.4,3.5) {vetoed};
    \node[ev] (s2) at (9.9,3.2) {};
    \node[label,above] at (9.9,3.5) {$s$};
    \node[evn] (n1) at (10.8,3.2) {};
    \node[label,blue!75!black,above] at (10.8,3.5) {$n$ (old)};
    \fill[green!12] (3.3,3.9) rectangle (6.3,4.18);
    \draw[win,green!45!black] (3.3,3.9) rectangle (6.3,4.18);
    \draw[green!45!black,thick] (3.3,3.8) -- (3.3,4.28);
    \node[label,green!45!black,above] at (4.8,4.18)
      {terminal window: accepted};
    \fill[red!10] (6.6,4.6) rectangle (9.6,4.88);
    \draw[win,red!60!black,dashed] (6.6,4.6) rectangle (9.6,4.88);
    \draw[red!60!black,thick] (6.6,4.5) -- (6.6,4.98);
    \node[label,red!60!black,above] at (8.1,4.88)
      {window containing $\mu$: discarded};
    \fill[blue!8] (9.9,3.9) rectangle (12.9,4.18);
    \draw[win,blue!75!black] (9.9,3.9) rectangle (12.9,4.18);
    \draw[blue!75!black,thick] (9.9,3.8) -- (9.9,4.28);
    \node[label,blue!75!black,above] at (11.4,4.18)
      {first window after the veto};
    \draw[->,thick] (-0.1,0.6) -- (10.2,0.6);
    \node[label,left,align=right] at (-0.3,0.7)
      {reset-segment axis\\ (live pieces stitched)};
    \filldraw[red!75!black] (4.5,0.6) circle (0.09);
    \node[label,red!75!black,below left] at (4.55,0.48) {reset seam};
    \draw[dashed,thin] (4.5,3.2) -- (4.5,0.75);
    \draw[dashed,thin] (9.3,3.2) -- (4.62,0.75);
    \draw[map] (9.9,3.2) -- (5.1,0.75);
    \draw[map] (12.9,3.2) -- (8.1,0.75);
    \node[ev] at (3.3,0.6) {};
    \node[label,above] at (3.3,0.68) {$e$};
    \fill[blue!8] (5.1,0.75) rectangle (8.1,1.03);
    \draw[win,blue!75!black] (5.1,0.75) rectangle (8.1,1.03);
    \node[ev] at (5.1,0.6) {};
    \node[label,below] at (5.1,0.5) {$s$};
    \node[evn] at (6.0,0.6) {};
    \node[label,blue!75!black,below] at (6.0,0.5) {$n$ (old)};
    \node[label,blue!75!black,above right] at (6.25,1.03)
      {first window after the seam: an $sn$ pair};
    \draw[->,thick,blue!75!black] (3.45,0.78) to[bend left=45] (6.0,1.1);
    \node[label,blue!75!black,above] at (4.4,1.75)
      {pending delayed event carries across the seam};
  \end{tikzpicture}}
  \caption{The reset boundary on the wall clock (top) and on the
    reset-segment axis (bottom).  Top: a muon at \(m\) opens the veto
    \([m,\,m+V]\) (gray), during which nothing is recorded (open circle);
    the guard \([m-T_c,\,m)\) (hatched) is excised in addition, purely so
    that no coincidence window straddles the excision.  Filled circles are
    recorded events, and every window has length \(T_c\).  The window
    opened by the recorded event inside the guard contains the muon and is
    discarded (red, dashed).  The terminal window, opened before the guard,
    is accepted (green) because it closes before \(m\); its content,
    including the part inside the guard, is read off the wall clock.  The
    first recorded event after the veto opens the first window beyond the
    veto (blue).  Bottom: the excised stretch collapses to a single point,
    the reset seam, the live pieces are stitched together (dotted lines map
    the first window), and window-opening geometry restarts at the seam.
    Delayed events still pending at the seam are not erased: the delayed
    capture \(n\) of the earlier prompt \(e\) falls inside the first window
    after the seam and turns it into an accidental \(sn\) pair.  That
    surviving pending population is the Markov state of
    Section~\ref{sec:exact-derivation}.}
  \label{fig:reset-timeline}
\end{figure*}

\paragraph{Reset boundaries and the reset-segment axis.}
The third ingredient of the model is a Poisson stream of \emph{reset
events}: occurrences that interrupt the coincidence machinery and force it
to restart.  Any window-discarding interruption of the stream can play this
role; in the reactor realization it is the muon veto.  Through-going muons
deposit large energy and are trailed by showers of induced activity, so each
muon opens an \emph{anti-coincidence veto}, an interval during which no
recorded event is accepted.  A coincidence window overlapping a veto would
mix vetoed and live time, so any window crossed by a muon is discarded.  The
derivation is accordingly written on the \emph{reset-segment axis}: starting
one window length \(T_c\) before each reset event and running to the end of
its veto, that stretch of the wall-clock axis is excised, the surrounding
pieces are stitched together, and the excised stretch collapses to a single
point, a \emph{reset boundary} (or reset seam), the boundary condition named
in the title (Fig.~\ref{fig:reset-timeline}).  Operationally, this is a
cut-and-stitch construction on the time line: the guarded and vetoed
stretch is removed, the surviving pieces are joined, and the join restarts
the coincidence-window geometry.  A neutron already awaiting capture is
not erased by this operation and may survive across the join.

Three time axes must be
kept distinct here.  The \emph{wall clock}
\(t\) is the raw recording time, on which vetoes and event times are first
defined.  The \emph{detector-live axis} deletes only the genuine veto
intervals \([m_i,\,m_i+V_i]\); it is the clock behind quoted veto live-time
fractions.  The reset-segment axis deletes, in addition, the guard
\([m_i-T_c,\,m_i)\) ahead of each reset: detector-live time excised purely so
that no window straddles a reset.  The guard is an analysis construction, so
the reset-segment axis is \emph{not} the detector's cumulative live time,
and rates on the three clocks differ by explicit factors: with raw reset
intensity \(R_\mu\) and mean veto length \(\bar V\), a fraction
\(e^{-R_\mu(T_c+\bar V)}\) of wall time survives to the reset-segment axis
and a fraction \(e^{-R_\mu \bar V}\) to the detector-live axis (Poisson
vacancy fractions, exact for independently marked veto lengths with
finite mean \(\bar V\)), so a
counting rate quoted per unit reset-segment time converts to per detector-live
time by the factor \(e^{-R_\mu T_c}\) and to per wall time by
\(e^{-R_\mu(T_c+\bar V)}\).  All rates in this paper are quoted per unit
reset-segment time.

A reset boundary restarts the window-opening geometry
and the completed-history schedule: no window contains the muon itself, the
completed history windows of Sec.~\ref{sec:history-chain} close before the seam,
and the window machinery starts from the same boundary geometry on its far
side.  It is not a regeneration point for the complete source state.  In
particular, a terminal accepted window may open
before the seam and extend into the excised pre-reset guard, provided it
closes before the physical reset event; such windows are evaluated by the
terminal current-window step rather than by the continuing history
kernel.  Meanwhile the population of delayed events still pending
from before the boundary can survive across it.  That surviving
cross-boundary population is a mark of the seam process and can be correlated
between adjacent seams.  It is exactly the memory promoted to a Markov state
in Section~\ref{sec:exact-derivation}.  The representation requires
\(R_\mu>0\): as \(R_\mu\to0\) seams become infinitely rare and the
history chain loses its seam-intensity source, while the physical coincidence
rates approach finite no-reset (tiling-window renewal) limits.  The
\(R_\mu\to0\) limit is therefore an analytic limit of the formulas rather
than an input to this reset-seam representation.

\paragraph{Reset-cluster lemma.}
This lemma is what allows the rest of the derivation to treat the reset
boundaries as a homogeneous Poisson process on the reset-segment axis, at
the raw wall-clock muon intensity and independent of everything else,
for every veto-length law satisfying the independence assumptions
below.  Its content is that overlapping excisions merge
into single seams at exactly the rate that compensates the excised time.
Precisely: on the reset-segment axis the reset boundaries are again a
homogeneous Poisson process, with the \emph{raw} wall-clock intensity of
the reset stream, for any non-negative veto-length law with finite mean,
the lengths independent and identically distributed (i.i.d.) and
independent of the reset locations and source streams.  This is the
busy-period regeneration property of an infinite-server queue with Poisson
arrivals and a general service-time law (M/G/\(\infty\) in Kendall's
notation), transcribed to veto excision (Sec.~\ref{subsec:toolkit} recalls
the queueing picture); a reader willing to accept the conclusion can pass
over the proof that follows.
Formally, let reset events form a Poisson process of wall intensity
\(R_\mu\), with veto lengths \(V_i \ge 0\) i.i.d.\ with finite mean,
independent of the positions and of the source streams.  Shift each reset to
its guard start and attach the total excision length,
\begin{equation}
  a_i=m_i-T_c, \qquad L_i=T_c+V_i .
  \label{eq:guard-start-marks}
\end{equation}
The pairs \(\{(a_i,L_i)\}\) form an independently marked Poisson process of
intensity \(R_\mu\), and the excised set is the Boolean union
\(\bigcup_i[a_i,a_i+L_i]\).  Reveal each mark when its guard start occurs.
In this marked filtration, the right endpoint of a connected Boolean cluster
is the end of the corresponding M/G/\(\infty\) busy period and is a stopping
time: at wall time \(t\), the guard starts and lengths already revealed
determine whether any marked interval remains active.  The strong Markov
property at that endpoint makes the wait to the next guard start independent
and \(\operatorname{Exp}(R_\mu)\).  This uncovered wait is exactly the
distance between consecutive seams after the intervening Boolean cluster is
collapsed.  Successive stitched-axis gaps are therefore independent
\(\operatorname{Exp}(R_\mu)\) variables, so the seam process is Poisson with
intensity \(R_\mu\).  This argument does not call the cluster endpoint a
stopping time in the unmarked reset-event filtration, which would not reveal
the veto lengths.

Three consequences fix the scope.  First, the veto-length mixture enters
only through the excised wall-time fraction
\(1-e^{-R_\mu(T_c+\bar V)}\).  Because the lemma requires only
independently marked veto lengths with finite mean, a graded veto policy
that assigns different hold-off durations by muon energy, topology, or
shower tag is already inside the model: the validation mixture's
\(\SI{400}{\micro s}\)/\SI{1}{s} shower split is exactly such a mark, and
any richer \(V_i\) law folds in the same way, through \(\bar V\) alone.
Second, each seam regenerates the geometry of the stitched axis but not the
cross-seam pending source state; that is why the pending-neutron history
chain of Sec.~\ref{sec:history-chain} is needed at all.  Third, what is
\emph{not} representable is a veto conditioned on the source events
themselves, such as a track-proximity veto applied only near the muon
trajectory: that correlates the excision with the events it excises and
breaks the independence the lemma rests on.  The symbol \(R_\mu\) is used
throughout for this common seam intensity, after its muon-veto
realization.  Finally, the finite-mean assumption makes the
one-dimensional Poisson Boolean clusters finite almost surely and leaves a
positive uncovered fraction, so collapsing the clusters produces
non-degenerate seams and live gaps; infinite-mean or possibly infinite
marks would require a separate qualification because an infinite cluster
or zero remaining live fraction can occur.

\paragraph{Dead time.}
After each recorded event the detector is blind for an interval \(T_0\).
Exactly how the blindness interacts with the window edge and with lost
events is a convention, and the derivation explicitly implements three
conventions.
In the default \emph{window-close} convention the interval is
non-paralyzable and truncated at the window edge, so the analyzer
re-arms at window close and no dead time leaks into the inter-window
gap.  In the strictly \emph{global non-paralyzable} convention the last
blind interval may cross the window edge into the following gap.  In
the strictly \emph{global paralyzable} convention every good event,
recorded or not, restarts the blindness.  The terminology follows classical
dead-time counters~\cite{Muller1973,Muller1991,Libert1978,Pomme2015}, but
the three conventions are a separate layer of the current/gap kernels; the
Boolean reset union of the lemma above is not asserted to implement those
counters.  The three names map onto
detector practice as follows.  Window-close describes dead time
imposed by the coincidence logic itself, an offline or trigger-level
analyzer that re-arms its own selection when it closes a window; it is
the natural default when \(T_0\) models analysis hold-off rather than
readout hold-off.  A fixed per-channel front-end dead time, set before
any window is defined, is the global non-paralyzable convention, and a
retriggerable hold-off is the paralyzable one.  An analysis whose
blindness comes from the readout should therefore use the global
kernels, not the default.  All three coincide
at \(T_0=0\).  At \(T_0>0\) their differences are themselves
exact analytical statements, quantified on the validation grid in
Sec.~\ref{subsec:prd-conventions}.  They are not small on the
scale of modern precision: at \(R_{\rm corr}=\SI{0.2}{Hz}\) and
\(T_c=\SI{400}{\micro s}\), the \(n\), \(ne\), and \(ns\) rates sit
\(1.89\times10^{-2}\) relative below window-close under the global
non-paralyzable convention at \(T_0=T_c/2\), and \(nn\) sits \(0.13\)
below.  The validation grid is
run both at \(T_0=0\) and, under the window-close convention, at a
small nonzero \(T_0=\SI{1}{\micro s}\) that exercises the dead-time
machinery directly; a dedicated campaign at
\(T_0=50\)--\(\SI{200}{\micro s}\) compares the three conventions,
each against its matched kernel.
This paper derives and proves, end to end, the window-close
construction of Secs.~\ref{sec:exact-derivation}--\ref{sec:propagator};
Sec.~\ref{subsec:general-KN} states compactly which object changes under
each of the other two conventions.

\paragraph{Energy dependence.}
Every rate in this paper is energy integrated: an event's reconstructed
energy enters only through its membership in the \(s\), \(e\), or \(n\)
class, decided by whatever cuts define those classes, and those
class-defining cuts are already inside \(R_s\), \(R_{\rm corr}\), and
\(\epsilon_n\).  The energy spectrum of a background is a separate
construction, taken from data or simulation (off-window pairs, per-class
singles spectra); what the present rates supply is the exact
normalization such a spectrum must carry.  Wherever the energy acceptance
factorizes from timing, this combination is exact: energy bins are then
independent marks on the recorded events, so a bin-resolved rate is the
energy-integrated sequence rate multiplied by the acceptance
probabilities of its members, with the window, dead-time, and veto
dynamics still driven by the full recorded stream and no kernel rerun.
On the delayed side the mark correlates with the lifetime component
through the capture energy, so the marking is applied per component,
which the component-resolved kernels support.

\paragraph{Validation configuration.}
The validation grid fixes
\begin{equation}
  R_s=\SI{50}{Hz},\quad R_\mu=\SI{200}{Hz},\quad \epsilon_n=0.8,
\end{equation}
with \((\tau_1,\tau_2)=(200,30)\,\si{\micro s}\) and \((f_1,f_2)=(0.8,0.2)\), and
scans four signal points, \(R_{\rm corr}=5\) or \(\SI{0.1}{Hz}\) crossed with
\(T_c=\SI{1500}{\micro s}\) or \(\SI{400}{\micro s}\), each run at zero dead time
\(T_0=0\) and repeated at a small nonzero \(T_0=\SI{1}{\micro s}\).  Every muon
opens a veto during which no event is recorded and which discards any
coincidence window it crosses: normal muons impose a
\(\SI{400}{\micro s}\) veto and shower muons a \(\SI{1}{s}\) veto with
probability \(5\times10^{-4}\), so the mean veto length is
\(\bar V=\SI{0.90}{ms}\), the detector-live fraction
\(e^{-R_\mu\bar V}=0.835\), and the reset-segment fraction
\(e^{-R_\mu(T_c+\bar V)}\) is \(0.619\) at \(T_c=\SI{1500}{\micro s}\) and
\(0.771\) at \(\SI{400}{\micro s}\).  The explicit veto lengths enter only
the toy Monte Carlo, through the excised live-time union; the
analytical reset-segment rates depend on the veto through \(R_\mu\)
alone, and on \(\bar V\) only through these clock conversions.

\paragraph{Streaming toy Monte Carlo.}
The same model is realized by a streaming toy Monte Carlo used as an independent
check (Sec.~\ref{sec:toymc}).  It draws the three Poisson processes (singles at
\(R_s\), correlated prompts at \(R_{\rm corr}\), muons at \(R_\mu\)), schedules for
each correlated prompt a delayed neutron detected with probability \(\epsilon_n\)
at a capture time sampled from~\eqref{eq:delay-density}, and processes the merged
stream in time order: it resolves the muon veto, opens and closes the coincidence
windows, imposes the dead time, and tallies the ordered \(s,e,n\) sequences and
their \(\Delta t\) distributions.  One stream of \(2\times10^{13}\)
events is generated per source-rate group and dead-time campaign, and
both \(T_c\) selections are applied to it.

\section{Exact rate derivation}
\label{sec:exact-derivation}

The calculation separates two kinds of bookkeeping.  Across windows, it
tracks the delayed events that were created earlier and are still
pending.  Inside the current window, it tracks the ordered sequence of
recorded events.  The first task is a Markov history chain, and the
second is a current-window propagator.

The quantity that couples one coincidence window to the next is the
population of pending neutrons alive at the reset seam, and an exact
treatment must sum its contributions from all prior windows.  The
natural way to do that is to promote the pending-neutron count \(h\)
to a discrete state vector and to introduce a transition kernel \(Q\)
(a transition rule represented as a matrix)
that propagates one inter-window gap-start opportunity through a completed
window to the next such opportunity.  The Neumann series
\((I-Q^T)^{-1}=\sum_{n=0}^{\infty}(Q^T)^n\) then sums over zero, one,
two, and more completed windows between the most recent reset seam and
the current opportunity; the population inherited from before that seam
enters through the seam seed \(p_0\).  This closed-form, all-orders cross-seam memory is the
structure that any treatment conditioning on a bounded backward
history must truncate (Sec.~\ref{subsec:why-approximation}).  The
physics of the history chain is
the subject of Sec.~\ref{sec:pending-physics}, and its solution is the
subject of Sec.~\ref{sec:history-chain}.  Inside each accepted window,
the time-ordered multiplicity integrals are then evaluated in closed
form by the current-window propagator of Sec.~\ref{sec:propagator},
written explicitly here for window depth \(k\in\{1,2,3\}\), the fold
number of the requested channel, and for
arbitrary \(K\)-component lifetime mixtures.  Higher \(k\) requires only
a longer block-bidiagonal chain and sufficient current-state headroom.

The transparent factorized
counterpart of this trigger/start/within-window/tail structure is
developed as a benchmark in Sec.~\ref{sec:event-placement}.  The
propagator keeps old and self neutrons in separate state slots,
carries unrecorded births and captures through event dead time, and
retains the full coupling among muons, singles, prompts, and neutrons
without linearizing in \(R_{\rm corr}\).  Throughout, \emph{exact}
refers to the analytical solution of the stochastic model specified in
Sec.~\ref{sec:model} on the retained history and current-window state spaces;
the two state-space truncations are controlled in
Sec.~\ref{subsec:truncation-contract}, and the two independent numerical
evaluations of the solution are compared in Sec.~\ref{subsec:prd-rates}.  The tools
involved (phase-type delay laws, sub-stochastic transition kernels,
matrix exponentials, Palm calculus) are standard in applied
probability but less common in particle physics;
Sec.~\ref{subsec:toolkit} explains why they are the natural language
for this problem and reviews their core concepts, and readers already
familiar with matrix-analytic methods can pass directly to
Sec.~\ref{sec:pending-physics}.

\subsection{Choice of framework and mathematical background}
\label{subsec:toolkit}

The rates of interest are ordered and conditioned: a \(k\)-fold rate
asks how often a window closes with exactly \(k\) recorded members
forming a prescribed ordered species sequence, under the veto and reset
boundaries of Sec.~\ref{sec:model} and with a pending-neutron
population inherited from earlier windows.  Once the reset-segment
construction has removed the veto intervals, every remaining source of
randomness in the model is an independent Poisson stream, an
independent discrete mark (daughter detection, component choice), or a
constant-hazard capture clock of the mixture~(\ref{eq:delay-density});
the spans \(T_c\) and \(T_0\) are fixed evolution times.  A system
built from competing memoryless clocks is a continuous-time Markov
chain in its occupancy configuration, here the vector of
pending-neutron counts per lifetime component, and for such chains
ordered multi-event probabilities, boundary-state populations, and
stationary intensities all have closed linear-algebra forms on the
retained finite state spaces.  Adopting this representation therefore
involves no expansion in \(R_{\rm corr}\), occupancy, or window
length: it is a change of representation, not an approximation.  The paragraphs
below review the tools this choice brings in and point to where each
acts.

\paragraph{Markov chains and matrix exponentials.}
A finite-state continuous-time Markov chain is specified by a
generator matrix \(A\): the off-diagonal entry \(A_{\alpha\beta}\) is
the rate of \(\alpha\to\beta\) transitions, and each diagonal entry is
minus the total exit rate of its state.  The occupation probabilities,
collected in a row vector \(p\), obey the master equation
\(\dot p=pA\), the same rate-equation formalism that describes a
radioactive decay chain, and the finite-time solution is the matrix
exponential \(p(t)=p(0)\,e^{At}\), the evolution operator of the
chain~\cite{MolerVanLoan2003,Higham2008}.  In this paper the two
generators \(A_{\rm vis}\) and \(A_{\rm blind}\) of
Sec.~\ref{sec:propagator} encode how the pending population evolves
while the detector is live and while it is blind; the active generator
is a \emph{sub}-generator, meaning that its propagator carries the
probability that no recorded event has yet occurred, and each recorded
event re-enters the evolution through an explicit event matrix.

\paragraph{Phase-type delay laws.}
For this calculation, the phase-type representation is a finite set of
capture clocks: each detected neutron draws a component \(i\) at
birth, which fixes the capture-time channel it follows, and then
captures with constant hazard \(1/\tau_i\).  The structural payoff is
memorylessness per component:
conditional on how many pending neutrons of each component exist,
their elapsed ages are irrelevant to the future, so the pending
population enters the calculation only through the count vector
\(h=(h_1,\ldots,h_K)\).  This is the property that closes the backward
history into the finite Markov state of
Sec.~\ref{sec:pending-physics}.  More generally, a phase-type
distribution is the absorption-time law of a finite
Markov chain; the family is dense, in the sense of weak convergence,
in the distributions on the positive half-line, and it is the standard
device for making a general waiting-time law exactly computable by
matrix calculus~\cite{Neuts1981,BladtNielsen2017}.  The capture
density~(\ref{eq:delay-density}) is the diagonal member of this
family, a mixture of pure exponentials.

\paragraph{Ordered windows as time-ordered products.}
The matrix product follows the detector event loop: evolve while no
recorded event occurs, insert the matrix for the next recorded
species, and repeat until the requested multiplicity is reached.
Inside one accepted window this alternation of free evolution
with recorded events has the structure of a time-ordered (Dyson-type)
expansion in time-dependent perturbation theory: the active propagator
\(V(t)=e^{A_{\rm vis}t}\) plays the free evolution, an event matrix
acts as the insertion for one recorded event, and a fold-\(k\) rate is
the term with one insertion per recorded follower, integrated over the
ordered arrival times.  Unlike a perturbative series, the expansion
terminates, because a fold-\(k\) channel contains exactly \(k\)
recorded events, and each ordered integral is evaluated in closed form
as one block of a larger matrix exponential (the Van Loan
construction~\cite{VanLoan1978,AlMohyHigham2011}), so neither
quadrature nor truncation in the number of insertions enters.  This
machinery is developed in Sec.~\ref{sec:propagator}.

\paragraph{Cross-window memory as a geometric series.}
Between consecutive reset seams the axis carries a random number of
completed windows, each updating the pending population.  Encoding one
inter-window gap plus one completed window as a transition kernel
\(Q\) on count vectors turns the sum over all intervening completed
windows into a Neumann series in \(Q\), the same all-orders geometric
resummation familiar from propagator expansions.  Convergence is
automatic because \(Q\) is \emph{sub-stochastic}: each row sums to at
most \(e^{-R_\mu T_c}<1\), the deficit collecting the ways the step is
instead terminated by the next reset, so long window chains between
seams are exponentially suppressed (Sec.~\ref{subsec:general-KN}).
This resummed memory is precisely the structure that a calculation
conditioning on a bounded backward history must truncate
(Sec.~\ref{subsec:why-approximation}).

\paragraph{Seam anchoring: Palm calculus and the reset process.}
Two point-process facts complete the toolkit.  First, a rate needs an
anchor.  The stationary visit intensities of
Sec.~\ref{sec:history-chain} are obtained by conditioning the
stationary process on a seam at the time origin, which defines the
Palm law of the seam process, and multiplying the expected number of
visits per seam by the seam intensity; the Campbell--Palm
formula~\cite{CoxIsham1980} is the rigorous form of this
renewal-reward logic.  In detector terms, the Palm construction means
choosing a typical reset seam, placing it at time zero, and counting
the window-opening opportunities before the next seam.  Second, the
seam process itself must be
identified.  Overlapping veto excisions merge exactly like overlapping
service intervals of an infinite-server (M/G/\(\infty\)) queue, and
the busy-period regeneration of that queue implies that the
stitched-axis seams again form a Poisson process at the raw intensity
\(R_\mu\); this is the reset-cluster lemma already stated in
Sec.~\ref{sec:model}.

Finally, the two comparison routes used in this paper retain
truncation or finite-sample uncertainties of their own.  Explicit
enumeration of event placements truncates the backward history
(Sec.~\ref{sec:event-placement} develops the transparent version of that
route; at \(T_0=0\) its relative residuals reach \(5\times10^{-5}\) on
the central pair channel and \(8\times10^{-3}\) on the most
history-sensitive one, and at nonzero \(T_0\) the omitted blind-interval
births produce larger, channel-dependent discrepancies, quantified in
Sec.~\ref{subsec:why-approximation}), and a Monte Carlo sample of size
\(N\) carries a statistical uncertainty scaling as \(N^{-1/2}\)
(Sec.~\ref{sec:toymc}).

\subsection{Physics of the pending-neutron chain}
\label{sec:pending-physics}

We build the chain from its simplest non-trivial case before giving the
general transition matrix.

\subsubsection{Worked two-state chain and Markov closure}
\label{subsec:decouples}

For one lifetime component and \(N_{\max}=1\), the history states are
\(h\in\{0,1\}\): \(h=0\) means that no old neutron is awaiting capture,
and \(h=1\) that exactly one is.  Define
\begin{align}
 d_h&=R_\mu+R_s+R_{\rm corr}+\frac{h}{\tau_1},\nonumber\\
 F&=e^{-R_\mu T_c},\qquad \sigma=e^{-T_c/\tau_1},
\end{align}
and
\begin{equation}
 \alpha=\epsilon_n f_1\sigma,\qquad
 \nu_1=R_{\rm corr}\epsilon_n f_1\tau_1(1-\sigma).
\end{equation}
Here \(F\) requires the history window to close before the next reset;
\(\sigma\) thins the old population; \(\alpha\) is the probability that the
opening \(e\) adds a daughter surviving to window close; and \(\nu_1\) is the
mean number of surviving daughters born inside the window.  Combining these
independent mechanisms gives
\begin{align}
 Q_{00}&=\frac{F e^{-\nu_1}}{d_0}
   \bigl[R_s+R_{\rm corr}(1-\alpha)\bigr],\nonumber\\
 Q_{01}&=\frac{F e^{-\nu_1}}{d_0}
   \Bigl[R_s\nu_1+R_{\rm corr}
   \{\alpha+(1-\alpha)\nu_1\}\Bigr],\nonumber\\
 Q_{10}&=\frac{F e^{-\nu_1}}{d_1}
   \Bigl[R_s(1-\sigma)
   +R_{\rm corr}(1-\sigma)(1-\alpha)\nonumber\\[-2pt]
   &\hspace{31mm}+\tau_1^{-1}\Bigr],\nonumber\\
 Q_{11}&=\frac{F e^{-\nu_1}}{d_1}
   \Bigl[R_s\{\sigma+(1-\sigma)\nu_1\}\nonumber\\[-2pt]
   &\quad+R_{\rm corr}\{\sigma(1-\alpha)+(1-\sigma)\alpha
   +(1-\sigma)(1-\alpha)\nu_1\}\nonumber\\[-2pt]
   &\quad+\nu_1/\tau_1\Bigr].
 \label{eq:worked-Q}
\end{align}
In words, \(Q_{01}\) fills the pending pool, \(Q_{10}\) empties it, and
\(Q_{11}\) keeps one pending neutron through several distinct
microscopic paths.
For example, the three correlated-source terms in \(Q_{11}\) say that
exactly one pending neutron remains because the old neutron alone survives,
the opener's daughter alone survives, or both disappear and one
window-internal daughter survives.  With \(p_0\) the pending-state
distribution at a reset seam (the reset-seam seed, given in closed form
by Eq.~(\ref{eq:seam-means})), the
visit-weight vector \(w=(w_0,w_1)^T\), the finite-cap approximation to
the long-run rate at which window-opening opportunities occur in each
state (Sec.~\ref{sec:history-chain}), obeys
\begin{align}
 (I-Q^T)w&=R_\mu p_0,\nonumber\\
 D&=(1-Q_{00})(1-Q_{11})-Q_{01}Q_{10},\nonumber\\
 w_h&=\frac{R_\mu}{D}
 \left[(1-Q_{1-h,1-h})p_{0h}
       +Q_{1-h,h}p_{0,1-h}\right].
 \label{eq:worked-stationary}
\end{align}
Thus \(w_1/w_0=O(R_{\rm corr}\epsilon_n\tau_1)\): the occupied history
state is suppressed by the small stationary pending population.

This two-state example also displays why the chain closes.  Singles, IBD
prompts, and resets are independent Poisson processes, while every daughter
is assigned a lifetime component at birth and then follows an independent
exponential capture clock.  Conditional on
\(h=(h_1,\ldots,h_K)\), the only inherited rate is therefore the
pure-death intensity \(\sum_i h_i/\tau_i\); all future source increments are
independent.  The component-wise birth and thinning kernels factorize before
the common cap is imposed, but the slots remain coupled through the shared
event race \(d_h\), reset termination, and the total-population cap.  The
Markov closure uses four model assumptions: (i) the \(s\), \(e\), and
seam streams are mutually independent homogeneous Poisson processes;
(ii) daughter production is an independent zero-or-one Bernoulli mark;
(iii) each detected daughter receives at birth a fixed label from the
finite exponential mixture~(\ref{eq:delay-density}) and thereafter has
hazard \(1/\tau_i\); (iv) the inter-window dead-time rule can be
integrated into \(Q\) without carrying an additional residual-blindness
state.  The displayed \(Q\) additionally assumes the window-close
non-paralyzable convention.  A continuing \(Q\) step closes before the
next seam; a window reaching the seam is terminal, and it is accepted
because it closes before the physical reset (Sec.~\ref{sec:model}).
Under these assumptions, \(h\) contains all information from the
backward history needed by the next step.
\subsubsection{General \texorpdfstring{\(K\)}{K},
  \texorpdfstring{\(N_{\max}\)}{Nmax}}
\label{subsec:general-KN}

With \(K\) lifetime components and maximum total pending \(N_{\max}\), the
state space is all tuples \(h=(h_1,\ldots,h_K)\) with \(\sum_i h_i\le N_{\max}\)
and dimension \(\binom{N_{\max}+K}{K}\)~\cite{Neuts1981,LatoucheRamaswami1999}.  The general transition entry is
\begin{equation}
  Q_{h,h'} = \frac{e^{-R_\mu T_c}}{d_h}
              \sum_{x\in\mathcal X(h)} R_x\; W_x(h,\,h'),
\end{equation}
where \(\mathcal X(h)=\{s,e\}\cup\{n_i:\,h_i\ge1\}\) (capture sums
here and in every capture operator below run over components with
positive occupancy), \(R_x\) is the rate of event type \(x\) in state
\(h\) (\(R_{n_i} = h_i/\tau_i\)), and \(W_x(h,h')\) is the probability of transitioning
from \(h\) to \(h'\) over the muon-free window of length \(T_c\) starting with
event type \(x\) (every accepted window is muon-free; the prefactor
\(e^{-R_\mu T_c}\) is the probability that the window also closes before
the next reset seam, making it a continuing step of the chain rather than
a terminal window).  Its closed form is most compact as a coefficient of a
multivariate probability-generating function: the notation
\([\mathbf z^{\,h'}]\), used below, takes the coefficient of
\(z_1^{h_1'}\cdots z_K^{h_K'}\), which is the probability that the
window ends with pending state \(h'\).  Define
\begin{align}
 \sigma_i&=e^{-T_c/\tau_i},\nonumber\\
 \nu_i&=R_{\rm corr}\epsilon_n f_i\tau_i(1-\sigma_i),\nonumber\\
 \widehat h_i&=h_i-\mathbf1_{\{x=n_i\}},
\end{align}
and
\begin{equation}
 S_x(\mathbf z)=
 \begin{cases}
  \begin{aligned}
   1&-\epsilon_n\sum_i f_i\sigma_i\\[-2pt]
    &+\epsilon_n\sum_i f_i\sigma_i z_i,
  \end{aligned}&x=e,\\
  1,&x\ne e .
 \end{cases}
\end{equation}
Then
\begin{equation}
\begin{split}
 W_x(h,h')
 ={}&[\mathbf z^{\,h'}]\,
 S_x(\mathbf z)
 \prod_i(1-\sigma_i+\sigma_i z_i)^{\widehat h_i}\\
 &\times
 \exp\!\left[\sum_i\nu_i(z_i-1)\right].
\end{split}
 \label{eq:history-transition-pgf}
\end{equation}
The three factors are, respectively, a possible daughter of the opening
\(e\), binomial thinning of the old population after removal of a triggering
\(n_i\), and independent Poisson populations of window-internal daughters
that survive to the close.  Numerical assembly enumerates only retained
tuples \(h'\), extracts these finite-cap coefficients, builds the sparse
matrix \(Q\), and solves Eq.~(\ref{eq:prd-visit-weights}).

Two structural properties make the chain well-posed, and both follow
directly from this form.  First, \(Q\) is \emph{sub-stochastic}: before
truncation the \(W_x(h,\cdot)\) are probability distributions over
arrival states, so each row of \(Q\) sums to exactly
\(e^{-R_\mu T_c}\,(d_h - R_\mu)/d_h \le e^{-R_\mu T_c} < 1\), with
\(d_h = R_\mu + \sum_x R_x\) the total exit rate of the inter-window
gap in state \(h\) (recall \(R_\mu > 0\)).  The deficit collects the
three ways a history step terminates: the gap ends at a reset seam
rather than at the next window-opening event, with probability
\(R_\mu/d_h\); the window opens but closes past the next seam, a
further factor \(1-e^{-R_\mu T_c}\); and, once the state space is
truncated, steps whose next gap-start state lies outside
\(\sum_i h_i' \le N_{\max}\).  The lost mass of this last class,
\(\delta_{\rm hist}(h)\equiv\sum_{\sum_ih_i'>N_{\max}}(Q_\infty)_{hh'}\),
is the computable truncation-overflow defect of the row.  Written out, the
row's full probability accounting reads
\begin{equation}
\begin{split}
  1 \;=\;{}&
  \underbrace{\frac{R_\mu}{d_h}}_{\substack{\text{reset before}\\\text{trigger}}}
  \;+\;
  \underbrace{\frac{d_h-R_\mu}{d_h}\bigl(1-e^{-R_\mu T_c}\bigr)}_{\substack{\text{trigger opens;}\\\text{seam before close}}}\\[2pt]
  &+\;
  \underbrace{e^{-R_\mu T_c}\frac{d_h-R_\mu}{d_h}}_{\text{row sum of }Q\text{ (pre-truncation)}},
\end{split}
\label{eq:row-accounting}
\end{equation}
an identity in \(d_h\), \(R_\mu\), and \(T_c\) alone that holds row by
row regardless of \(N_{\max}\); truncation then further splits the
last term into the kept mass and the row's overflow defect
\(\delta_{\rm hist}(h)\), so that the retained row sum of \(Q\) is
\(e^{-R_\mu T_c}(d_h-R_\mu)/d_h - \delta_{\rm hist}(h)\) and the total
row deficit is the sum of the first two terms of
Eq.~(\ref{eq:row-accounting}) and \(\delta_{\rm hist}(h)\).
Second, sub-stochasticity makes
the visit-weight system~(\ref{eq:prd-visit-weights}) uniquely solvable:
the row-sum bound gives \(\lVert Q \rVert_\infty \le e^{-R_\mu T_c} < 1\),
so the spectral radius of \(Q\) (and of \(Q^T\)) is below one, the
Neumann series \((I-Q^T)^{-1} = \sum_{n\ge0} (Q^T)^n\) converges, and
the stationary weights exist, are unique, and are non-negative because
every term of the series applied to the non-negative source
\(R_\mu p_0\) is non-negative.  Thus the coefficient construction above,
sub-stochasticity, and the existence, uniqueness, and positivity of \(w\)
complete the history-chain specification.

\paragraph{Scope: the other two dead-time conventions.}
Readers using only the default window-close convention may skip this
paragraph; its key point is that the global non-paralyzable convention
changes only the history kernel \(Q\), while the paralyzable convention
requires a larger state description.
Everything above, and every generator of
Sec.~\ref{sec:propagator}, is built under the window-close convention
assumed for the displayed \(Q\) in Sec.~\ref{subsec:decouples}: dead
time re-arms at window close and never leaks past the seam.  Under the
strictly global non-paralyzable convention, exactly one object of the
construction changes: the history-chain transition matrix \(Q\).  A
recorded event within \(T_0\) of the window edge leaves a residual
blind stretch that now survives into the following inter-window gap
instead of being clipped there, so the gap's live time, and hence the
supply of the next window's opener, is reduced; the within-window
propagator \(G^{(k)}\), the event matrices, and the trigger embedding
of Sec.~\ref{sec:propagator} are untouched, since the change is entirely
a property of what happens \emph{between} windows.  The modified
kernel still has every off-diagonal entry non-negative and a row sum
that cannot exceed the window-close row sum of
(\ref{eq:row-accounting}) (added blindness can only remove
window-opening probability, never add it), so \(\lVert Q^{\rm G}
\rVert_\infty \le e^{-R_\mu T_c}<1\) and every conclusion of this
subsection, sub-stochasticity and the existence, uniqueness, and
positivity of \(w\), carries over unchanged; only the numerical value
of \(w\) differs.  This is the mechanism, quantified in
Sec.~\ref{subsec:prd-conventions}, behind the global non-paralyzable
convention's \(1.89\times10^{-2}\) relative depletion of the \(n\),
\(ne\), and \(ns\) rates at \(T_0=T_c/2\) (\(R_{\rm corr}=\SI{0.2}{Hz}\),
\(T_c=\SI{400}{\micro s}\)).  Under the strictly global paralyzable
convention a second change stacks on top of the first: every good
event, recorded or not, restarts the blind interval, so the exit rate
\(d_h\) no longer summarizes the gap's competition on its own: a
rejected, blind-arriving event still resets the clock without
appearing in \(h\).  A single finite occupancy state can no longer
carry this information, and the construction is replaced by an
explicit sum over accept/reject mark sequences within a window, whose
number of terms grows exponentially with \(T_c/T_0\) for \(T_0>0\),
rather than polynomially as for the first two conventions; the
validation grid therefore evaluates this convention at
\(T_c/T_0\le4\) (Sec.~\ref{subsec:prd-conventions}).  This paper proves the
window-close construction and states exactly what changes for the other
conventions; Sec.~\ref{subsec:prd-conventions} summarizes the matched
toy-Monte-Carlo checks for all three.

\subsection{Backward pending-neutron chain}
\label{sec:history-chain}

The history state \(h=(h_1,\ldots,h_K)\) is the old pending-neutron
population at a window-opening opportunity at the start of an inter-window
gap, either immediately after a reset seam or after a completed history
window, truncated to \(\sum_i h_i\le N_{\max}\).  Reset seams supply the
boundary source for the chain.  With stationary cross-seam initialization,
the reset pending
distribution is the product of Poisson distributions with means
\begin{equation}
  \mu_i=R_{\rm corr}\epsilon_n f_i\tau_i ,
  \qquad
  p_{0,\infty}(h)=\prod_i e^{-\mu_i}\frac{\mu_i^{h_i}}{h_i!},
  \label{eq:seam-means}
\end{equation}
and the calculation uses its conditional restriction to the retained
states \(\mathcal H_N=\{h:\sum_ih_i\le N\}\) at history cap \(N\)
(production takes \(N=N_{\max}\)),
\((p_{0,N})_h=p_{0,\infty}(h)/Z_N\) with
\(Z_N=\sum_{h\in\mathcal H_N}p_{0,\infty}(h)\), together with the
killed principal restriction \(Q_N\) of the untruncated kernel
\(Q_\infty\) to \(\mathcal H_N\).  The stationary
opportunity-visit intensities \(w_h\), with units of rate, solve
\begin{equation}
  (I-Q^T)w=R_\mu p_0 ,
  \label{eq:prd-visit-weights}
\end{equation}
where \((Q,p_0,w)\) stands for \((Q_\infty,p_{0,\infty},w_\infty)\) on
the untruncated space and for \((Q_N,p_{0,N},w_N)\) in the calculation.
These weights are the only information from the previous windows that
the current-window calculation consumes.

\textbf{Physical interpretation.}
A useful picture is a ledger of neutrons that have been produced but
have not yet captured: the ledger at a gap-start opportunity is the
only information from earlier windows needed by the next window.
The history state \(h\) counts the \emph{old} pending neutrons.  At a
gap-start opportunity these are all pending daughters that cannot be the
self partner of a prompt recorded in the current window, including
daughters born in earlier windows and in excised stretches; during the
current window, daughters of prompts that fire while the detector is
blind are likewise assigned to the old slots.  By contrast \emph{self} neutrons,
introduced in Sec.~\ref{sec:propagator}, are born from IBD events
recorded inside the current window.  Old and self daughters of component \(i\) share the same
residual hazard \(1/\tau_i\), so the two slot families are not needed
for the survival law; they preserve parentage, so that the capture
operators \(E_n^{\rm old}\) and \(E_n^{\rm self}\) of
Sec.~\ref{sec:propagator} implement the True/False projection of the
\(en\) channel.  The doubled state space is the minimal extension that
carries this distinction.  Readers interested in the rate construction rather than
the point-process proof may take Eq.~(\ref{eq:prd-visit-weights}) as
the seam-anchored balance equation and continue to
Sec.~\ref{sec:propagator}.

\emph{Palm-intensity derivation.}  The following argument makes the
seam-anchored renewal-reward logic of Sec.~\ref{subsec:toolkit}
precise.  Let \(\{S_j\}\) be the stationary reset-seam
point process on the reset-segment axis.  By the reset-cluster lemma of
Sec.~\ref{sec:model}, it has intensity \(R_\mu\) and is independent of the
source streams.  Under its Palm law \(\mathbb P^0\), place a seam at
\(S_0=0\), let \(S_1\) be the next seam, and let \(M_h[S_0,S_1)\) count
window-opening opportunity-state visits in state \(h\) between them,
including the visit immediately after \(S_0\) and each visit after a
completed history window.  The Campbell--Palm intensity
formula~\cite{CoxIsham1980} gives
\begin{equation}
  w_h=R_\mu\,\mathbb E^0\!\left[M_h[S_0,S_1)\right].
  \label{eq:palm-visit-intensity}
\end{equation}
The pending state at the Palm seam has marginal distribution
\(p_{0,\infty}\); the argument is written for the untruncated triple
\((Q,p_0,w)=(Q_\infty,p_{0,\infty},w_\infty)\), and its finite-cap
version substitutes \((Q_N,p_{0,N},w_N)\).
Conditional on that state, future source increments are independent and the
within-segment evolution is Markov with substochastic step \(Q\).  Therefore
the first opportunity visit is in state \(h\) with probability \((p_0)_h\),
the second with probability \([Q^Tp_0]_h\), the third with probability
\([(Q^T)^2p_0]_h\), and so on.  Hence
\begin{equation}
  \begin{aligned}
    \mathbb E^0\!\left[M_h[S_0,S_1)\right]
    &=\bigl[(I+Q^T+(Q^T)^2+\cdots)\,p_0\bigr]_h \\
    &=\bigl[(I-Q^T)^{-1}p_0\bigr]_h ,
  \end{aligned}
  \label{eq:renewal-reward-visits}
\end{equation}
where the Neumann-series argument of Sec.~\ref{subsec:general-KN} supplies both
convergence and finite mean reward.  Substitution into
(\ref{eq:palm-visit-intensity}) yields
\begin{equation}
  w = R_\mu\,(I-Q^T)^{-1}p_0
    \quad\Longleftrightarrow\quad
    (I-Q^T)w = R_\mu\,p_0 ,
\end{equation}
recovering~(\ref{eq:prd-visit-weights}).  On the untruncated state
space, \(w_\infty\) is exactly the long-run intensity, per unit
reset-segment time, of window-opening opportunity-state visits in state
\(h\); at finite \(N_{\max}\), \(w_N\) is its killed finite-cap
approximation with the conditional seed \(p_{0,N}\), and
Sec.~\ref{subsec:truncation-contract} bounds \(w_\infty-\widetilde w_N\).
Neither is the rate of windows that actually open.  This argument requires stationarity and
ergodicity, the Palm marginal \(p_0\), and Markov evolution conditional
on that mark.  It does not require pending-state marks at adjacent seams
to be independent.

\subsection{Current-window propagator}
\label{sec:propagator}

The master formula of this paper follows the logic of a detector event
loop: a rate of available history states, a selection of the species
that opens the window, a propagation of the requested recorded
followers, and a sum over what remains unobserved at the close.
When the visit weights of Sec.~\ref{sec:history-chain} are combined with the
within-window matrix constructed in this section, every \(k\)-fold
coincidence rate computed in this paper takes the matrix master form
\begin{equation}
  R^{(k)}_{c_1 \cdots c_k}
   \;=\; w^{\!\top} J_{c_1}\,
         G^{(k)}(c_2, \ldots, c_k)\, \mathbf{1},
  \label{eq:exact-master}
\end{equation}
where \(w\) is the gap-start opportunity-visit intensity vector of
(\ref{eq:prd-visit-weights}), living on the history states
\(\{h:\sum_i h_i\le N_{\max}\}\); \(J_{c_1}\) is the \emph{trigger
embedding operator}, a rectangular matrix from the history space to the
doubled (old/self) current space; \(G^{(k)}\) is the within-window
propagator on the doubled space; and \(\mathbf{1}\) is the all-ones
column vector that contracts the post-window state distribution to a
scalar.  Writing \(\iota(h)\) for the doubled state with old occupancy
\(h\) and no self neutrons, and \(d_h\) for the total exit rate of the
inter-window gap in state \(h\), the embedding rows are
\begin{align}
  (J_s)_{h,\alpha} &= \frac{R_s}{d_h}\,
      \delta_{\alpha,\,\iota(h)},\\
  (J_e)_{h,\alpha} &= \frac{R_{\rm corr}}{d_h}
      \Bigl[(1-\epsilon_n)\,\delta_{\alpha,\,\iota(h)}
      +\epsilon_n\textstyle\sum_i f_i\,
        \delta_{\alpha,\,\iota(h)+e_i^{\rm self}}\Bigr],\\
  (J_n)_{h,\alpha} &= \frac{1}{d_h}\textstyle\sum_i
      \dfrac{h_i}{\tau_i}\,
      \delta_{\alpha,\,\iota(h)-e_i^{\rm old}}.
  \label{eq:trigger-embedding}
\end{align}
The actual species-resolved and total window-opening intensities are
\begin{equation}
  \Lambda_c=w^{\!\top}J_c\mathbf{1},\qquad
  \Lambda_{\rm open}
   =w^{\!\top}(J_s+J_e+J_n)\mathbf{1}.
  \label{eq:opener-intensities}
\end{equation}
Thus \(w^{\!\top}J_{c_1}\), rather than \(w^{\!\top}\) alone, is the
rate-weighted current-state
distribution at the instant the window opens: the gap ends with an
event of species \(c_1\) in competition with the other exit channels of
\(d_h\); a triggering \(e\) deposits its own self neutron already at
this step, while recorded followers add theirs through the event
matrices; and a triggering \(n\) removes the captured old neutron.  At
nonzero dead time the leading blind factor \(e^{A_{\rm blind}T_0}\) is
the first factor of \(G^{(k)}\), acting immediately after the
embedding.  The construction below specifies \(G^{(k)}\)
through the active and blind generators \(A_{\rm vis}\), \(A_{\rm blind}\)
and the event matrices
\(\{E_s, E_e, E_n^{\rm old}, E_n^{\rm self}\}\), and shows that
\(G^{(k)}\) reduces in turn to a single block-matrix exponential plus
boundary terms for \(k=1\), \(k=2\), and \(k=3\).
Equation~(\ref{eq:exact-master}) is the matrix generalization of the
scalar four-factor form (\ref{eq:event-placement-skeleton}): the row
vector \(w^{\!\top}J_{c_1}\) plays the role of trigger weight and
window-start distribution, the within-window
exponential plays the role of the propagator, and the all-ones column
\(\mathbf{1}\) plays the role of the tail projection, the sum over the
final unobserved states.

\emph{One-state sanity check.}  For \(\epsilon_n=0\), so that only
the state \(h=0\) is populated, let
\(R_p=R_s+R_{\rm corr}\).  Then \(p_0=1\),
\(d_0=R_\mu+R_p\), and
\(Q_{00}=e^{-R_\mu T_c}R_p/(R_\mu+R_p)\), so
\begin{equation}
  \begin{aligned}
    w_0&=\frac{R_\mu(R_\mu+R_p)}
              {R_\mu+R_p-R_pe^{-R_\mu T_c}},\\
    \Lambda_{\rm open}
      &=\frac{R_\mu R_p}
              {R_\mu+R_p-R_pe^{-R_\mu T_c}}.
  \end{aligned}
  \label{eq:k0-opener-check}
\end{equation}
At \(R_\mu=\SI{200}{Hz}\), \(R_p=\SI{55}{Hz}\), and
\(T_c=\SI{1.5}{\milli\second}\), these give
\(w_0=\SI{238.034}{Hz}\) and
\(\Lambda_{\rm open}=\SI{51.3407}{Hz}\).  The latter, obtained by
the factor \(R_p/(R_\mu+R_p)\) in
Eq.~(\ref{eq:opener-intensities}), is the actual total opening
intensity.

Inside the current window we double the state vector.  Old neutrons and
self neutrons, born from an \(e\) recorded in the current window, are stored
separately: a current state
\(\alpha=(o_1,\ldots,o_K;\,m_1,\ldots,m_K)\) carries \(o_i\) old and
\(m_i\) self pending neutrons of lifetime component \(i\), truncated to
the total population \(\sum_i(o_i+m_i)\le C\) with
\begin{equation}
  C = N_{\max}+H,
  \label{eq:current-cap}
\end{equation}
a state space of dimension \(\binom{C+2K}{2K}\).  The independent current
headroom \(H\) absorbs pending neutrons added after the window opens: the
trigger's own self neutron, those of recorded followers, and blind-interval
births.  We use \(H=3\), which preserves the
\(C=N_{\max}+3\) construction, but \(H\) can now be raised without changing
the history cap.  Any transition that would leave the truncated set is
dropped; the dropped mass is the current-window row defect quantified in
Sec.~\ref{subsec:truncation-contract}.

The active generator \(A_{\rm vis}\) has only an exit-rate
diagonal,
\begin{equation}
  (A_{\rm vis})_{\alpha\alpha}
  =-\left(R_s+R_{\rm corr}+\lambda_{\rm old}(\alpha)
  +\lambda_{\rm self}(\alpha)\right),
  \label{eq:avis}
\end{equation}
with \(\lambda_{\rm old}(\alpha)=\sum_i o_i/\tau_i\) and
\(\lambda_{\rm self}(\alpha)=\sum_i m_i/\tau_i\), because any visible
event leaves the current active interval.  Which visible event ended
the interval, and what it does to the state, is carried by the event
matrices (\(e_i^{\rm old}\), \(e_i^{\rm self}\) denote unit occupancy
increments of the two slot families):
\begin{align}
  (E_s)_{\alpha\beta} &= R_s\,\delta_{\beta,\alpha},
  \nonumber\\
  (E_e)_{\alpha\beta} &= R_{\rm corr}(1-\epsilon_n)\,
      \delta_{\beta,\alpha}
      +R_{\rm corr}\,\epsilon_n\textstyle\sum_i f_i\,
       \delta_{\beta,\,\alpha+e_i^{\rm self}},
  \nonumber\\
  (E_n^{\rm old})_{\alpha\beta} &= \textstyle\sum_i
      \dfrac{o_i}{\tau_i}\,\delta_{\beta,\,\alpha-e_i^{\rm old}},
  \nonumber\\
  (E_n^{\rm self})_{\alpha\beta} &= \textstyle\sum_i
      \dfrac{m_i}{\tau_i}\,\delta_{\beta,\,\alpha-e_i^{\rm self}},
  \label{eq:event-matrices}
\end{align}
with \(E_n=E_n^{\rm old}+E_n^{\rm self}\) when the origin ledger is
not consulted.  A recorded \(s\) changes nothing; a recorded \(e\)
adds its own neutron to a self slot with probability
\(\epsilon_n f_i\), or nothing with probability \(1-\epsilon_n\); a
recorded \(n\) removes the captured neutron from the slot family it
occupied.  Event-matrix rows, like generator rows, drop self-raising
entries beyond the cap~(\ref{eq:current-cap}).

During event dead time the detector records nothing while the pending
population keeps evolving.  The blind generator has off-diagonal
entries
\begin{align}
  (A_{\rm blind})_{\alpha\beta}
   ={}& R_{\rm corr}\,\epsilon_n \textstyle\sum_i f_i\,
       \delta_{\beta,\,\alpha+e_i^{\rm old}}
     + \textstyle\sum_i\dfrac{o_i}{\tau_i}\,
       \delta_{\beta,\,\alpha-e_i^{\rm old}}
  \nonumber\\
   &+ \textstyle\sum_i\dfrac{m_i}{\tau_i}\,
       \delta_{\beta,\,\alpha-e_i^{\rm self}}
  \qquad(\beta\ne\alpha),
  \label{eq:ablind}
\end{align}
and diagonal
\((A_{\rm blind})_{\alpha\alpha}
 =-(R_{\rm corr}\epsilon_n+\lambda_{\rm old}+\lambda_{\rm self})\).
The bookkeeping is deliberately asymmetric: an IBD prompt arriving
inside a blind interval is not recorded, so its surviving neutron
joins the \emph{old} population (it can never be claimed as the
recorded partner of a recorded prompt), while captures inside the
blind interval remove their pending neutron without producing a
recorded \(n\).  Uncorrelated singles and neutronless IBD events
change no state and leave no record, so they do not appear in
\(A_{\rm blind}\) at all.  On the untruncated state space every
off-diagonal entry is balanced by the diagonal, whereas a finite cap drops
outward birth entries.  Thus
\begin{equation}
  e^{A^{(\infty)}_{\rm blind}s}\,\mathbf{1}=\mathbf{1},
  \qquad
  e^{A^{(C)}_{\rm blind}s}\,\mathbf{1}\leq\mathbf{1},
  \label{eq:blind-ones}
\end{equation}
where the finite-cap deficit is precisely the probability mass lost through
the dropped birth entries.

\paragraph{Explicit kernels for one, two, and three recorded events.}
Write \(V(t)=e^{A_{\rm vis}t}\) for the active propagator,
\(B=e^{A_{\rm blind}T_0}\) for the full blind factor, and
\(L=T_c-T_0\) for the active span remaining after the trigger's own
blind interval.  The assembly rule is mechanical: reading the sequence
left to right, every recorded event contributes its event matrix
followed by a blind factor \(B\) clipped at the window edge; between
recorded events the state evolves with \(V\); at the window edge the
state is contracted against \(\mathbf{1}\).  After the terminal recorded
event, no later record or post-window state is queried.  The remaining blind
evolution is therefore marginalized in the untruncated process, where its
total probability is one, before any finite-cap representation is introduced.
This is why the \emph{terminal} blind factor clipped at the window edge is
absent below; its removal does not rely on the finite-cap inequality in
(\ref{eq:blind-ones}).  For \(T_c>T_0\),
\begin{align}
  G^{(1)} ={}& B\,V(L),
  \label{eq:G1}\\
  G^{(2)}(c_2) ={}& B\biggl[\,\int_0^{a_2}\!\!V(s)\,E_{c_2}\,B\,
                  V(L-T_0-s)\,ds
  \nonumber\\
          &\hphantom{B\biggl[\,}
                  +\int_{a_2}^{L}\!\!V(s)\,E_{c_2}\,ds\biggr],
  \label{eq:G2}\\
  G^{(3)}(c_2,c_3) ={}& B\biggl[\,
      \iint_{\mathcal{D}}
      V(s)\,E_{c_2}\,B\,V(s')\,E_{c_3}\,B\,
  \nonumber\\
          &\hphantom{B\biggl[}\qquad\times
      V(L-2T_0-s-s')\,ds\,ds'
  \nonumber\\
          &\hphantom{B\biggl[\,}
      +\iint_{\mathcal{D}'}
      V(s)\,E_{c_2}\,B\,V(s')\,E_{c_3}\,ds\,ds'\biggr],
  \label{eq:G3}
\end{align}
with \(a_2=\max(0,L-T_0)\) and domains
\(\mathcal{D}=\{s,s'\ge0,\ s+s'\le L-2T_0\}\) and
\(\mathcal{D}'=\{s,s'\ge0,\ L-2T_0<s+s'\le L-T_0\}\); an integral
with an empty domain vanishes.  Each bracket is a full-interval bulk term
plus a clipped boundary term.  In \(G^{(2)}\) a follower at active
time \(s\le L-T_0\) is trailed by its complete blind interval and the
remaining active evolution, while a follower within \(T_0\) of the
edge is trailed by a clipped blind stretch that contracts away.  The
same splitting produces the two domains of \(G^{(3)}\): in
\(\mathcal{D}\) both followers fit full blind intervals inside the window;
in \(\mathcal{D}'\) the second follower sits within \(T_0\) of the
edge (a first follower within \(T_0\) of the edge leaves no room for a
recorded second follower and correctly contributes nothing to fold
three).  The degenerate cases obey the same rule: for
\(T_c\le T_0\) the whole post-trigger window is blind and no later state is
queried, so analytic terminal marginalization gives \(G^{(1)}=I\) and
\(G^{(k\ge2)}=0\).  This identity is imposed before finite-cap projection and
does not assert mass conservation for \(e^{A^{(C)}_{\rm blind}T_c}\).  For
\(T_0<T_c\le 2T_0\) the bulk term of
\(G^{(2)}\) is empty (\(a_2=0\)) and the boundary term runs over
\(0<s\le L\).
At \(T_0=0\), \(B=I\) and the boundary terms vanish, their domains
having zero width.

The time-ordered integrals are evaluated without quadrature as single
block-matrix exponentials~\cite{VanLoan1978,Higham2008,MolerVanLoan2003,AlMohyHigham2011}: with
\(M=E_{c_2}B\),
\begin{equation}
  \int_0^L e^{A_{\rm vis}(L-u)}\, M\, e^{A_{\rm vis}u}\,du
\end{equation}
is the upper-right block of
\begin{equation}
  \exp\left[
  \begin{pmatrix}
    A_{\rm vis} & M\\
    0 & A_{\rm vis}
  \end{pmatrix}L
  \right],
\end{equation}
and the double integrals of \(G^{(3)}\) use the corresponding
three-block construction, with a zero generator in the last diagonal
block for the clipped boundary strip.  This evaluates the ordered
waiting-time integrals in the matrix-exponential calculus of
phase-type distributions~\cite{Higham2008,CoxIsham1980,Neuts1981,LatoucheRamaswami1999}.

\paragraph{Truth projection of the \(en\) channel.}
The True/False split of the \(en\) channel on the validation grid is a
projection inside the same construction: substituting
\(E_n^{\rm self}\) or \(E_n^{\rm old}\) for \(E_{c_2}\) in
(\ref{eq:G2}) gives the rate at which the delayed partner is, or is
not, the trigger's own neutron.  What makes this a truth statement is
the old/self asymmetry built into (\ref{eq:trigger-embedding}),
(\ref{eq:event-matrices}), and (\ref{eq:ablind}): in a two-fold
\(en\) window the only recorded prompt is the trigger itself and
blind-interval births are booked old, so at the capture the self slots
hold the trigger's own pending neutron or nothing.

\paragraph{Where the muon enters.}
No factor of \(R_\mu\) appears in \(A_{\rm vis}\), \(A_{\rm blind}\),
or any event matrix.  This is a statement of the time basis, not an
approximation: all rates are quoted per unit reset-segment time
(Sec.~\ref{sec:model}), and on that axis every accepted window is
muon-free by construction.  If the next seam arrives before a history
window closes, that window is terminal and generates no further chain
step; it may extend into the excised guard, but it closes before the
physical reset (Sec.~\ref{sec:model}).  The reset therefore enters only
through the \(R_\mu\) term of the gap race
\(d_h=R_\mu+R_s+R_{\rm corr}+\sum_ih_i/\tau_i\) and through the
survival factor \(e^{-R_\mu T_c}\) carried by every entry of the
continuing chain step \(Q\) (Sec.~\ref{subsec:general-KN}).  Rates per detector-live time and per
wall time follow from the fixed conversion factors
\(e^{-R_\mu T_c}\) and \(e^{-R_\mu(T_c+\bar V)}\) of
Sec.~\ref{sec:model}.

\subsection{Minimal example}
\label{subsec:minimal-example}

For \(K=1\) (hence \(f_1=1\)), \(N_{\max}=1\), and \(T_0=0\), write
\(R_p=R_s+R_{\rm corr}\).  The history space is \(h\in\{0,1\}\), with
\(d_0=R_p+R_\mu\) and \(d_1=d_0+\tau^{-1}\); the trigger-reachable current
states are \((0;0),(1;0),(0;1),(1;1)\), where the entries are old and self
occupancies.  In this basis the trigger embeddings are
\begin{align}
 J_s&=\begin{pmatrix}R_s/d_0&0&0&0\\
                     0&R_s/d_1&0&0\end{pmatrix},\nonumber\\
 J_n&=\begin{pmatrix}0&0&0&0\\
                 (\tau d_1)^{-1}&0&0&0\end{pmatrix},
 \nonumber\\
 J_e&=R_{\rm corr}
 \begin{pmatrix}
  (1-\epsilon_n)/d_0&0&\epsilon_n/d_0&0\\
  0&(1-\epsilon_n)/d_1&0&\epsilon_n/d_1
 \end{pmatrix}.
 \label{eq:minimal-J}
\end{align}
The distinction between opportunity visits and actual openings is now
explicit:
\begin{align}
 \Lambda_s&=R_s\sum_{h=0,1}\frac{w_h}{d_h},&
 \Lambda_e&=R_{\rm corr}\sum_{h=0,1}\frac{w_h}{d_h},\nonumber\\
 \Lambda_n&=\frac{w_1}{\tau d_1},&
 \Lambda_{\rm open}&=\Lambda_s+\Lambda_e+\Lambda_n .
 \label{eq:minimal-opener-rates}
\end{align}
A triggering \(e\) deposits its self neutron in \(J_e\), while a triggering
\(n\) consumes the old neutron in \(J_n\).

The active generator is diagonal,
\((A_{\rm vis})_{\alpha\alpha}=-(R_p+(o+m)/\tau)\), so the one-fold
master formula reduces to
\begin{align}
 R_e^{(1)}
 &=\sum_{h=0,1}w_h\frac{R_{\rm corr}}{d_h}
   \Bigl[(1-\epsilon_n)e^{-(R_p+h/\tau)T_c}
   \nonumber\\[-2pt]
 &\hspace{30mm}
   +\epsilon_n e^{-(R_p+(h+1)/\tau)T_c}\Bigr],\nonumber\\
 R_n^{(1)}
 &=w_1\frac{e^{-R_p T_c}}{\tau d_1}.
 \label{eq:minimal-fold1}
\end{align}
For a two-fold \(en\) window opened from \(h=0\), the self-capture
contribution is the scalar member of \(G^{(2)}\),
\begin{equation}
 R_{en}^{(2)}\big|_{h=0}^{\rm True}
 =\frac{w_0R_{\rm corr}\epsilon_n}{d_0}
  \int_0^{T_c}
  e^{-(R_p+\tau^{-1})t}\,
  \frac{e^{-R_p(T_c-t)}}{\tau}\,dt .
 \label{eq:minimal-true}
\end{equation}
The analogous integrals with \(E_n^{\rm old}\), available from the
states \((1;0)\) and \((1;1)\), give the False partner.  Restoring the lifetime
mixture, dead time, and larger caps changes the matrix dimensions but not
this assembly:
\(w^T\!\to J_{c_1}\to G^{(k)}\to\mathbf1\).
\subsection{Finite-cap error bound}
\label{subsec:truncation-contract}

The history cap \(N_{\max}\) and current cap
\(C=N_{\max}+H\) are the only state-space approximations in
Eq.~(\ref{eq:exact-master}).  In this subsection \(N\) denotes a generic
history cap with current cap \(C_N=N+H\); production uses
\(N=N_{\max}\).  With
\(\mu=\sum_iR_{\rm corr}\epsilon_nf_i\tau_i\), the discarded reset-seed
probability at cap \(N\) is the Poisson tail
\begin{equation}
 \delta_{\rm init}
 =\Pr\{\operatorname{Pois}(\mu)>N\}
 \le e^{-\mu}\left(\frac{e\mu}{N+1}\right)^{N+1},
 \label{eq:occupancy-tail-bound}
\end{equation}
the Chernoff tail bound, valid whenever the tail threshold satisfies
\(\mu\le N+1\); here the much stronger condition
\(N_{\max}\gg\mu\) holds.
With \(\mathcal H_N\) the retained history states, let
\(r_N^{(k)}=J_{c_1,N}G_N^{(k)}(c_2,\ldots,c_k)\mathbf1\) be the
requested-channel reward per history visit at cap \(N\), so that
\(R_N^{(k)}=w_N^{\!\top}r_N^{(k)}\).  Define the resolvent factor
\(A\) and the finite-cap visit mass \(W_N\); together with
\(\delta_{\rm init}\), the remaining two defects are
\(\delta_{\rm hist}\) and \(\delta_{\rm curr}^{(k)}\):
\begin{align}
 A&=\frac{1}{1-e^{-R_\mu T_c}},\qquad
 W_N=\lVert w_N\rVert_1,\nonumber\\
 \delta_{\rm hist}
 &=\max_{h\in\mathcal H_N}
   \sum_{h'\notin\mathcal H_N}(Q_\infty)_{hh'},\nonumber\\
 \delta_{\rm curr}^{(k)}
 &=\left\lVert
 r_\infty^{(k)}\big|_{\mathcal H_N}-r_N^{(k)}
 \right\rVert_\infty .
 \label{eq:finite-cap-pieces}
\end{align}
Their propagation is controlled by two inequalities:
\begin{align}
 \left\lVert(I-Q^T)^{-1}\right\rVert_1
 &\le A,\nonumber\\
 \left\lVert\Delta Q^T\widetilde w_N\right\rVert_1
 &\le\delta_{\rm hist}W_N ,
 \label{eq:finite-cap-proof-core}
\end{align}
where tildes denote zero extension to a larger cap and
\(\Delta Q=Q_{N+1}-\widetilde Q_N\).  The second bound is weighted:
new boundary rows may have order-one sums, but they carry zero weight in
\(\widetilde w_N\); each retained row can restore at most its own overflow
mass.  Subtracting the visit equations on their common enlarged space gives
the exact perturbation identity
\begin{equation}
\begin{split}
 w_{N+1}-\widetilde w_N
 ={}&(I-Q_{N+1}^T)^{-1}
 \Bigl[\Delta Q^T\widetilde w_N\\
 &\quad+R_\mu
 (p_{0,N+1}-\widetilde p_{0,N})\Bigr],
\end{split}
\label{eq:resolvent-perturbation-w}
\end{equation}
and hence
\begin{equation}
\begin{split}
 \lVert w_{N+1}-\widetilde w_N\rVert_1
 \le A\Bigl[
 \delta_{\rm hist}W_N
 +R_\mu\lVert p_{0,N+1}-\widetilde p_{0,N}\rVert_1
 \Bigr].
\end{split}
\label{eq:hist-defect-propagation}
\end{equation}
The channel-rate difference has a separate current-operator term:
\begin{equation}
\begin{split}
 |R_{N+1}^{(k)}-R_N^{(k)}|
 \le{}&
 \lVert w_{N+1}-\widetilde w_N\rVert_1
 \lVert r_{N+1}^{(k)}\rVert_\infty\\
 &+W_N
 \lVert r_{N+1}^{(k)}-\widetilde r_N^{(k)}\rVert_\infty .
\end{split}
\label{eq:channel-rate-truncation}
\end{equation}
Every component of \(r^{(k)}\), at any finite cap and for the
untruncated operator, is a channel probability conditional on the
history state, so \(\lVert r_N^{(k)}\rVert_\infty\le1\),
\(\lVert r_\infty^{(k)}\rVert_\infty\le1\), and hence also
\(\lVert r_\infty^{(k)}|_{\mathcal H_N}\rVert_\infty\le1\).  For the
conditional,
renormalized seed used in the calculation,
\(\lVert p_{0,\infty}-\widetilde p_{0,N}\rVert_1
=2\delta_{\rm init}\).  Applying
Eqs.~(\ref{eq:resolvent-perturbation-w})--(\ref{eq:channel-rate-truncation})
directly between cap \(N\) and the infinite operator therefore gives
\begin{equation}
 \boxed{\;
 |R_\infty^{(k)}-R_N^{(k)}|
 \le 2R_\mu A\delta_{\rm init}
     +A\delta_{\rm hist}W_N
     +\delta_{\rm curr}^{(k)}W_N .\;}
 \label{eq:finite-cap-bound}
\end{equation}
The three terms respectively bound initialization, propagated history
overflow, and current-window overflow, all in absolute-rate units.  For the
window-close convention each quantity is isolated, so
Eq.~(\ref{eq:finite-cap-bound}) is complete.  For the global
non-paralyzable kernel, the raw history deficit is not itself a truncation
error: it combines probability flowing beyond the retained history cap with
physical opportunity loss from a blind interval that extends past window
close.  We therefore report it only as a conservative history-deficit
envelope, not as \(\delta_{\rm hist}\).

The current-window defect \(\delta_{\rm curr}^{(k)}\) is defined
through the untruncated reward and is not evaluated directly; for
\(k\le3\) and \(H\ge k\), a conservative current-headroom envelope
for the window-close and global non-paralyzable paths is
\begin{equation}
 \delta_{\rm curr}^{(k)}
 \le
 \Pr\!\left\{
 \operatorname{Pois}(R_{\rm corr}\epsilon_n kT_0)>H-k
 \right\},
 \label{eq:current-headroom-envelope}
\end{equation}
which makes the current-window term of Eq.~(\ref{eq:finite-cap-bound})
computable for these channels.  For the window-close convention this
completes the three-term bound; for the global non-paralyzable
convention the history contribution remains the conservative envelope
described above.  It reserves one self-pending slot for each recorded event and
bounds births during at most \(kT_0\) of blind time; captures can only
reduce the population.  It does not apply to a paralyzable blind interval, whose
unrecorded events can extend the blind measure.

At the largest validation-grid source rate,
\(\mu=6.64\times10^{-4}\), \(N_{\max}=4\) gives
\(\delta_{\rm init}\le6.1\times10^{-18}\).  Raising
\(N_{\max}=4\to5\) changes no channel by more than
\(6.7\times10^{-16}\,\mathrm{Hz}\), or \(3.7\times10^{-9}\) relative, on
the \(T_0=0\) and \(T_0=\SI{1}{\micro s}\) grids, and by no more than
\(1.7\times10^{-14}\,\mathrm{Hz}\), or \(4.1\times10^{-8}\) relative, at
\(T_0=\SI{100}{\micro s}\) in any of the three dead-time conventions.
Table~\ref{tab:cap-scan} lists the complete scan by \(T_0\) and
dead-time convention.
\begin{table}[t]
 \centering
 \caption{Largest change of any exact rate when the history cap
 \(N_{\max}\) is raised at fixed current headroom \(H=3\), listed by
 \(T_0\) and dead-time convention (WC window-close, G global
 non-paralyzable, GP global paralyzable; all three coincide at
 \(T_0=0\)).  Each row block takes the maximum over the four
 configurations (\(R_{\rm corr}=5\) or \SI{0.1}{Hz}, \(T_c=1500\) or
 \SI{400}{\micro s}), except that the GP block covers only the two
 \(T_c=\SI{400}{\micro s}\) configurations; each configuration
 contributes 41 rates, the 39 ordered one-, two-, and three-fold
 sequences plus the True and False components of \(en\), so the scan
 compares 738 rates at each \(N_{\max}\), with \(\Delta R=R_{N+1}-R_N\).  The largest relative change is
 always in the pure-capture \(nnn\) channel; production uses
 \(N_{\max}=4\).}
 \label{tab:cap-scan}
 \footnotesize
 \setlength{\tabcolsep}{3.5pt}
 \begin{tabular}{llcc}
  \toprule
  \(T_0\), convention & Cap step & max \(|\Delta R|\) (Hz) & max \(|\Delta R|/R_{N+1}\)\\
  \midrule
  \(T_0=0\) & \(2\to3\) & \(5.8\times10^{-8}\) & \(1.0\)\\
   & \(3\to4\) & \(4.5\times10^{-12}\) & \(7.8\times10^{-5}\)\\
   & \(4\to5\) & \(6.7\times10^{-16}\) & \(3.3\times10^{-9}\)\\
  \midrule
  \(T_0=\SI{1}{\micro s}\), WC & \(2\to3\) & \(5.7\times10^{-8}\) & \(9.7\times10^{-1}\)\\
   & \(3\to4\) & \(4.8\times10^{-12}\) & \(8.2\times10^{-5}\)\\
   & \(4\to5\) & \(6.7\times10^{-16}\) & \(3.7\times10^{-9}\)\\
  \midrule
  \(T_0=\SI{100}{\micro s}\), WC & \(2\to3\) & \(5.0\times10^{-8}\) & \(1.4\times10^{-1}\)\\
   & \(3\to4\) & \(1.0\times10^{-11}\) & \(6.4\times10^{-5}\)\\
   & \(4\to5\) & \(7.1\times10^{-15}\) & \(1.5\times10^{-8}\)\\
  \midrule
  \(T_0=\SI{100}{\micro s}\), G & \(2\to3\) & \(4.0\times10^{-7}\) & \(1.3\times10^{-1}\)\\
   & \(3\to4\) & \(9.7\times10^{-11}\) & \(1.1\times10^{-4}\)\\
   & \(4\to5\) & \(1.6\times10^{-14}\) & \(3.7\times10^{-8}\)\\
  \midrule
  \(T_0=\SI{100}{\micro s}\), GP & \(2\to3\) & \(3.9\times10^{-7}\) & \(2.7\times10^{-1}\)\\
   & \(3\to4\) & \(9.6\times10^{-11}\) & \(1.5\times10^{-4}\)\\
   & \(4\to5\) & \(1.7\times10^{-14}\) & \(4.1\times10^{-8}\)\\
 \bottomrule
 \end{tabular}
\end{table}
The unit relative change at \(N_{\max}=2\to3\) is the kinematic truncation
zero of \(nnn\) at \(T_0=0\): a window starting with at most two old
neutrons and no recorded \(e\) cannot contain three captures.  At
\(T_0>0\) a correlated prompt that fires inside a blind interval is
unrecorded, but its pending daughter can capture and be recorded once
the detector re-arms, supplying the third capture; \(nnn\) at
\(N_{\max}=2\) is then no longer zero and the relative change is below
unity.  With an unnormalized seed,
all four discarded mechanisms (the seed tail, history overflow, and the
birth entries dropped from the blind generator and from the event
matrices) remove non-negative path contributions, so
rates increase monotonically toward their infinite-cap values; conditioning
the seed adds only a relative \(O(\delta_{\rm init})\) displacement.

\subsection{Exact observable definitions}
\label{subsec:pair-efficiency}

Hereafter, an analytical rate without a qualifier is the ordered-propagator
result of Eq.~(\ref{eq:exact-master}); superscripts ``fact'' and ``LO'' mark
only the two comparison calculations below.  In particular, the three central
selection efficiencies are exact rate ratios: \(\epsilon_{\rm pair}\)
and \(\epsilon_{\rm mult}\) are defined for \(R_{\rm corr}>0\), while
\(\epsilon_{\rm singles}\) is defined for \(R_s>0\); in addition,
\(\epsilon_{\rm mult}\) requires \(\epsilon_n\Phi(T_0,T_c)>0\),
equivalently \(\epsilon_n>0\) and \(T_0<T_c\) (for \(T_0\ge T_c\) the
True-pair rate vanishes and \(\epsilon_{\rm mult}\) is undefined),
\begin{align}
 \epsilon_{\rm pair}
 &\equiv \frac{R_{en}^{\rm True}}{R_{\rm corr}},
 \label{eq:pair-efficiency}\\
 \epsilon_{\rm mult}
 &\equiv
 \frac{\epsilon_{\rm pair}}
      {\epsilon_n\Phi(T_0,T_c)},
 \label{eq:mult-efficiency}\\
 \epsilon_{\rm singles}
 &\equiv \frac{R_s^{\rm rec}}{R_s},
 \label{eq:singles-efficiency}
\end{align}
where \(\Phi(t_a,t_b)=\int_{t_a}^{t_b}\rho(t)\,dt\) and
\(R_s^{\rm rec}\) is the exact accepted one-fold \(s\)-window rate.

\section{Factorized comparison}
\label{sec:event-placement}

All central numerical results use Eq.~(\ref{eq:exact-master}).  The secondary
construction in this section instead factors each one- or two-fold rate into
trigger, opening-history, live-tail, and delayed-event terms.  It is useful
because every physical loss remains visible, but it is not a corollary of the
matrix solution and is not the source of the quoted rates.  For a trigger
\(x\in\{s,e,n\}\) and tail label \(y\in\{i,s,e,n\}\), where \(i\) denotes no
recorded follower, write
\begin{equation}
 R_{xy}^{\rm fact}=R_x\,P_{\rm start}^{(x)}
        \,\mathcal T_{x\to y}\,\mathcal D_{x\to y},
 \label{eq:event-placement-skeleton}
\end{equation}
where the factors are the trigger intensity, clean-opening probability,
prompt-like tail occupancy, and delayed-event condition.  An uppercase
\(N\) in a start superscript denotes an old pending neutron and is distinct
from the recorded label \(n\).  For a neutron trigger the donor intensity,
the parent-source intensity associated with the pending neutron, is
\(R_{\rm corr}\), while the capture-at-opening weight is carried by
\(P_{\rm start}^{(N)}\).

\subsection{Primitives and composition rule}
\label{subsec:building-blocks}

With \(T_w=T_c-T_0\), define
\begin{align}
 P_t(r,t)&=r e^{-rt},&
 P_0(r,t)&=e^{-rt},\nonumber\\
 \Phi(t_a,t_b)
 &=\int_{t_a}^{t_b}\rho(t)\,dt,\nonumber\\
 &=\sum_i f_i\!\left(e^{-t_a/\tau_i}-e^{-t_b/\tau_i}\right),
 \label{eq:factorized-primitives}\\
 \Phi_w&=\Phi(T_0,T_c),&
 \Phi(0,\infty)&=1 .
 \label{eq:Phi-norm}
\end{align}
\(P_t\) is a first-arrival density, \(P_0\) is a zero-arrival probability,
and \(\Phi(t_a,t_b)\) is the unconditional probability that a neutron
captures at an age between \(t_a\) and \(t_b\).  Every bracket below is
written in birth-to-capture ages, so a pending neutron born earlier
needs no survival conditioning: each component is memoryless, but the
mixture is not, and a neutron known to have survived to age \(a\)
carries the updated weights \(f_ie^{-a/\tau_i}/\sum_jf_je^{-a/\tau_j}\).
The independent \(s\) and \(e\) streams
therefore give the joint live-tail survival
\(P_0(R_s+R_{\rm corr},T_w)\).  Conditional on the trigger and the pending
content at opening, this Poisson factor and the daughter-capture factor are
exactly independent.  The approximation enters only through how much
opening history and pending content are retained, and through the
trigger-only dead-time correction at \(T_0>0\).

\subsection{Window-start geometry}
\label{subsec:window-start}

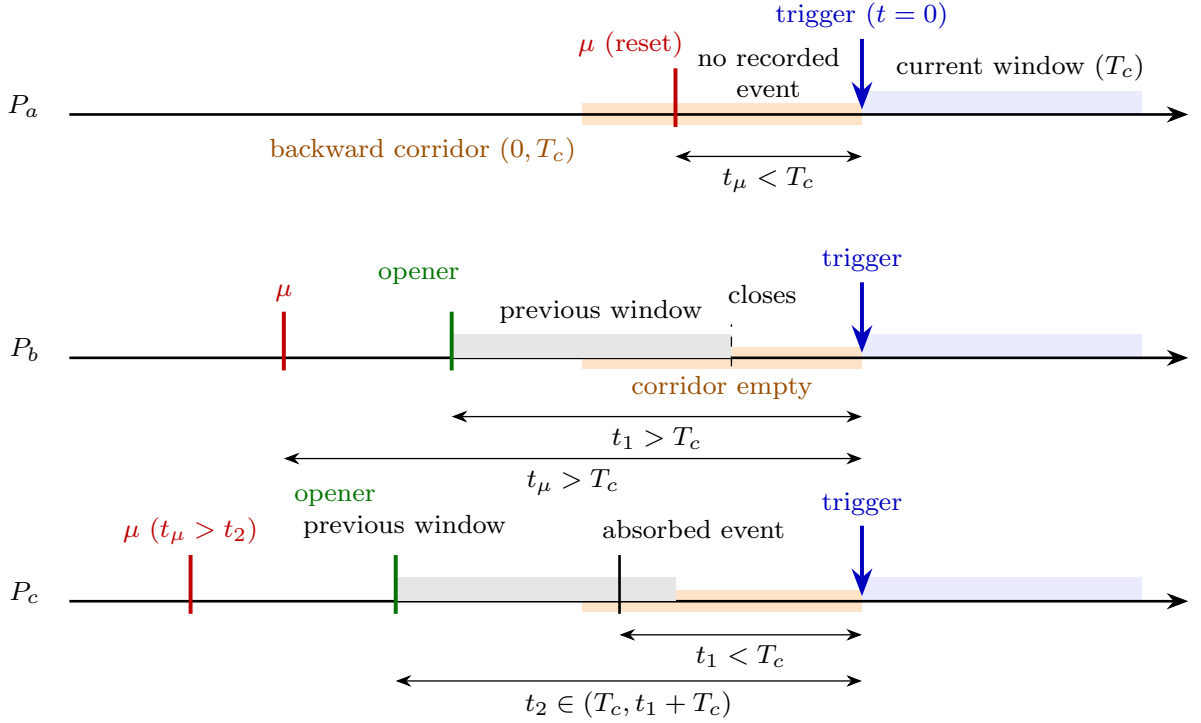
\begin{figure*}[t]
 \centering
 \resizebox{0.88\textwidth}{!}{%
 \begin{tikzpicture}[
   >=Stealth,yscale=0.9,
   muon/.style={red!75!black,thick,line width=1.2pt},
   trigger/.style={blue!75!black,thick,line width=1.2pt},
   opener/.style={green!45!black,thick,line width=1.2pt},
   dim/.style={<->,thin},
   label/.style={font=\footnotesize}]
  \foreach \y/\name in {0/{$P_c$},2.9/{$P_b$},5.8/{$P_a$}} {
   \fill[orange!22] (5.4,\y-0.13) rectangle (8.4,\y+0.13);
   \fill[blue!8] (8.4,\y) rectangle (11.4,\y+0.28);
   \draw[->,thick] (-0.1,\y) -- (11.9,\y);
   \node[label,left] at (-0.3,\y+0.1) {\name};
  }
  \node[label,above] at (10.1,6.08) {current window (\(T_c\))};
  \draw[muon] (6.4,5.65) -- (6.4,6.35);
  \node[label,red!75!black,above left] at (6.6,6.35) {\(\mu\) (reset)};
  \draw[trigger,->] (8.4,6.7) -- (8.4,5.86);
  \node[label,blue!75!black,above] at (8.4,6.7) {trigger (\(t=0\))};
  \node[label,align=center] at (7.4,6.3) {no recorded\\event};
  \draw[dim] (6.4,5.3) -- (8.4,5.3)
    node[midway,below,label] {\(t_\mu<T_c\)};
  \node[label,orange!60!black,below left] at (5.45,5.65)
    {backward corridor \((0,T_c)\)};
  \fill[gray!20] (4.0,2.9) rectangle (7.0,3.18);
  \node[label,above] at (5.6,3.18) {previous window};
  \draw[dashed] (7.0,2.8) -- (7.0,3.3);
  \node[label,above right] at (6.85,3.42) {closes};
  \draw[opener] (4.0,2.75) -- (4.0,3.45);
  \node[label,green!45!black,above left] at (4.2,3.65) {opener};
  \draw[muon] (2.2,2.75) -- (2.2,3.45);
  \node[label,red!75!black,above] at (2.2,3.45) {\(\mu\)};
  \draw[trigger,->] (8.4,3.8) -- (8.4,2.96);
  \node[label,blue!75!black,above] at (8.4,3.8) {trigger};
  \node[label,orange!60!black] at (6.9,2.55) {corridor empty};
  \draw[dim] (4.0,2.2) -- (8.4,2.2)
    node[midway,below,label] {\(t_1>T_c\)};
  \draw[dim] (2.2,1.7) -- (8.4,1.7)
    node[midway,below,label] {\(t_\mu>T_c\)};
  \fill[gray!20] (3.4,0) rectangle (6.4,0.28);
  \node[label,above left] at (4.7,0.62) {previous window};
  \draw[opener] (3.4,-0.15) -- (3.4,0.55);
  \node[label,green!45!black,above left] at (3.3,1.0) {opener};
  \draw[thick] (5.8,-0.15) -- (5.8,0.55);
  \node[label,above right] at (5.5,0.62) {absorbed event};
  \draw[muon] (1.2,-0.15) -- (1.2,0.55);
  \node[label,red!75!black,above] at (1.2,0.55) {\(\mu\) (\(t_\mu>t_2\))};
  \draw[trigger,->] (8.4,0.9) -- (8.4,0.06);
  \node[label,blue!75!black,above] at (8.4,0.9) {trigger};
  \draw[dim] (5.8,-0.4) -- (8.4,-0.4)
    node[midway,below,label] {\(t_1<T_c\)};
  \draw[dim] (3.4,-0.95) -- (8.4,-0.95)
    node[midway,below,label] {\(t_2\in(T_c,t_1+T_c)\)};
 \end{tikzpicture}}
 \caption{The three opening-history cases of Eq.~(\ref{eq:Pstart-decomp}),
 read backward from the proposed trigger at \(t=0\) (blue), whose window
 (blue) has length \(T_c\).  The orange band is the backward corridor
 \((0,T_c)\); \(t_\mu\), \(t_1\), \(t_2\) are backward distances.  \(P_a\):
 the most recent reset (red) lies inside the corridor and nothing was
 recorded between it and the trigger.  \(P_b\): the reset lies beyond
 \(T_c\) and the corridor holds no recorded event; a completed window
 (gray, opened at \(t_1>T_c\)) may precede the trigger, and may even extend
 into the corridor as long as none of its member events lies there; or no
 completed window may exist at all.  \(P_c\): the corridor does hold an event, at
 \(t_1<T_c\), but it was absorbed by an earlier window opened at
 \(t_2\in(T_c,\,t_1+T_c)\) that closed before the trigger; the reset lies
 beyond that opener.  Since \(t_1\) is the nearest recorded event and is
 absorbed, the trigger still opens its own window.}
 \label{fig:event-placement-three-cases}
\end{figure*}

Work backward from the proposed trigger at \(t=0\)
(Fig.~\ref{fig:event-placement-three-cases}).  Its clean-opening
history is partitioned into three mutually exclusive cases:
\begin{itemize}
 \item \(P_a\): the most recent reset is within \(T_c\), and the trigger is
       the first recorded event after it;
 \item \(P_b\): the reset is farther back than \(T_c\), and the backward
       corridor \((0,T_c)\) is empty;
 \item \(P_c\): that corridor is nonempty, but its nearest event was
       absorbed in an earlier window that closed before the trigger.
\end{itemize}
Thus
\begin{equation}
 P_{\rm start}=P_a+P_b+P_c .
 \label{eq:Pstart-decomp}
\end{equation}
\paragraph{Exhaustiveness of the three cases.}
Condition on the trigger opening its own window, rather than falling inside
an earlier one, and let \(t_\mu\) be the backward distance to the most recent
reset.  Since \(R_\mu>0\), \(t_\mu\) is almost surely finite and positive.
Exactly one of \(t_\mu\le T_c\) and \(t_\mu>T_c\) holds; in the latter set,
the corridor \((0,T_c)\) is either empty or nonempty.  These alternatives are
mutually exclusive and exhaustive.  It remains only to identify them with
the three cases above.

If \(t_\mu\le T_c\), an unabsorbed opener at backward distance
\(t_1<t_\mu\) would have a window extending to
\(t_1-T_c<t_\mu-T_c\le0\), across the proposed trigger.  This contradicts
the conditioning, so the trigger is the first recorded event after the
reset, precisely \(P_a\).  If \(t_\mu>T_c\) and the corridor is empty, the
definition is precisely \(P_b\).  Finally, suppose \(t_\mu>T_c\) and let
\(t_1\in(0,T_c)\) be the nearest corridor event.  It cannot be an unabsorbed
opener for the same reason, so an earlier event at
\(t_2\in(T_c,t_1+T_c)\) opened a window that absorbed \(t_1\) and closed
before the trigger.  Because an accepted completed window cannot cross the
most recent reset, \(t_\mu>t_2\); this is \(P_c\).  Each converse follows
from the corresponding definition.  Thus the three cases form an exact
geometric partition, up to zero-probability boundary events.  Their
evaluations below are approximate in all three cases: they place only
the finitely many prompts written out explicitly, and only the stated
\(N\)-marked configurations, as examined in
Sec.~\ref{subsec:why-approximation}.

A correlated prompt born at backward distance \(u\) must not leave a
recorded old neutron between the relevant exclusion boundary and the end of
the proposed window.  The clean and recorded-old-neutron brackets are
\begin{align}
 \bar{\mathcal B}(u,d)
 &=1-\epsilon_n\!\left[
   \Phi(u-d,u+T_c)-\Phi(u,u+T_0)\right],
 \label{eq:clean-bracket}\\
 \mathcal B_N(u)
 &=\epsilon_n\!\left[
   \Phi(u,u+T_c)-\Phi(u,u+T_0)\right].
 \label{eq:recorded-bracket}
\end{align}
The subtraction restores captures in the trigger's blind slice, which are
unrecorded and harmless.  Geometry fixes the exclusion boundary,
\begin{equation}
 d^*=\min(t_\mu,T_c,t_{\rm close}),
 \label{eq:exclusion-start}
\end{equation}
where \(t_{\rm close}\) is the backward distance to the most recent completed
window close, or infinity if none exists.  A capture before the most recent
reset is erased; one inside the completed window is absorbed; one between
\(d^*\) and the trigger pre-empts the opening; and one in the current live
tail produces the recorded-old-neutron variant.

For case \(a\), the most recent reset is at
\(t_\mu\in[0,T_c]\), and the nearest \(s\) and \(e\) both lie beyond it:
\begin{equation}
\begin{split}
 P_a=\int_0^{T_c}\!dt_\mu\,P_t(R_\mu,t_\mu)
 &\int_{t_\mu}^{\infty}\!dt_s\,P_t(R_s,t_s)\\[-2pt]
 {}\times&
 \int_{t_\mu}^{\infty}\!dt_i\,P_t(R_{\rm corr},t_i)
 \bar{\mathcal B}(t_i,t_\mu).
\end{split}
\label{eq:Pa-explicit}
\end{equation}
For case \(b\), let \(t_s\) and \(t_i\) be the backward distances to
the nearest single and to the nearest correlated prompt; both exceed
\(T_c\) because the corridor is empty.  Then
\begin{equation}
\begin{split}
 P_b={}&\int_{T_c}^{\infty}\!dt_\mu\,P_t(R_\mu,t_\mu)
 \int_{T_c}^{\infty}\!dt_s\,P_t(R_s,t_s)\\
 &\times\int_{T_c}^{\infty}\!dt_i\,P_t(R_{\rm corr},t_i)\,
 \bar{\mathcal B}\bigl(t_i,d^*(t_\mu,t_s,t_i)\bigr),
\end{split}
\label{eq:Pb-explicit}
\end{equation}
where the exclusion boundary follows Eq.~(\ref{eq:exclusion-start}) with
the last completed window read from the two tracked prompts: if neither
lies between the reset and the trigger, \(d^*=T_c\); otherwise the
farther of those that do opens a window, the nearer one is absorbed by
it when it falls within \(T_c\) after it and opens its own window
otherwise, and \(d^*=\min(T_c,t_{\rm close})\) with \(t_{\rm close}\)
the close of the last window so formed.  According to the arrangement,
\(d^*\) is therefore \(\min(T_c,t_i-T_c)\), \(\min(T_c,t_s-T_c)\), or
\(T_c\).  For case \(c\), the absorbed event is at
\(t_1\in(0,T_c)\), the opener at
\(t_2\in(T_c,t_1+T_c)\), and the reset lies beyond it.  Summing the four
species assignments gives
\begin{equation}
\begin{split}
 P_c={}&\int_0^{T_c}\!dt_1
 \int_{T_c}^{t_1+T_c}\!dt_2
 \int_{t_2}^{\infty}\!dt_\mu\,P_t(R_\mu,t_\mu)\Bigl\{\\
 &P_t(R_s,t_1)P_t(R_s,t_2-t_1)
 \Bigl[\int_{t_2}^{\infty}\!dt_i\,P_t(R_{\rm corr},t_i)\\
 &\quad+\int_{t_1}^{t_2}\!dt_i\,P_t(R_{\rm corr},t_i)
          \bar{\mathcal B}(t_i,t_2-T_c)\Bigr]\\
 &+P_t(R_s,t_1)P_t(R_{\rm corr},t_2)
          \bar{\mathcal B}(t_2,t_2-T_c)\\
 &+P_t(R_{\rm corr},t_1)P_t(R_s,t_2)
          \bar{\mathcal B}(t_1,t_2-T_c)\\
 &+P_t(R_{\rm corr},t_1)P_t(R_{\rm corr},t_2-t_1)
   \int_{t_1}^{\infty}\!dt_s\,P_t(R_s,t_s)\\
 &\qquad\times
   \bar{\mathcal B}(t_1,t_2-T_c)
   \bar{\mathcal B}(t_2,t_2-T_c)\Bigr\}.
\end{split}
\label{eq:Pc-explicit}
\end{equation}
The terms correspond to \((s,s)\), \((s,e)\), \((e,s)\), and \((e,e)\)
assignments of the absorbed event and opener.  The nearest-occurrence
densities enforce their ordering; only one absorbed event and one completed
previous window are resolved, and placements in which an old-neutron
capture is the absorbed event or the opener are not resolved at all (they
enter the exact chain through the capture operators of
Sec.~\ref{sec:propagator}).

The geometry does not distinguish an \(s\) trigger from an \(e\) trigger, so
\(P_{\rm start}^{(s)}=P_{\rm start}^{(e)}=P_{\rm start}\).  To obtain a
window containing one old neutron, replace one clean bracket by
\(\mathcal B_N\).  For example,
\begin{equation}
\begin{split}
 P_a^{(x,N)}=\int_0^{T_c}\!dt_\mu\,P_t(R_\mu,t_\mu)
 &\int_{t_\mu}^{\infty}\!dt_s\,P_t(R_s,t_s)\\[-2pt]
 {}\times&
 \int_{t_\mu}^{\infty}\!dt_i\,P_t(R_{\rm corr},t_i)
 \mathcal B_N(t_i),
\end{split}
\label{eq:PaN-explicit}
\end{equation}
and cases \(b,c\) use the same substitution.  When two clean brackets occur,
only one is replaced; two recorded old neutrons are a higher-order pending
configuration, and the unbracketed prompt beyond the opener in the
two-singles term of Eq.~(\ref{eq:Pc-explicit}) is likewise left
unresolved.  The scale is
\begin{equation}
 P_{\rm start}^{(s,N)}=P_{\rm start}^{(e,N)}
 \sim P_{\rm start}\epsilon_nR_{\rm corr}T_c\,
 \Phi(T_c,T_c+T_w).
\label{eq:Pstart-eN}
\end{equation}

For an old neutron that itself opens the window, the start factor is
normalized per correlated donor,
\(P_{\rm start}^{(N)}=P_a^{(N)}+P_b^{(N)}+P_c^{(N)}\).  The first two
pieces illustrate the construction:
\begin{align}
 P_a^{(N)}
 &=\int_0^{T_c}\!dt_\mu\,P_t(R_\mu,t_\mu)
   \int_{t_\mu}^{\infty}\!dt_s\,P_t(R_s,t_s)\nonumber\\
 &\quad\times
   \left[\epsilon_n\int_{t_\mu}^{\infty}\!dt_n\,\rho(t_n)\right]
   \int_{t_\mu}^{\infty}\!dt_i\,P_t(R_{\rm corr},t_i)
   \bar{\mathcal B}(t_i,t_\mu),
 \label{eq:Pstart-N}\\
 P_b^{(N)}&=P_b\,\epsilon_n\Phi(T_c,\infty).
 \label{eq:Pstart-N-b}
\end{align}
In case \(c\) the donor prompt takes one of three positions in the
chained geometry of Eq.~(\ref{eq:Pc-explicit}): it is the absorbed
nearest event, the opener, or a further correlated prompt inside the
previous window; its capture density is integrated over its birth
position in each placement, the remaining roles are filled by the
\(s/e\) placements, and all other tracked prompts and the reset are
resolved as in Eqs.~(\ref{eq:Pb-explicit}) and (\ref{eq:Pc-explicit})
(Supplemental Sec.~I lists the placement classes).  A second old neutron in the live
tail defines \(P_{\rm start}^{(N,N)}\) by the same
\(\bar{\mathcal B}\to\mathcal B_N\) substitution.  In the
limit \(\epsilon_n\to0\), every \(N\)-marked start factor vanishes and
\(\bar{\mathcal B}\to1\), recovering the ordinary Poisson-then-tail
factorization.  At fixed \(\epsilon_n\), the limit \(R_{\rm corr}\to0\)
instead sends \(P_{\rm start}^{(s,N)}\), \(P_{\rm start}^{(e,N)}\), and
\(P_{\rm start}^{(N,N)}\) to zero, while the per-donor
\(P_{\rm start}^{(N)}\) approaches a finite isolated-donor limit; all
neutron-triggered rates nevertheless vanish through their prefactor
\(R_{\rm corr}\).  The proof above establishes that the three cases exhaust
the opening histories; Eqs.~(\ref{eq:Pa-explicit})--(\ref{eq:Pc-explicit})
then resolve the retained arrangements used in their numerical evaluation.

\subsection{One- and two-fold rate dictionary}
\label{subsec:thirteen-rates}

Applying Eq.~(\ref{eq:event-placement-skeleton}) by trigger species gives the
complete factorized dictionary.  For an \(s\) trigger,
\begin{align}
 R_{ss}^{\rm fact}
 &=R_sP_{\rm start}^{(s)}
   R_sT_w e^{-(R_s+R_{\rm corr})T_w},
 \label{eq:Rss}\\
 R_{si}^{\rm fact}
 &=R_sP_{\rm start}^{(s)}e^{-(R_s+R_{\rm corr})T_w},
 \label{eq:Rsi}\\
 R_{se}^{\rm fact}
 &=R_sP_{\rm start}^{(s)}e^{-R_sT_w}\nonumber\\
 &\quad\times\int_{T_0}^{T_c}\!dt\,P_t(R_{\rm corr},t)
   P_0(R_{\rm corr},T_c-t)\nonumber\\
 &\quad\times
   \left[1-\epsilon_n\Phi(0,T_c-t)\right],
 \label{eq:Rse}\\
 R_{sn}^{\rm fact}
 &=R_sP_{\rm start}^{(s,N)}
   e^{-(R_s+R_{\rm corr})T_w}.
 \label{eq:Rsn}
\end{align}
The time-dependent bracket in \(R_{se}\) requires the follower prompt's
daughter not to capture in the remaining tail.  The old-neutron capture
weight of \(R_{sn}\) is already inside \(P_{\rm start}^{(s,N)}\).

For an \(e\) trigger, the one-fold and non-\(en\) two-fold rates are
\begin{align}
 R_{ei}^{\rm fact}
 &=R_{\rm corr}P_{\rm start}^{(e)}
   e^{-(R_s+R_{\rm corr})T_w}\nonumber\\
 &\quad\times
   \left[(1-\epsilon_n)+\epsilon_n(1-\Phi_w)\right],
 \label{eq:Rei}\\
 R_{es}^{\rm fact}
 &=R_{\rm corr}P_{\rm start}^{(e)}
   R_sT_w e^{-(R_s+R_{\rm corr})T_w}\nonumber\\
 &\quad\times(1-\epsilon_n\Phi_w),
 \label{eq:Res}\\
 R_{ee}^{\rm fact}
 &=R_{\rm corr}P_{\rm start}^{(e)}e^{-R_sT_w}
   (1-\epsilon_n\Phi_w)\nonumber\\
 &\quad\times\int_{T_0}^{T_c}\!dt\,P_t(R_{\rm corr},t)
   P_0(R_{\rm corr},T_c-t)\nonumber\\
 &\quad\times
   \left[1-\epsilon_n\Phi(0,T_c-t)\right].
 \label{eq:Ree}
\end{align}
The two branches in \(R_{ei}\) are an undetected daughter and a detected
daughter capturing outside the live tail.  The constant bracket in
\(R_{es}\) and \(R_{ee}\) prevents the trigger's own daughter from turning
the window into a three-fold sequence.

For a neutron trigger, every correlated decay is a potential donor and
\(P_{\rm start}^{(N)}\) supplies the per-donor opening weight:
\begin{align}
 R_{ni}^{\rm fact}
 &=R_{\rm corr}P_{\rm start}^{(N)}
   e^{-(R_s+R_{\rm corr})T_w},
 \label{eq:Rn}\\
 R_{ns}^{\rm fact}
 &=R_{\rm corr}P_{\rm start}^{(N)}
   R_sT_w e^{-(R_s+R_{\rm corr})T_w},
 \label{eq:Rns}\\
 R_{ne}^{\rm fact}
 &=R_{\rm corr}P_{\rm start}^{(N)}e^{-R_sT_w}
   \int_{T_0}^{T_c}\!dt\,P_t(R_{\rm corr},t)\nonumber\\
 &\quad\times P_0(R_{\rm corr},T_c-t)
   \nonumber\\
 &\quad\times
   \left[1-\epsilon_n\Phi(0,T_c-t)\right],
 \label{eq:Rne}\\
 R_{nn}^{\rm fact}
 &=R_{\rm corr}P_{\rm start}^{(N,N)}
   e^{-(R_s+R_{\rm corr})T_w}.
 \label{eq:Rnn}
\end{align}
Together with the \(en\) split below, these equations cover every one- and
two-fold \(s/e/n\) sequence.

\subsection{The True and False \texorpdfstring{\(en\)}{en} components}
\label{subsec:ren-anatomy}

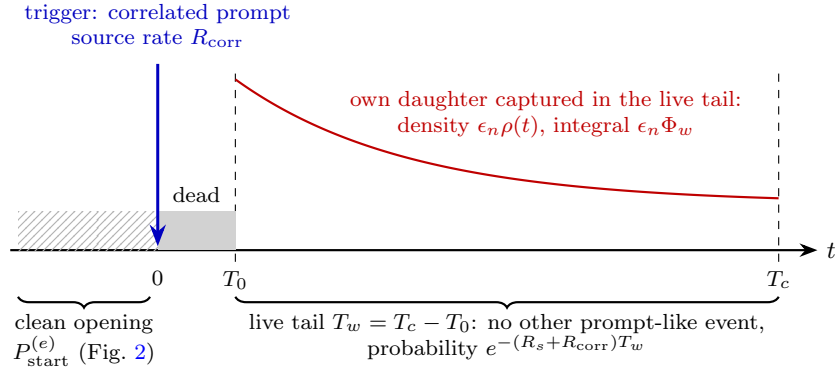
\begin{figure*}[t]
 \centering
 \resizebox{0.62\textwidth}{!}{%
 \begin{tikzpicture}[>=Stealth,label/.style={font=\footnotesize}]
  \draw[->,thick] (-1.9,0) -- (8.5,0) node[right] {\(t\)};
  \fill[pattern=north east lines, pattern color=gray!60]
    (-1.8,0) rectangle (0,0.5);
  \draw[dashed] (1.0,-0.15) -- (1.0,2.3);
  \draw[dashed] (8.0,-0.15) -- (8.0,2.3);
  \node[label,below] at (0.0,-0.15) {\(0\)};
  \node[label,below] at (1.0,-0.15) {\(T_0\)};
  \node[label,below] at (8.0,-0.15) {\(T_c\)};
  \fill[gray!35] (0,0) rectangle (1.0,0.5);
  \node[label,above] at (0.5,0.5) {dead};
  \draw[blue!75!black,->,thick,line width=1.1pt] (0,2.4) -- (0,0.05);
  \node[label,blue!75!black,above,align=center] at (0,2.5)
    {trigger: correlated prompt\\source rate \(R_{\rm corr}\)};
  \draw[red!75!black,thick,domain=1.0:8.0,smooth,samples=80]
    plot (\x,{0.6+1.6*exp(-(\x-1.0)*0.45)});
  \node[label,red!75!black,align=center] at (5.0,1.75)
    {own daughter captured in the live tail:\\
     density \(\epsilon_n\rho(t)\), integral \(\epsilon_n\Phi_w\)};
  \draw[decorate,decoration={brace,mirror,amplitude=4pt},thick]
    (1.0,-0.6) -- (8.0,-0.6);
  \node[label,below,align=center] at (4.5,-0.7)
    {live tail \(T_w=T_c-T_0\): no other prompt-like event,\\
     probability \(e^{-(R_s+R_{\rm corr})T_w}\)};
  \draw[decorate,decoration={brace,mirror,amplitude=4pt},thick]
    (-1.8,-0.6) -- (-0.1,-0.6);
  \node[label,below,align=center] at (-0.95,-0.7)
    {clean opening\\\(P_{\rm start}^{(e)}\) (Fig.~\ref{fig:event-placement-three-cases})};
 \end{tikzpicture}}
 \caption{The four factors of \(R_{en,\mathrm{fact}}^{\rm True}\) in
 Eq.~(\ref{eq:Ren-true}), laid out in time around the
 trigger: the clean-opening history before it (hatched;
 \(P_{\rm start}^{(e)}\) from Fig.~\ref{fig:event-placement-three-cases}),
 the trigger's source rate \(R_{\rm corr}\) (arrow), the requirement that
 no other prompt-like event fall in the live tail \(T_w=T_c-T_0\) after
 the dead-time slice (gray), and the capture of the trigger's own daughter
 in that tail, with density \(\epsilon_n\rho(t)\) integrating to
 \(\epsilon_n\Phi_w\).}
 \label{fig:event-placement-anatomy}
\end{figure*}

The central pair separates according to the parent of the recorded neutron:
\begin{align}
 R_{en,\mathrm{fact}}^{\rm True}
 &=R_{\rm corr}P_{\rm start}^{(e)}
   e^{-(R_s+R_{\rm corr})T_w}\epsilon_n\Phi_w,
 \label{eq:Ren-true}\\
 R_{en,\mathrm{fact}}^{\rm False}
 &=R_{\rm corr}P_{\rm start}^{(e,N)}
   e^{-(R_s+R_{\rm corr})T_w}(1-\epsilon_n\Phi_w),
 \label{eq:Ren-false}\\
 R_{en,\mathrm{fact}}^{\rm Total}
 &=R_{en,\mathrm{fact}}^{\rm True}
   +R_{en,\mathrm{fact}}^{\rm False}.
 \label{eq:Ren-total}
\end{align}
The True formula reads directly as four independent factors
(Fig.~\ref{fig:event-placement-anatomy}): the prompt
fires at \(R_{\rm corr}\), opens cleanly with
\(P_{\rm start}^{(e)}\), sees no other prompt-like event in the live tail,
and has its own detected daughter capture there.  In the False term the
recorded member instead comes from the old population contained in
\(P_{\rm start}^{(e,N)}\), while the trigger's own daughter must not appear
in the tail.  An old neutron must be born in an earlier correlated source,
survive the previous window and intervening reset structure, and then capture
in the current live tail.  This explains both the scaling in
Eq.~(\ref{eq:Pstart-eN}) and the observed hierarchy: over the validation grid
the True component exceeds the False one by \(4.4\)--\(6.8\) orders of
magnitude.

\subsection{Selection efficiencies in the factorized picture}

The factorized estimates of the exact ratios
(\ref{eq:pair-efficiency})--(\ref{eq:singles-efficiency}) are
\begin{align}
 \epsilon_{\rm pair}^{\rm fact}
 &=\frac{R_{en,\mathrm{fact}}^{\rm True}}{R_{\rm corr}},
 \nonumber\\
 \epsilon_{\rm mult}^{\rm fact}
 &=P_{\rm start}^{(e)}e^{-(R_s+R_{\rm corr})T_w},
 \nonumber\\
 \epsilon_{\rm singles}^{\rm fact}
 &=\frac{R_{si}^{\rm fact}}{R_s}.
 \label{eq:factorized-efficiencies}
\end{align}
Only the True rate enters \(\epsilon_{\rm pair}\): adding False pairs would
count an accidental background as signal efficiency.  All three ratios use
reset-segment time, so the veto live-time fraction is not included.  For
example, at \(R_{\rm corr}=\SI{0.1}{Hz}\),
\(T_c=\SI{1500}{\micro s}\), and \(T_0=0\), the exact values satisfy
\[
 \epsilon_{\rm pair}=0.6965
 \simeq \epsilon_n\Phi_w\epsilon_{\rm mult}
 =0.8\times0.9996\times0.8710 .
\]
For the validation veto mixture,
\(\epsilon_\mu=e^{-R_\mu\bar V}=0.8353\); the same pair probability becomes
\(0.5160\) per detector-live time after multiplying by
\(e^{-R_\mu T_c}\), and \(0.4310\) per wall time after multiplying by
\(e^{-R_\mu(T_c+\bar V)}\).  These clock factors must be restored before
comparing efficiencies quoted under different live-time conventions.

\subsection{Two-fold time-shape taxonomy}
\label{subsec:dt-shapes}

The same factors determine each differential delay density
(Table~\ref{tab:dt-shapes}).  Up to the small
prompt-like survival variation across the live tail, the tail species fixes
one of three primitive shapes:
\begin{itemize}
 \item \(y=s\), for \(ss,es,ns\): approximately flat, from homogeneous
       singles arrivals;
 \item \(y=e\), for \(se,ee,ne\): rising, because
       \(1-\epsilon_n\Phi(0,T_c-\Delta t)\) increases toward the window edge;
 \item \(y=n\), for \(sn,en,nn\): falling; the trigger's own daughter
       in \(en\) follows \(\epsilon_n\rho(\Delta t)\).  For a donor born
       \(u\) before the trigger, the unnormalized old-neutron contribution
       is \(\epsilon_n\rho(u+\Delta t)\); conditional on that neutron
       being pending at the trigger, its normalized residual density is
       \(\rho(u+\Delta t)/S(u)\) with \(S(u)=\sum_if_ie^{-u/\tau_i}\),
       and averaging over \(u\) gives a positive combination of the same
       exponentials tilted toward the longer lifetimes.
\end{itemize}
\begin{table}[t]
 \centering
 \caption{Dominant factorized \(\Delta t\) shape of every two-fold
 sequence.  Small old-neutron and mixed-source components do not alter the
 qualitative classes.  \(S(u)=\sum_if_ie^{-u/\tau_i}\) is the survival
 probability of a pending neutron to age \(u\).}
 \label{tab:dt-shapes}
 \footnotesize
 \setlength{\tabcolsep}{4pt}
 \begin{tabular}{lll}
  \toprule
  Pairs & Shape & Controlling factor\\
  \midrule
  \(ss,es,ns\) & flat & singles arrival \(R_s\)\\
  \(se,ee,ne\) & rising &
    \(1-\epsilon_n\Phi(0,T_c-\Delta t)\)\\
  \(sn,en,nn\) & falling & \(\rho(\Delta t)\); \(\rho(u+\Delta t)/S(u)\) for an old\\
   & & neutron pending at age \(u\)\\
  \bottomrule
 \end{tabular}
\end{table}
The factorized shapes are interpretive.  Figure~\ref{fig:prd-dt} instead
compares the exact binned \(G^{(2)}\) densities with the independent event
stream at all four configurations.

\subsection{Controlled limits of the factorization}
\label{subsec:why-approximation}

The factorized construction retains the nearest reset and only the finite
completed-window history generated by the event placements of
Eqs.~(\ref{eq:Pa-explicit})--(\ref{eq:PaN-explicit}) and their
neutron-start analogues.  Relative to the exact chain it omits three
identifiable layers:
\begin{enumerate}
 \item \emph{Finite opening-event inventory and depth.}  The backward
 histories use only this finite event inventory as openers or absorbed
 events; histories in which an old-neutron capture opens or is absorbed
 in a previous window, further prompt occurrences beyond the displayed
 placements, previous windows with several absorbed prompts, and
 completed-window chains that require further openers are not
 resolved.  Their leading
 corrections carry another window-occupancy factor
 \((R_s+R_{\rm corr})T_c\) or a capture-opening weight, while the exact
 Neumann series includes an arbitrary number of completed windows with
 openers of every species.
 \item \emph{Higher pending multiplicity.}  The \(N\)-marked variants
 retain at most one recorded old-neutron follower in the current live
 tail, in addition to a possible old-neutron trigger; configurations
 with two or more such followers are omitted.  The history state
 \(h=(h_1,\ldots,h_K)\) instead sums pending populations of every retained
 size, with the finite-cap error bounded in
 Sec.~\ref{subsec:truncation-contract}.
 \item \emph{Follower dead-time microstructure.}  At \(T_0>0\), the
 factorization treats the trigger's blind slice but not blind intervals
 opened by recorded followers, nor unrecorded prompts born inside a blind
 interval whose daughters appear later.  The exact \(A_{\rm blind}\) and
 event matrices retain both effects.
\end{enumerate}
At \(T_0=0\), the total and True \(en\) rates agree with the exact values to
at most \(5\times10^{-5}\) relative on the grid (Supplemental
Table~S1).  Old-neutron-fed channels
are less protected because the finite-history truncation enters their
leading term: \(R_{en}^{\rm False}\) and \(R_{sn}\) reach
\(6\times10^{-4}\), the \(n\), \(ne\), and \(ns\) channels
\(9\times10^{-4}\), and \(R_{nn}\) \(8\times10^{-3}\).  At \(T_0=\SI{1}{\micro s}\), the omitted
unrecorded-prompt supply lowers the tiny False component by \(4.6\)--\(12.5\%\),
while the Total stays at the \(5\times10^{-5}\) level.  Other two-fold
channels are less accurate at this nonzero dead time, since the
follower blind intervals decide directly whether a pair stays two-fold:
across the grid the largest relative discrepancies are \(11.5\%\) for
\(sn\), \(3.0\%\) for \(ee\), \(2.9\%\) for \(ne\), \(2.8\%\) for
\(se\), and \(2.0\%\) for \(nn\).  The factorization is
therefore an interpretable approximation and a cross-check, not a
precision substitute for Eq.~(\ref{eq:exact-master}).

The earlier leading-order calculation~\cite{YuWangChen2015} evaluates
the central pair channel, with prompt efficiency mapped to unity and the
same reset-segment rate normalization, as
\begin{equation}
 R_{en}^{\rm LO}
 =R_{\rm corr}\,
  P_{\rm start}^{\rm LO}(R_s,R_\mu;T_c)\,
  \epsilon_n\Phi(0,T_c)\,e^{-R_sT_c},
 \label{eq:twin-ren}
\end{equation}
where \(P_{\rm start}^{\rm LO}=P_a^{\rm LO}+P_b^{\rm LO}+P_c^{\rm LO}\)
is its window-start probability, with \(P_a^{\rm LO}\), \(P_b^{\rm LO}\),
and \(P_c^{\rm LO}\) the three terms of Eqs.~(1)--(3) of
Ref.~\cite{YuWangChen2015} and their sum its Eq.~(4);
Eq.~(\ref{eq:twin-ren}) is its Eq.~(10).  Its opening factor and its
tail-emptiness factor are built from the singles and reset streams
alone, an approximation justified there by \(R_{\rm corr}\ll R_s\), so
correlated prompts do not compete as window openers or followers, and
Eq.~(\ref{eq:twin-ren}) contains only the trigger's own delayed member,
with no \(en^{\rm False}\) component from an old pending neutron.  The treatment
carries no event dead time, so the comparison is made at \(T_0=0\),
where the window-close, global non-paralyzable, and global paralyzable
conventions coincide.  For the channels it defines (the one-fold \(s\) and \(e\) channels,
\(R_{ss}\), \(R_{es}\), \(R_{en}\), \(R_{sss}\), \(R_{ess}\), and the
ordering-summed \(R_{ens}+R_{esn}\)), these omissions give an
approximately common relative excess
\(\delta_{\rm LO}=(R^{\rm LO}-R^{\rm exact})/R^{\rm exact}\), controlled by
\(R_{\rm corr}T_c\).  For \(R_{en}\), \(\delta_{\rm LO}\) is
\(0.415\%\) at \(R_{\rm corr}=\SI{5}{Hz}\),
\(T_c=\SI{400}{\micro s}\), and \(1.38\%\) at
\(T_c=\SI{1500}{\micro s}\).  Supplemental Sec.~II (Table~S2) compares
these eight channels with the exact rates.
\label{subsec:twin-reach}
\section{Toy Monte Carlo}
\label{sec:toymc}

The validation generator is an exact event-stream simulation in the spirit of
Gillespie sampling~\cite{Gillespie1977}.  It samples the next uncorrelated
single, IBD prompt, muon, and pending neutron capture, then forwards the event
to one or more window analyzers.  The generator does not manually reset or
inject pending neutrons at a muon.  Delayed neutron captures are ordinary
future events in the stream.  It samples exactly the stochastic model
of Sec.~\ref{sec:model}, so agreement validates the analytical
solution of that model, its boundary and dead-time conventions
included, and not the model's fidelity to any real detector; the
latter is a data comparison outside this paper's scope.  The analyzer applies the muon veto, discards
good detector events inside the veto, aborts any still-open coincidence
candidate crossed by a muon, imposes the event dead time in the
convention under test, records exact multiplicity-one, -two, and
-three windows together with an aggregate overflow counter for
windows of multiplicity four or more, and fills nine \(\Delta t\)
histograms for \(ss,se,\ldots,nn\).

The production validation generated \(2\times10^{13}\) events for each
raw source-rate group, once for the \(T_0 = 0\) grid and once for its
\(T_0 = \SI{1}{\micro s}\) window-close companion.  This exposure is a
computing-budget choice: it fixes the statistical resolution of every
comparison in Sec.~\ref{sec:results}, not the reach of the calculation,
whose rates carry no statistical error, and a larger sample would only
tighten the same tests.  The two \(T_c\)
selections at fixed \(R_{\rm corr}\) were
analyzed from the same raw stream, matching the detector prescription that the
window length is a selection parameter rather than a new physical event
sample.  Each shard is analyzed independently, and the shard results are
merged into sequence rates and two-fold time distributions.  The
generator carries absolute time at 113-bit precision and evaluates
window-boundary predicates in anchor-relative coordinates, so that
time-axis rounding does not bias boundary outcomes at the
parts-per-million resolution of the campaign.

\section{Results}
\label{sec:results}

Unless a factorized or leading-order qualifier is shown, every analytical
rate, efficiency, and time density in this section is obtained from the exact
ordered propagator, Eq.~(\ref{eq:exact-master}).  The independent toy Monte
Carlo of Section~\ref{sec:toymc} tests those predictions over the full rate
grid, the two-fold time densities, matched dead-time conventions, and
aggregate high multiplicity.

\subsection{Rate comparison and pulls}
\label{subsec:prd-rates}

\begin{figure*}[t]
  \centering
  \includegraphics[width=\textwidth]{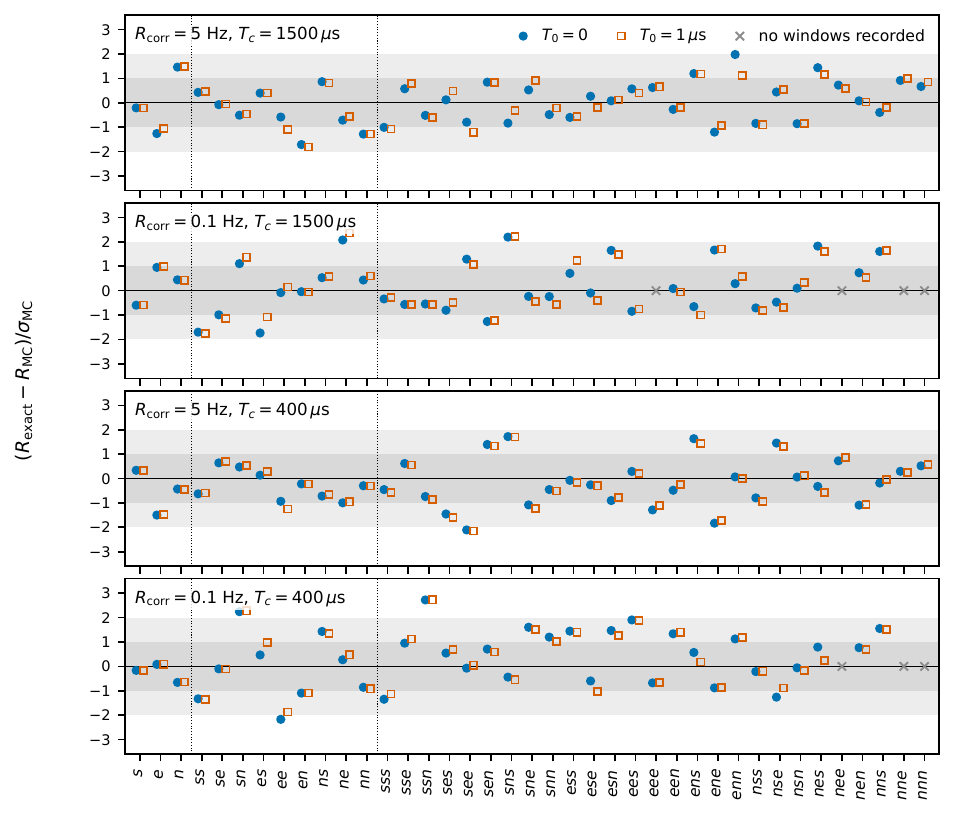}
  \caption{Pulls \((R_{\rm exact}-R_{\rm MC})/\sigma_{\rm MC}\) of the
  exact rates of Table~\ref{tab:exact-rates-grid} against the merged
  \(2\times10^{13}\)-event toy-MC campaigns, for all 39 ordered sequences
  (dotted lines separate the one-, two-, and three-fold groups) at the
  four validation configurations (one panel each), at \(T_0=0\) (filled
  circles) and at \(T_0=\SI{1}{\micro s}\) in the toy-MC rearm
  convention matched to the window-close kernel (open squares).  Shaded
  bands mark \(\pm1\) and \(\pm2\).  A gray cross marks a channel that
  recorded no windows (below one expected count at this exposure) and
  carries no pull.  The two dead-time settings were analyzed on the same
  event-stream ensembles, so their pulls are correlated channel by
  channel.}
  \label{fig:prd-pulls-exact}
\end{figure*}

Table~\ref{tab:exact-rates-grid} lists the exact rate of every ordered
sequence, together with the analytical True/False split of \(en\), at
the eight full-statistics campaign setups: the four validation
configurations (\(R_s=\SI{50}{Hz}\), \(R_\mu=\SI{200}{Hz}\),
\(\epsilon_n=0.8\); \(R_{\rm corr}\) and \(T_c\) as labeled) at
\(T_0=0\) and at \(T_0=\SI{1}{\micro s}\) in the window-close
convention.  Figure~\ref{fig:prd-pulls-exact} compares these rates
channel by channel with the \(2\times10^{13}\)-event toy-MC campaigns,
as pulls \((R_{\rm exact}-R_{\rm MC})/\sigma_{\rm MC}\) formed after all
shards have been merged.  Each dead-time setting covers 156 channels,
one per ordered sequence and configuration (four configurations times
39 sequences); the seven
three-fold channels at \(R_{\rm corr}=\SI{0.1}{Hz}\) that recorded
zero counts carry no pull, leaving 149 channels with nonzero toy-MC
statistical error.  For those, the maximum absolute one-, two-, and
three-fold pulls are 1.50, 2.24, and 2.73 at \(T_0=0\), and 1.48,
2.36, and 2.73 at \(T_0=\SI{1}{\micro s}\).  The largest percentage differences occur only
in very rare channels with zero or order-one toy-MC counts; the
high-rate physical channels, including \(s\), \(e\), and \(en\), agree
within their toy-MC statistical errors.  An independent
adaptive-quadrature evaluation of the same kernels, which does not use
the matrix propagator, converges on all 82 analytical entries (41 per
configuration) of the two \(T_c=\SI{400}{\micro s}\) configurations and
reproduces the toy-MC comparison of their 78 observable sequence rates;
the largest difference between the two evaluations is
\(\SI{4.04e-8}{Hz}\) absolute and \(7.2\times10^{-3}\) relative, the
latter on a three-fold rate of order \(\SI{1e-7}{Hz}\).
The toy-MC rates and
statistical errors behind every marker are tabulated in the Supplemental
Material: the one- and two-fold channels at both the \(10^{10}\)-event
and the \(2\times10^{13}\)-event exposures in Supplemental
Tables~S3--S6, and the three-fold channels at \(2\times10^{13}\) events
in Supplemental Tables~S7 and~S8.

\begin{table*}[t]
  \centering
  \caption{Exact rates (Hz) of every ordered sequence at the eight
    full-statistics campaign setups: the four validation configurations
    (\(R_s=\SI{50}{Hz}\), \(R_\mu=\SI{200}{Hz}\), \(\epsilon_n=0.8\);
    columns abbreviate \(R_{\rm corr}\) in Hz over \(T_c\) in
    \si{\micro s}) at \(T_0=0\) and at \(T_0=\SI{1}{\micro s}\) in the
    window-close convention, all from the ordered propagator,
    Eq.~(\ref{eq:exact-master}).  The \(en\) rate is also split into its
    True and False components (the recorded neutron is the trigger's own
    daughter or an old pending neutron, Section~\ref{subsec:ren-anatomy});
    the split is analytical and was not tagged in these campaigns.  These
    are the predictions compared with the toy MC in
    Fig.~\ref{fig:prd-pulls-exact}.}
  \label{tab:exact-rates-grid}
  \scriptsize
  \setlength{\tabcolsep}{2pt}
  \begin{tabular}{lcccccccc}
    \toprule
    & \multicolumn{4}{c}{\(T_0=0\)} &
    \multicolumn{4}{c}{\(T_0=\SI{1}{\micro s}\)} \\
    \cmidrule(lr){2-5}\cmidrule(lr){6-9}
    Channel & \(5/1500\) & \(0.1/1500\) & \(5/400\) & \(0.1/400\) &
    \(5/1500\) & \(0.1/1500\) & \(5/400\) & \(0.1/400\) \\
    \midrule
    \(s\) & \(42.9676\) & \(43.5524\) & \(47.8852\) & \(48.0817\) & \(42.9698\) & \(43.5545\) & \(47.8876\) & \(48.0841\) \\
    \(e\) & \(0.860872\) & \(0.0174518\) & \(1.37246\) & \(0.0275618\) & \(0.897172\) & \(0.0181876\) & \(1.41294\) & \(0.0283746\) \\
    \(n\) & \(0.139720\) & \(0.00278097\) & \(0.515352\) & \(0.0103076\) & \(0.139727\) & \(0.00278111\) & \(0.515380\) & \(0.0103082\) \\
    \midrule
    \(ss\) & \(3.22257\) & \(3.26643\) & \(0.957703\) & \(0.961633\) & \(3.22075\) & \(3.26458\) & \(0.955409\) & \(0.959325\) \\
    \(se\) & \(0.0929666\) & \(0.00188464\) & \(0.0468023\) & \(0.000939886\) & \(0.0956497\) & \(0.00193903\) & \(0.0475436\) & \(0.000954769\) \\
    \(sn\) & \(0.00135209\) & \(2.69127\times 10^{-5}\) & \(0.00441448\) & \(8.83075\times 10^{-5}\) & \(0.00151617\) & \(3.02413\times 10^{-5}\) & \(0.00455900\) & \(9.12113\times 10^{-5}\) \\
    \(es\) & \(0.0645654\) & \(0.00130888\) & \(0.0274492\) & \(0.000551237\) & \(0.0674156\) & \(0.00136666\) & \(0.0283574\) & \(0.000569472\) \\
    \(ee\) & \(0.00186262\) & \(7.55190\times 10^{-7}\) & \(0.00134142\) & \(5.38771\times 10^{-7}\) & \(0.00200065\) & \(8.11148\times 10^{-7}\) & \(0.00140935\) & \(5.66051\times 10^{-7}\) \\
    \(en\) & \(3.43591\) & \(0.0696530\) & \(3.41618\) & \(0.0686015\) & \(3.40001\) & \(0.0689249\) & \(3.37614\) & \(0.0677970\) \\
    \(en^{\rm True}\) & \(3.43588\) & \(0.0696530\) & \(3.41606\) & \(0.0686015\) & \(3.39998\) & \(0.0689249\) & \(3.37601\) & \(0.0677969\) \\
    \(en^{\rm False}\) & \(2.70897\times 10^{-5}\) & \(1.07841\times 10^{-8}\) & \(0.000126525\) & \(5.06205\times 10^{-8}\) & \(3.20103\times 10^{-5}\) & \(1.27695\times 10^{-8}\) & \(0.000135699\) & \(5.42980\times 10^{-8}\) \\
    \(ns\) & \(0.0104790\) & \(0.000208572\) & \(0.0103070\) & \(0.000206153\) & \(0.0104731\) & \(0.000208454\) & \(0.0102824\) & \(0.000205658\) \\
    \(ne\) & \(0.000302305\) & \(1.20341\times 10^{-7}\) & \(0.000503697\) & \(2.01491\times 10^{-7}\) & \(0.000311030\) & \(1.23813\times 10^{-7}\) & \(0.000511677\) & \(2.04681\times 10^{-7}\) \\
    \(nn\) & \(4.78647\times 10^{-5}\) & \(1.90483\times 10^{-8}\) & \(0.000234965\) & \(9.41477\times 10^{-8}\) & \(4.81726\times 10^{-5}\) & \(1.91708\times 10^{-8}\) & \(0.000235415\) & \(9.43261\times 10^{-8}\) \\
    \midrule
    \(sss\) & \(0.120846\) & \(0.122491\) & \(0.00957703\) & \(0.00961633\) & \(0.120543\) & \(0.122183\) & \(0.00948299\) & \(0.00952184\) \\
    \(sse\) & \(0.00428032\) & \(8.67716\times 10^{-5}\) & \(0.000560742\) & \(1.12609\times 10^{-5}\) & \(0.00437333\) & \(8.86567\times 10^{-5}\) & \(0.000564138\) & \(1.13290\times 10^{-5}\) \\
    \(ssn\) & \(1.33986\times 10^{-5}\) & \(2.66693\times 10^{-7}\) & \(3.02790\times 10^{-5}\) & \(6.05718\times 10^{-7}\) & \(2.60588\times 10^{-5}\) & \(5.23390\times 10^{-7}\) & \(3.32148\times 10^{-5}\) & \(6.64702\times 10^{-7}\) \\
    \(ses\) & \(0.00269217\) & \(5.45763\times 10^{-5}\) & \(0.000375303\) & \(7.53686\times 10^{-6}\) & \(0.00279768\) & \(5.67149\times 10^{-5}\) & \(0.000382040\) & \(7.67211\times 10^{-6}\) \\
    \(see\) & \(0.000100573\) & \(4.07769\times 10^{-8}\) & \(2.28719\times 10^{-5}\) & \(9.18631\times 10^{-9}\) & \(0.000106624\) & \(4.32298\times 10^{-8}\) & \(2.35984\times 10^{-5}\) & \(9.47800\times 10^{-9}\) \\
    \(sen\) & \(0.229290\) & \(0.00464822\) & \(0.0489692\) & \(0.000983381\) & \(0.226437\) & \(0.00459036\) & \(0.0480011\) & \(0.000963930\) \\
    \(sns\) & \(8.80084\times 10^{-5}\) & \(1.75176\times 10^{-6}\) & \(5.80105\times 10^{-5}\) & \(1.16043\times 10^{-6}\) & \(9.88178\times 10^{-5}\) & \(1.97109\times 10^{-6}\) & \(5.97028\times 10^{-5}\) & \(1.19446\times 10^{-6}\) \\
    \(sne\) & \(2.65438\times 10^{-6}\) & \(1.05668\times 10^{-9}\) & \(3.20495\times 10^{-6}\) & \(1.28223\times 10^{-9}\) & \(3.06252\times 10^{-6}\) & \(1.22172\times 10^{-9}\) & \(3.35213\times 10^{-6}\) & \(1.34131\times 10^{-9}\) \\
    \(snn\) & \(2.26484\times 10^{-7}\) & \(9.01341\times 10^{-11}\) & \(9.81272\times 10^{-7}\) & \(3.93218\times 10^{-10}\) & \(2.34675\times 10^{-7}\) & \(9.34020\times 10^{-11}\) & \(9.92555\times 10^{-7}\) & \(3.97718\times 10^{-10}\) \\
    \(ess\) & \(0.00242120\) & \(4.90831\times 10^{-5}\) & \(0.000274492\) & \(5.51237\times 10^{-6}\) & \(0.00252949\) & \(5.12781\times 10^{-5}\) & \(0.000283129\) & \(5.68578\times 10^{-6}\) \\
    \(ese\) & \(8.57580\times 10^{-5}\) & \(3.47701\times 10^{-8}\) & \(1.60717\times 10^{-5}\) & \(6.45506\times 10^{-9}\) & \(9.16885\times 10^{-5}\) & \(3.71744\times 10^{-8}\) & \(1.68234\times 10^{-5}\) & \(6.75690\times 10^{-9}\) \\
    \(esn\) & \(0.0284014\) & \(0.000575755\) & \(0.0193539\) & \(0.000388650\) & \(0.0280661\) & \(0.000568948\) & \(0.0190195\) & \(0.000381928\) \\
    \(ees\) & \(5.39387\times 10^{-5}\) & \(2.18691\times 10^{-8}\) & \(1.07567\times 10^{-5}\) & \(4.32035\times 10^{-9}\) & \(5.86798\times 10^{-5}\) & \(2.37913\times 10^{-8}\) & \(1.13910\times 10^{-5}\) & \(4.57504\times 10^{-9}\) \\
    \(eee\) & \(2.01503\times 10^{-6}\) & \(1.63396\times 10^{-11}\) & \(6.55543\times 10^{-7}\) & \(5.26587\times 10^{-12}\) & \(2.23416\times 10^{-6}\) & \(1.81165\times 10^{-11}\) & \(7.02803\times 10^{-7}\) & \(5.64545\times 10^{-12}\) \\
    \(een\) & \(0.00516832\) & \(2.09546\times 10^{-6}\) & \(0.00210542\) & \(8.45613\times 10^{-7}\) & \(0.00534578\) & \(2.16740\times 10^{-6}\) & \(0.00214497\) & \(8.61490\times 10^{-7}\) \\
    \(ens\) & \(0.229292\) & \(0.00464822\) & \(0.0489697\) & \(0.000983381\) & \(0.226439\) & \(0.00459036\) & \(0.0480016\) & \(0.000963930\) \\
    \(ene\) & \(0.00685969\) & \(2.78120\times 10^{-6}\) & \(0.00263700\) & \(1.05909\times 10^{-6}\) & \(0.00696987\) & \(2.82586\times 10^{-6}\) & \(0.00263192\) & \(1.05704\times 10^{-6}\) \\
    \(enn\) & \(0.000108124\) & \(4.30412\times 10^{-8}\) & \(0.000314950\) & \(1.25995\times 10^{-7}\) & \(0.000132753\) & \(5.30433\times 10^{-8}\) & \(0.000329398\) & \(1.31808\times 10^{-7}\) \\
    \(nss\) & \(0.000392963\) & \(7.82147\times 10^{-6}\) & \(0.000103070\) & \(2.06153\times 10^{-6}\) & \(0.000391977\) & \(7.80180\times 10^{-6}\) & \(0.000102059\) & \(2.04127\times 10^{-6}\) \\
    \(nse\) & \(1.39186\times 10^{-5}\) & \(5.54066\times 10^{-9}\) & \(6.03484\times 10^{-6}\) & \(2.41408\times 10^{-9}\) & \(1.42210\times 10^{-5}\) & \(5.66103\times 10^{-9}\) & \(6.07141\times 10^{-6}\) & \(2.42868\times 10^{-9}\) \\
    \(nsn\) & \(4.72410\times 10^{-7}\) & \(1.88001\times 10^{-10}\) & \(1.60481\times 10^{-6}\) & \(6.43038\times 10^{-10}\) & \(5.09322\times 10^{-7}\) & \(2.02694\times 10^{-10}\) & \(1.61797\times 10^{-6}\) & \(6.48283\times 10^{-10}\) \\
    \(nes\) & \(8.75429\times 10^{-6}\) & \(3.48488\times 10^{-9}\) & \(4.03910\times 10^{-6}\) & \(1.61574\times 10^{-9}\) & \(9.09740\times 10^{-6}\) & \(3.62144\times 10^{-9}\) & \(4.11163\times 10^{-6}\) & \(1.64473\times 10^{-9}\) \\
    \(nee\) & \(3.27040\times 10^{-7}\) & \(2.60374\times 10^{-12}\) & \(2.46153\times 10^{-7}\) & \(1.96934\times 10^{-12}\) & \(3.46716\times 10^{-7}\) & \(2.76037\times 10^{-12}\) & \(2.53972\times 10^{-7}\) & \(2.03187\times 10^{-12}\) \\
    \(nen\) & \(0.000745606\) & \(2.96804\times 10^{-7}\) & \(0.000527065\) & \(2.10815\times 10^{-7}\) & \(0.000736331\) & \(2.93110\times 10^{-7}\) & \(0.000516649\) & \(2.06646\times 10^{-7}\) \\
    \(nns\) & \(3.11744\times 10^{-6}\) & \(1.24062\times 10^{-9}\) & \(3.09449\times 10^{-6}\) & \(1.23992\times 10^{-9}\) & \(3.13362\times 10^{-6}\) & \(1.24705\times 10^{-9}\) & \(3.08130\times 10^{-6}\) & \(1.23460\times 10^{-9}\) \\
    \(nne\) & \(9.40049\times 10^{-8}\) & \(7.48204\times 10^{-13}\) & \(1.70849\times 10^{-7}\) & \(1.36914\times 10^{-12}\) & \(9.71757\times 10^{-8}\) & \(7.73438\times 10^{-13}\) & \(1.73032\times 10^{-7}\) & \(1.38660\times 10^{-12}\) \\
    \(nnn\) & \(1.07554\times 10^{-8}\) & \(8.56043\times 10^{-14}\) & \(5.78822\times 10^{-8}\) & \(4.64152\times 10^{-13}\) & \(1.09680\times 10^{-8}\) & \(8.72961\times 10^{-14}\) & \(5.83284\times 10^{-8}\) & \(4.67717\times 10^{-13}\) \\
    \bottomrule
  \end{tabular}
\end{table*}

\begin{figure*}[t]
  \centering
  \includegraphics[width=\textwidth]{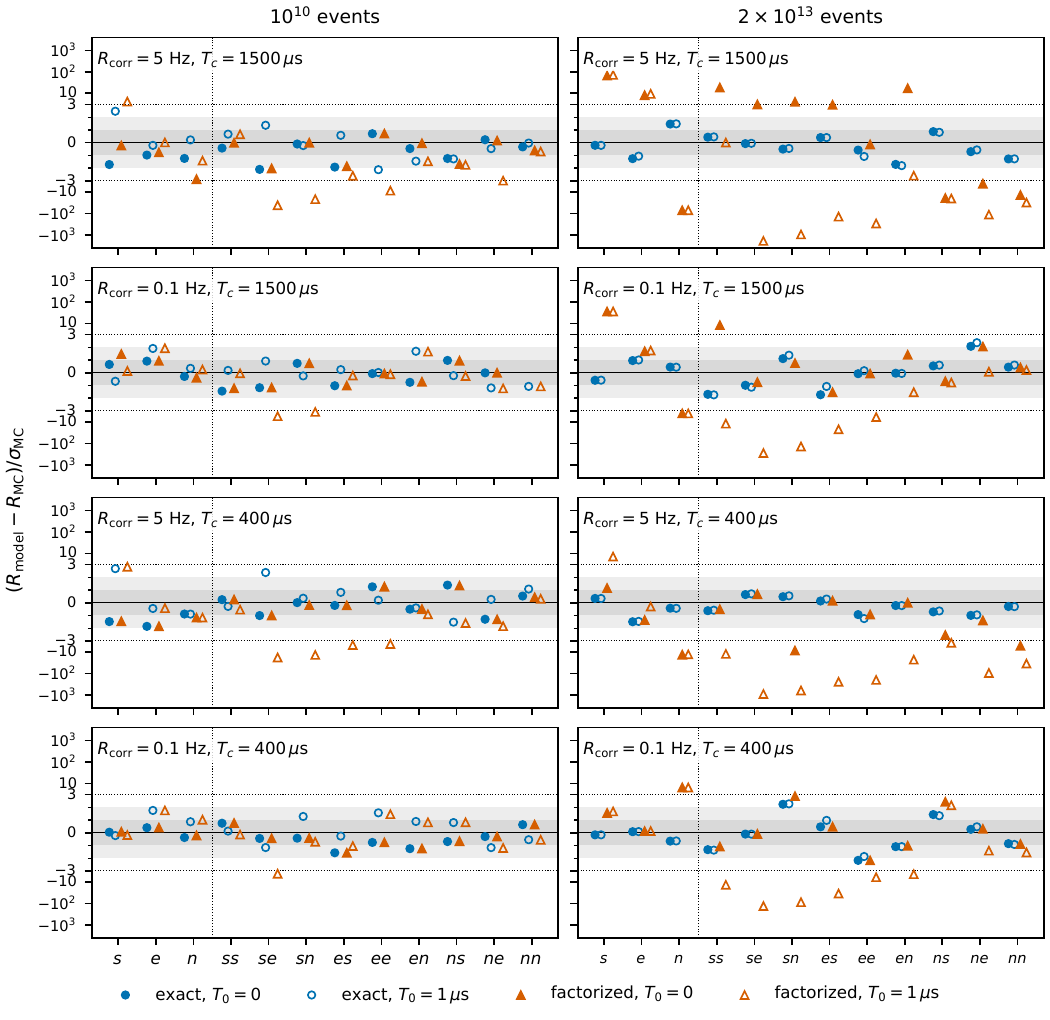}
  \caption{Exact (circles) and factorized (triangles) pulls
  \((R_{\rm model}-R_{\rm MC})/\sigma_{\rm MC}\) of the twelve directly
  measured one- and two-fold channels at the four validation
  configurations (rows), at the \(10^{10}\)-event (left) and
  \(2\times10^{13}\)-event (right) exposures, at \(T_0=0\) (filled) and
  \(T_0=\SI{1}{\micro s}\) (open).  The ordinate is linear within
  \(\pm3\) (dotted lines; shaded bands at \(\pm1\) and \(\pm2\)) and
  logarithmic beyond.  The rates behind every marker are listed in
  Supplemental Tables~S3--S6.  At \(10^{10}\) events and \(T_0=0\) both
  constructions are consistent with the toy MC; at \(2\times10^{13}\)
  events, and at \(T_0=\SI{1}{\micro s}\) already at \(10^{10}\) events,
  the factorization residuals of Section~\ref{subsec:why-approximation}
  are resolved while the exact pulls stay within \(\pm3\).}
  \label{fig:prd-pulls-factorized}
\end{figure*}

Figure~\ref{fig:prd-pulls-factorized} places the factorized
construction of Section~\ref{sec:event-placement} beside the exact
treatment for the twelve directly measured one- and two-fold channels,
at the \(10^{10}\)-event exposure and at the full \(2\times10^{13}\)-event
exposure.  At \(10^{10}\) events and \(T_0=0\) the two constructions
are statistically indistinguishable: every factorized pull is below
\(2.9\) in magnitude and every exact pull below \(2.2\).  At
\(2\times10^{13}\) events the per-channel resolution tightens by a
factor of about forty-five and the factorization residuals of
Section~\ref{subsec:why-approximation} are resolved: the old-neutron
layer moves the factorized \(n\) rate at \(R_{\rm corr}=\SI{5}{Hz}\),
\(T_c=\SI{1500}{\micro s}\) to \(-72\,\sigma_{\rm MC}\), and the
few-\(10^{-5}\) high-rate layer moves \(s\) at the same configuration
to \(+66\,\sigma_{\rm MC}\), while every exact pull stays below \(2.3\).
At \(T_0=\SI{1}{\micro s}\) the follower blind intervals that the
factorization omits shift its two-fold rates by up to \(11.5\%\), for
\(sn\) at \(R_{\rm corr}=\SI{0.1}{Hz}\), \(T_c=\SI{1500}{\micro s}\).
The better-measured \(se\) channel at \(R_{\rm corr}=\SI{5}{Hz}\),
\(T_c=\SI{1500}{\micro s}\), whose factorized rate is \(2.8\%\) low,
gives \(-42\,\sigma_{\rm MC}\) at \(10^{10}\) events and
\(-1.9\times10^{3}\,\sigma_{\rm MC}\) at the full exposure, whereas the
exact pulls stay below \(2.7\) and \(2.4\), respectively.  This is the
controlled sense in which the high-precision approximation fails.

The channel-by-channel statistical resolution of the
\(2\times10^{13}\)-event campaign deserves an explicit statement,
because it is the resolution to which the accidental two-fold rates
are actually verified.  Table~\ref{tab:twofold-resolution} lists the relative
one-sigma statistical resolution \(1/\sqrt{N}\) of every two-fold
channel in the \(T_0=0\) campaign at the four configurations.  The accidental channels built
from at least one high-rate stream (\(ss\), \(se\), \(es\), \(sn\),
\(ns\)) are verified to between \(2.5\times10^{-6}\) and
\(8.7\times10^{-4}\) relative (that is, at the \(0.0003\%\) to
\(0.09\%\) level), and the correlated \(en\) channel to
\(2.2\times10^{-6}\)--\(1.7\times10^{-5}\).  The rare accidental
combinations of two correlated-source events are statistics-limited:
\(ee\) reaches \(0.52\%\), \(ne\) \(1.3\%\), and \(nn\) only \(3.3\%\)
at the low-IBD long-window point, where the \(2\times10^{13}\)-event
stream yields 929 \(nn\) windows.  Any tightening of these last
channels scales as the square root of the campaign size
(Sec.~\ref{sec:toymc}).

\begin{table}[t]
  \centering
  \caption{Relative one-sigma statistical resolution
    \(1/\sqrt{N}\) (in percent) of each two-fold channel in the
    \(T_0=0\), \(2\times10^{13}\)-event campaign, per configuration (columns abbreviate
    \(R_{\rm corr}\) in Hz and \(T_c\) in \(\si{\micro s}\)).  The
    \(en\) channel is dominated by its True component; the False
    (accidental) part is analytical and was not tagged separately in
    this campaign.}
  \label{tab:twofold-resolution}
  \footnotesize
  \setlength{\tabcolsep}{5pt}
  \begin{tabular}{lcccc}
    \toprule
    Channel & \(5/1500\) & \(0.1/1500\) & \(5/400\) & \(0.1/400\) \\
    \midrule
    \(ss\) & 0.00025 & 0.00025 & 0.00042 & 0.00041 \\
    \(se\) & 0.0015 & 0.010 & 0.0019 & 0.013 \\
    \(sn\) & 0.012 & 0.087 & 0.0062 & 0.043 \\
    \(es\) & 0.0018 & 0.012 & 0.0025 & 0.017 \\
    \(ee\) & 0.011 & 0.52 & 0.011 & 0.55 \\
    \(en\) & 0.00025 & 0.0017 & 0.00022 & 0.0015 \\
    \(ns\) & 0.0045 & 0.031 & 0.0040 & 0.028 \\
    \(ne\) & 0.026 & 1.3 & 0.018 & 0.90 \\
    \(nn\) & 0.066 & 3.3 & 0.027 & 1.3 \\
    \bottomrule
  \end{tabular}
\end{table}

The 149 channels with nonzero toy-MC statistical error have an
RMS pull of \(1.004\) and a maximum absolute pull of \(2.73\) on the
\(T_0 = 0\) grid, and \(0.994\) and \(2.73\) on the
\(T_0 = \SI{1}{\micro s}\) grid.  Because channels and window choices
share event streams, the
channel pulls are correlated, so these are descriptive statistics
rather than a global goodness-of-fit test.

\begin{figure*}[t]
  \centering
  \includegraphics[width=\textwidth]{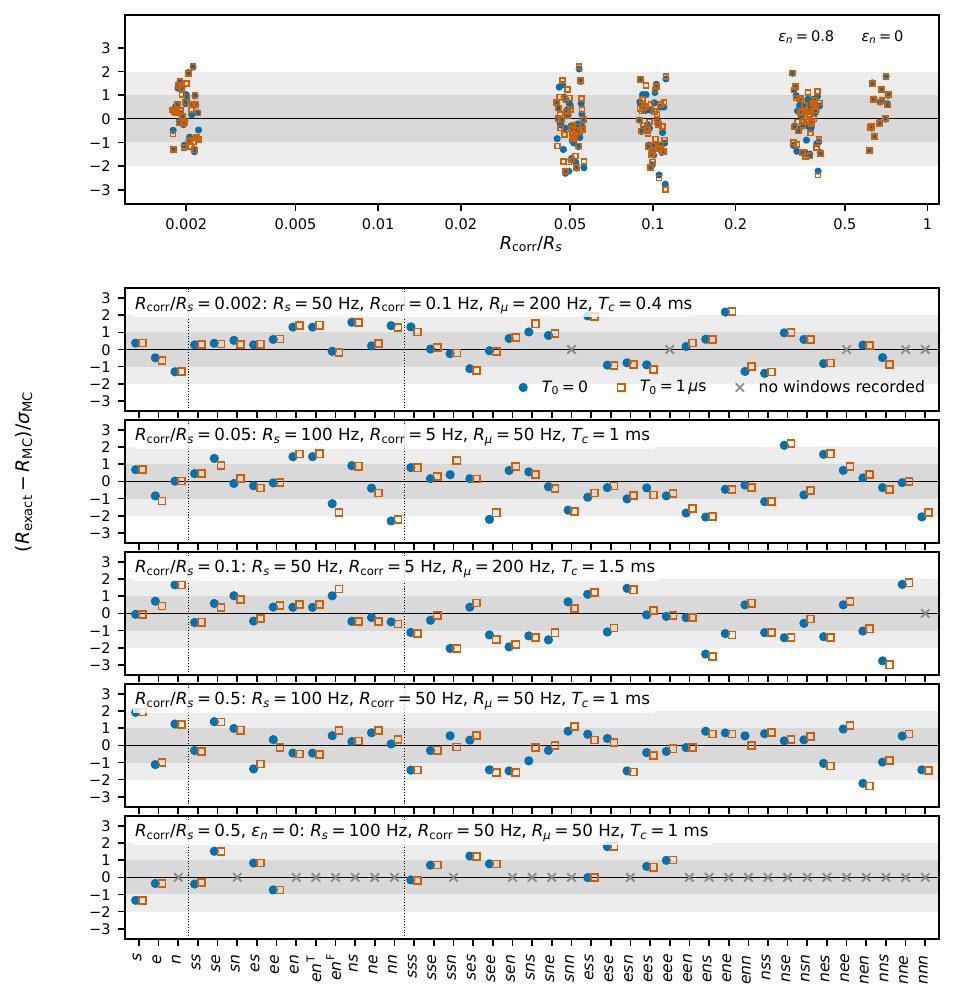}
  \caption{Pulls \((R_{\rm exact}-R_{\rm MC})/\sigma_{\rm MC}\) of the
  exact kernels against the toy-MC streams at the five additional
  parameter points, for all 39 ordered sequences and the truth-tagged
  True and False components of \(en\) (\(en^{\rm T}\), \(en^{\rm F}\)),
  at \(T_0=0\) (filled circles) and at \(T_0=\SI{1}{\micro s}\) in the
  window-close convention (open squares); \(\sigma_{\rm MC}\) is the
  standard error estimated from the shard-to-shard rate dispersion.
  Top: every pull against \(R_{\rm corr}/R_s\), the channels of one
  point spread over a narrow band around its ratio; the two points at
  \(R_{\rm corr}/R_s=0.5\), which differ only in \(\epsilon_n\), are
  drawn on either side of the tick.  Below: one panel per point in
  order of increasing \(R_{\rm corr}/R_s\), labeled with its parameters
  (\(\epsilon_n=0.8\) unless stated, capture mixture as in
  Section~\ref{sec:model}).  The \(R_s=\SI{100}{Hz}\) points use
  \SI{400}{\micro s} muon vetoes only; the \(R_s=\SI{50}{Hz}\) points use
  the \SI{400}{\micro s}/\SI{1}{s} shower split of the validation grid.
  The exposures are \(8\times10^{11}\) events at
  \(R_{\rm corr}/R_s=0.002\), \(1.8\times10^{10}\) at \(0.05\), and
  \(1.2\times10^{10}\) at \(0.1\) and at both points at \(0.5\).  Shaded
  bands mark \(\pm1\) and \(\pm2\).  A gray cross marks a channel
  without a pull: at \(\epsilon_n=0\) every \(n\)-bearing channel is
  identically zero in both the kernel and the simulation, and six rare
  three-fold channels at the two \(R_s=\SI{50}{Hz}\) points recorded no
  windows at either \(T_0\).}
  \label{fig:prd-pulls-closure}
\end{figure*}

A second comparison covers five additional parameter points, spanning
\(R_{\rm corr}/R_s\) from \(0.002\) to \(0.5\) and including
shower-muon reset conditions and a point with no delayed component.
In these simulated streams each neutron retains its parent identity,
so the \(en\) channel enters as its True and False components
separately.  Figure~\ref{fig:prd-pulls-closure} shows
the pulls of the exact kernels against these streams at \(T_0 = 0\) and
at \(T_0 = \SI{1}{\micro s}\) under the window-close convention, first
against \(R_{\rm corr}/R_s\) and then channel by channel at each point
in order of increasing ratio.  Of
the 356 channel rates not constrained to zero, 344 have nonzero toy-MC
errors and reproduce the exact kernels within statistics, with a
largest absolute pull of \(3.0\) and no trend with the ratio; the
remaining 12 (six rare three-fold channels at the two
\(R_s=\SI{50}{Hz}\) points, at both \(T_0\)) recorded no windows and
carry no pull.  All 54 rates required to vanish are exactly zero.  The
complete tables are Supplemental Tables~S9--S13, in the same order.

\begin{figure*}[tp]
  \centering
  \includegraphics[width=\textwidth]{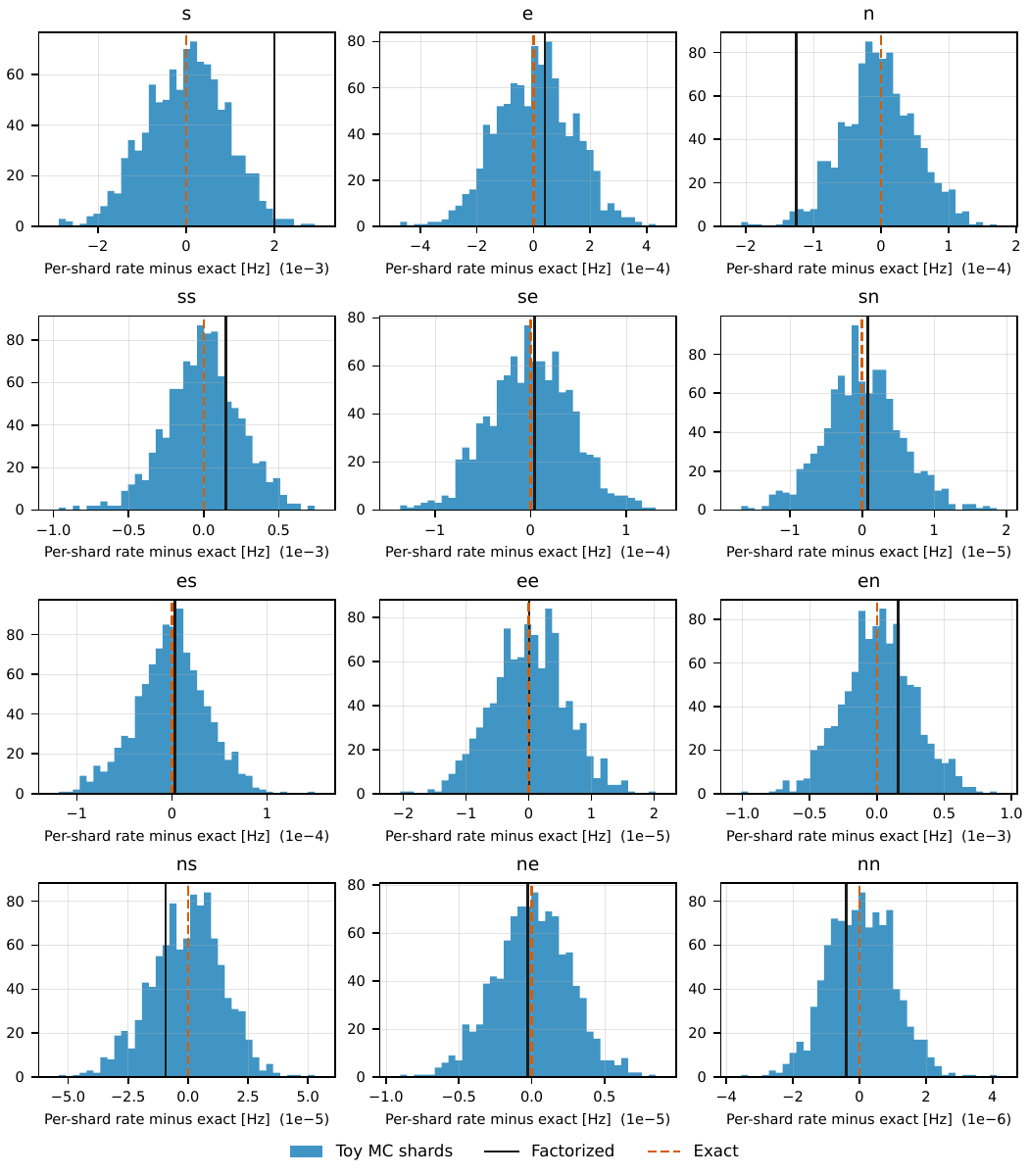}
  \caption{Per-shard distributions of the rate residual
  \(R_{\rm shard}-R_{\rm exact}\) for the twelve directly measurable
  channels at \(R_{\rm corr}=\SI{5}{Hz}\), \(T_c=\SI{1500}{\micro s}\),
  over the 1000 production shards of the \(2\times10^{13}\)-event
  campaign at \(T_0=0\), where the three dead-time conventions coincide.
  The exact ordered-propagator prediction sits at zero (dashed) and the
  secondary factorized comparison at \(R_{\rm fact}-R_{\rm exact}\)
  (solid).  The exact line is the prediction tested by the shard
  distributions; the two analytical lines coincide to better than the
  line width in most panels, and the visible \(s\) and \(n\) offsets are
  the largest factorized pulls of Fig.~\ref{fig:prd-pulls-factorized}
  seen at the per-shard resolution.}
  \label{fig:prd-shard-1}
\end{figure*}

A complementary view of the same \(T_0=0\) statistics is given by the
distribution of rates across the independent subsamples (shards) of
the production run.  Figure~\ref{fig:prd-shard-1} and Supplemental Figs.~S1--S3
show, for each configuration, the per-shard rate or count distribution
of the twelve directly measurable channels (the three single labels and
the nine ordered pairs; the True/False split of \(en\) is analytical
and was not tagged in this campaign).  The exact dashed line is the
ordered-propagator prediction under test; the factorized solid line is the
secondary comparison defined by Eq.~(\ref{eq:event-placement-skeleton}).
High-rate channels populate an approximately Gaussian histogram of the
per-shard rate residual \(R_{\rm shard}-R_{\rm exact}\) around zero;
channels whose per-shard expectation drops below 25 events (the \(ne\)
and \(nn\) channels at \(R_{\rm corr}=\SI{0.1}{Hz}\), Supplemental
Figs.~S2 and S3) are shown as raw count histograms against the Poisson
law implied by the exact rate.  In most panels the
factorized and exact lines coincide to better than the line width,
consistent with the \(10^{-4}\)-level agreement quantified in
Section~\ref{sec:event-placement}; their small resolved differences are
quantified in Fig.~\ref{fig:prd-pulls-factorized} and Supplemental
Tables~S3--S6, while
Sec.~\ref{subsec:thirteen-rates} gives the factorized formula for every
plotted channel.  The two figures resolve the same differences on
different scales: a shard holds one thousandth of the campaign, so the
width of a shard histogram is \(\sqrt{1000}\approx32\) merged
statistical errors \(\sigma_{\rm MC}\).  The \(+66\,\sigma_{\rm MC}\)
and \(-72\,\sigma_{\rm MC}\) factorized offsets of \(s\) and \(n\) at
\(R_{\rm corr}=\SI{5}{Hz}\), \(T_c=\SI{1500}{\micro s}\) therefore
appear in Fig.~\ref{fig:prd-shard-1} as displacements of about two
histogram widths, and every smaller offset in
Fig.~\ref{fig:prd-pulls-factorized} lies well inside its histogram.

\subsection{Selection efficiencies: exact values against the
  factorized approximation}
\label{subsec:prd-efficiencies}

The exact rate ratios in
Eqs.~(\ref{eq:pair-efficiency})--(\ref{eq:singles-efficiency}) define the
three selection efficiencies used here.  Table~\ref{tab:pair-efficiency}
evaluates their ordered-propagator values on the validation grid; the
factorized columns are secondary estimates from
Eq.~(\ref{eq:factorized-efficiencies}).  The
multiplicity cost is set mainly by the window length
(\(\epsilon_{\rm mult} \approx 0.86\) at \(T_c = \SI{1500}{\micro s}\)
against \(\approx 0.96\) at \(T_c = \SI{400}{\micro s}\)), with a
rate-driven reduction of about one percentage point between the low-
and high-IBD configurations at the long window: at
\(R_{\rm corr} = \SI{5}{Hz}\) the source population crowds its own
selection.  At \(T_0 > 0\) the pair efficiency drops through the blind
slice of the capture spectrum (the analytic time acceptance
\(\epsilon_n\,\Phi(T_0,T_c)\) falls from \(0.7996\) to \(0.7912\) at
\(T_c = \SI{1500}{\micro s}\) and from \(0.7134\) to \(0.7049\) at
\(T_c = \SI{400}{\micro s}\)) while \(\epsilon_{\rm mult}\) is almost
unchanged: the dead time removes capture-time acceptance, not
multiplicity acceptance.

\begin{figure*}[t]
  \centering
  \includegraphics[width=\textwidth]{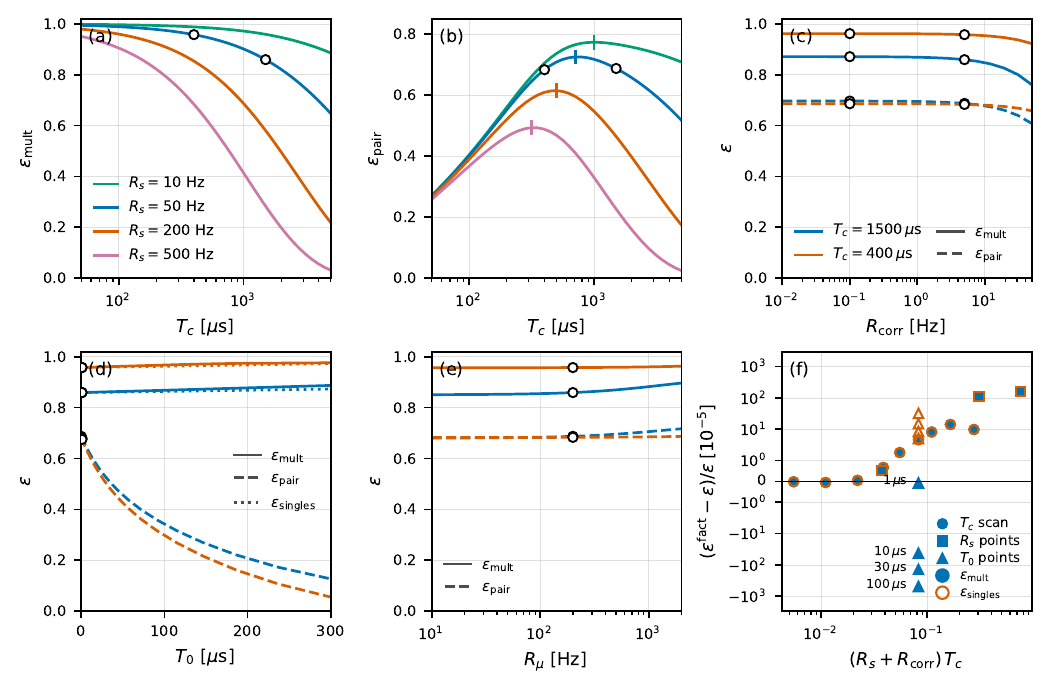}
  \caption{Exact selection efficiencies along the parameters that
    drive the multiplicity selection, all other parameters at the
    validation-grid values (\(R_s=\SI{50}{Hz}\),
    \(R_{\rm corr}=\SI{5}{Hz}\), \(R_\mu=\SI{200}{Hz}\), \(\epsilon_n=0.8\),
    \(T_c=\SI{1500}{\micro s}\), \(T_0=0\)); open circles mark the
    validation configurations that lie on a scanned line, and solid,
    dashed, and dotted curves are \(\epsilon_{\rm mult}\),
    \(\epsilon_{\rm pair}\), and \(\epsilon_{\rm singles}\).  (a)
    \(\epsilon_{\rm mult}\) and (b) \(\epsilon_{\rm pair}\) versus the
    window length for four singles rates, the ticks in (b) marking the
    maximum of each curve; (c) both efficiencies versus the correlated
    rate and (e) versus the reset rate at the two validation windows;
    (d) both efficiencies and \(\epsilon_{\rm singles}\) versus the
    event dead time under the window-close convention.  (f) Relative
    deviation of the factorized estimates of
    Eq.~(\ref{eq:factorized-efficiencies}) from the exact values,
    \(\epsilon_{\rm mult}\) (filled) and \(\epsilon_{\rm singles}\)
    (open), against the window occupancy \((R_s+R_{\rm corr})T_c\) for
    the window-length scan at \(R_s=\SI{50}{Hz}\) (circles), the
    singles-rate points at \(T_c=\SI{1500}{\micro s}\) (squares), and the
    dead-time points at \(T_c=\SI{1500}{\micro s}\) (triangles at
    \(T_0=1\), \(10\), \(30\), and \SI{100}{\micro s}, labeled); the
    vertical scale is linear within \(\pm10^{-5}\) and logarithmic
    beyond.  All
    efficiencies are reset-segment-axis ratios; the veto live-time
    factors are not included.}
  \label{fig:prd-efficiency-scan}
\end{figure*}

Figure~\ref{fig:prd-efficiency-scan} follows the three efficiencies
away from the grid, along the parameters that drive the multiplicity
selection; Supplemental Table~S14 lists the values.  The window
occupancy \((R_s+R_{\rm corr})T_c\) is the dominant variable:
\(\epsilon_{\rm mult}\) falls from \(0.989\) at \(T_c=\SI{100}{\micro s}\)
to \(0.647\) at \SI{5}{ms} for \(R_s=\SI{50}{Hz}\), and at fixed
\(T_c=\SI{1500}{\micro s}\) from \(0.959\) at \(R_s=\SI{10}{Hz}\) to
\(0.283\) at \SI{500}{Hz} (panel a).  Because the capture acceptance
\(\epsilon_n\Phi(0,T_c)\) rises with \(T_c\) while the multiplicity cost
falls, the pair efficiency has a maximum in \(T_c\) (panel b): near
\SI{1.0}{ms} for \(R_s=\SI{10}{Hz}\), \SI{0.7}{ms} at \SI{50}{Hz},
\SI{0.5}{ms} at \SI{200}{Hz}, and \SI{0.3}{ms} at \SI{500}{Hz}, so the
two validation windows straddle the \SI{50}{Hz} optimum.  The
correlated rate is a further occupancy term: raising \(R_{\rm corr}\)
from \(0.1\) to \SI{50}{Hz} lowers \(\epsilon_{\rm mult}\) from
\(0.871\) to \(0.760\) at \(T_c=\SI{1500}{\micro s}\) (panel c).  Dead
time acts on the two efficiencies differently (panel d): at
\(T_c=\SI{1500}{\micro s}\), \(\epsilon_{\rm pair}\) falls from \(0.687\)
to \(0.207\) as \(T_0\) grows to \SI{200}{\micro s}, following the
capture acceptance, while \(\epsilon_{\rm mult}\) rises from \(0.859\)
to \(0.878\), because the blind interval also removes events that
would spoil the window, and \(\epsilon_{\rm singles}\) separates from
\(\epsilon_{\rm mult}\) as discussed below.  On the reset-segment axis
the reset rate works in the opposite direction to the source rates
(panel e): a reset boundary restarts the window sequence, so a prompt
that follows a boundary within \(T_c\) cannot lie inside an earlier
window, and the shortened look-back makes a clean start more likely;
\(\epsilon_{\rm mult}\) rises from \(0.851\) to \(0.897\) between
\(R_\mu=10\) and \SI{2000}{Hz} at \(T_c=\SI{1500}{\micro s}\), close to
the value expected from a look-back of mean length
\((1-e^{-R_\mu T_c})/R_\mu\).  The veto live-time factors of
Section~\ref{sec:model}, which this axis excludes by construction,
remain the dominant muon cost in a detector-live or wall-clock
normalization.

\begin{table*}[t]
  \centering
  \caption{Exact selection efficiencies on the validation grid under
    the window-close dead-time convention, compared with the
    factorized event-placement estimates of
    Eq.~(\ref{eq:factorized-efficiencies}).  The central exact columns
    are the ordered-propagator rate ratios of
    Eqs.~(\ref{eq:pair-efficiency})--(\ref{eq:singles-efficiency}), defined on
    the reset-segment axis.  The
    deviation columns give
    \(\delta = ({\rm factorized} - {\rm exact})/{\rm exact}\) in units
    of \(10^{-5}\); \(\delta\) of \(\epsilon_{\rm mult}\) equals
    \(\delta\) of \(\epsilon_{\rm pair}\) exactly, because both sides
    share the analytic time acceptance \(\epsilon_n\,\Phi(T_0, T_c)\)
    of Eq.~(\ref{eq:mult-efficiency}).  Total-convention pair values are
    above the True ones by less than \(5\times10^{-5}\) relative on
    this grid.
    Rates in Hz, times in \(\si{\micro s}\).}
  \label{tab:pair-efficiency}
  \footnotesize
  \setlength{\tabcolsep}{4.5pt}
  \begin{tabular}{lccccccccc}
    \toprule
    & & \multicolumn{2}{c}{\(\epsilon_{\rm pair}\)}
      & \multicolumn{2}{c}{\(\epsilon_{\rm mult}\)}
      & \multicolumn{2}{c}{\(\epsilon_{\rm singles}\)}
      & \multicolumn{2}{c}{\(\delta\;[10^{-5}]\)} \\
    \cmidrule(lr){3-4}\cmidrule(lr){5-6}\cmidrule(lr){7-8}\cmidrule(lr){9-10}
    Configuration & \(T_0\) & exact & fact. & exact & fact.
                  & exact & fact. & pair & singles \\
    \midrule
    \(R_{\rm corr}=5\),   \(T_c=1500\) & 0 & 0.687177 & 0.687209 & 0.859351 & 0.859391 & 0.859351 & 0.859391 & \(+4.64\) & \(+4.64\) \\
    \(R_{\rm corr}=0.1\), \(T_c=1500\) & 0 & 0.696530 & 0.696547 & 0.871047 & 0.871069 & 0.871047 & 0.871069 & \(+2.46\) & \(+2.46\) \\
    \(R_{\rm corr}=5\),   \(T_c=400\) & 0 & 0.683211 & 0.683212 & 0.957703 & 0.957704 & 0.957703 & 0.957704 & \(+0.05\) & \(+0.05\) \\
    \(R_{\rm corr}=0.1\), \(T_c=400\) & 0 & 0.686015 & 0.686016 & 0.961633 & 0.961634 & 0.961633 & 0.961634 & \(+0.10\) & \(+0.10\) \\
    \midrule
    \(R_{\rm corr}=5\),   \(T_c=1500\) & 1 & 0.679996 & 0.679995 & 0.859439 & 0.859438 & 0.859395 & 0.859438 & \(-0.08\) & \(+5.03\) \\
    \(R_{\rm corr}=0.1\), \(T_c=1500\) & 1 & 0.689249 & 0.689232 & 0.871135 & 0.871112 & 0.871091 & 0.871112 & \(-2.53\) & \(+2.47\) \\
    \(R_{\rm corr}=5\),   \(T_c=400\) & 1 & 0.675201 & 0.675168 & 0.957803 & 0.957757 & 0.957753 & 0.957757 & \(-4.82\) & \(+0.40\) \\
    \(R_{\rm corr}=0.1\), \(T_c=400\) & 1 & 0.677969 & 0.677936 & 0.961730 & 0.961682 & 0.961681 & 0.961682 & \(-4.90\) & \(+0.11\) \\
    \bottomrule
  \end{tabular}
\end{table*}

The exact values expose a structural identity.  At \(T_0 = 0\) the
multiplicity cost of the pair equals the singles efficiency,
\(\epsilon_{\rm mult} = \epsilon_{\rm singles}\), to all computed
digits: both are the probability of a clean window start followed by a
window otherwise empty of recorded events, and at zero dead time the
required delayed member of the pair does not disturb that exposure.
At \(T_0 > 0\) the two split, by a relative
\(\sim (R_s + R_{\rm corr})\, T_0\) (about \(5\times10^{-5}\) at
\(T_0 = \SI{1}{\micro s}\)): the recorded delayed event re-arms the
dead time inside the pair's window and blinds part of its own tail, so
slightly fewer third events can spoil the pair than can spoil a
singles window of the same length.

The factorized estimates track all three efficiencies to within
\(5.1\times10^{-5}\) relative everywhere on this grid, the same level as
the underlying \(en\) rate agreement of
Sec.~\ref{subsec:why-approximation}.  Two limitations of the
approximation become visible at this magnification.  The deviation is
largest at the long window at \(T_0 = 0\) (\(+4.6\times10^{-5}\) at
high rate) and flips sign at \(T_0 = \SI{1}{\micro s}\), where the
factorized dead-time correction models only the trigger blind interval.  The
factorized construction also cannot split \(\epsilon_{\rm mult}\) from
\(\epsilon_{\rm singles}\) at all: its two columns are identical by
construction at every \(T_0\), so the exact \(T_0 > 0\) splitting
above is precisely a factorization-breaking effect.  The two efficiencies
remain distinct observables: \(\epsilon_{\rm mult}\) controls the pair
selection, whereas \(\epsilon_{\rm singles}\) controls inversion from
recorded one-fold windows to the underlying singles rate.

Panel (f) of Fig.~\ref{fig:prd-efficiency-scan} extends this comparison
beyond the grid.  At \(T_0=0\) the
relative deviation of the factorized efficiencies from the exact values
follows the window occupancy: below \(10^{-6}\) up to
\((R_s+R_{\rm corr})T_c\approx0.02\), \(4.6\times10^{-5}\) at the grid
point (occupancy \(0.083\)), \(1.4\times10^{-4}\) at \(T_c=\SI{3}{ms}\)
(\(0.17\)) and \(9.8\times10^{-5}\) at \SI{5}{ms} (\(0.28\)), so the
window-length dependence is not monotonic at long windows, and
\(1.1\times10^{-3}\) and \(1.6\times10^{-3}\) at \(R_s=200\) and
\SI{500}{Hz} (\(0.31\) and \(0.76\)); these are the
occupancy-suppressed history layers listed in
Sec.~\ref{subsec:why-approximation}.  At \(T_0>0\) the omitted follower
blind intervals dominate instead.  At \(T_c=\SI{1500}{\micro s}\) and
\(T_0=\SI{1}{\micro s}\) their negative contribution nearly cancels the
positive zero-dead-time offset, leaving a multiplicity (and pair)
deviation of \(-8\times10^{-7}\); from \SI{10}{\micro s} to
\SI{100}{\micro s} the deviation grows from \(-4.3\times10^{-4}\) to
\(-5.0\times10^{-3}\), in magnitude roughly \((R_s+R_{\rm corr})T_0\),
while the singles deviation reaches only \(3.2\times10^{-4}\).

\subsection{Two-fold time densities}
\label{subsec:prd-dt-densities}

\begin{figure*}[t]
  \centering
  \includegraphics[width=0.49\textwidth]{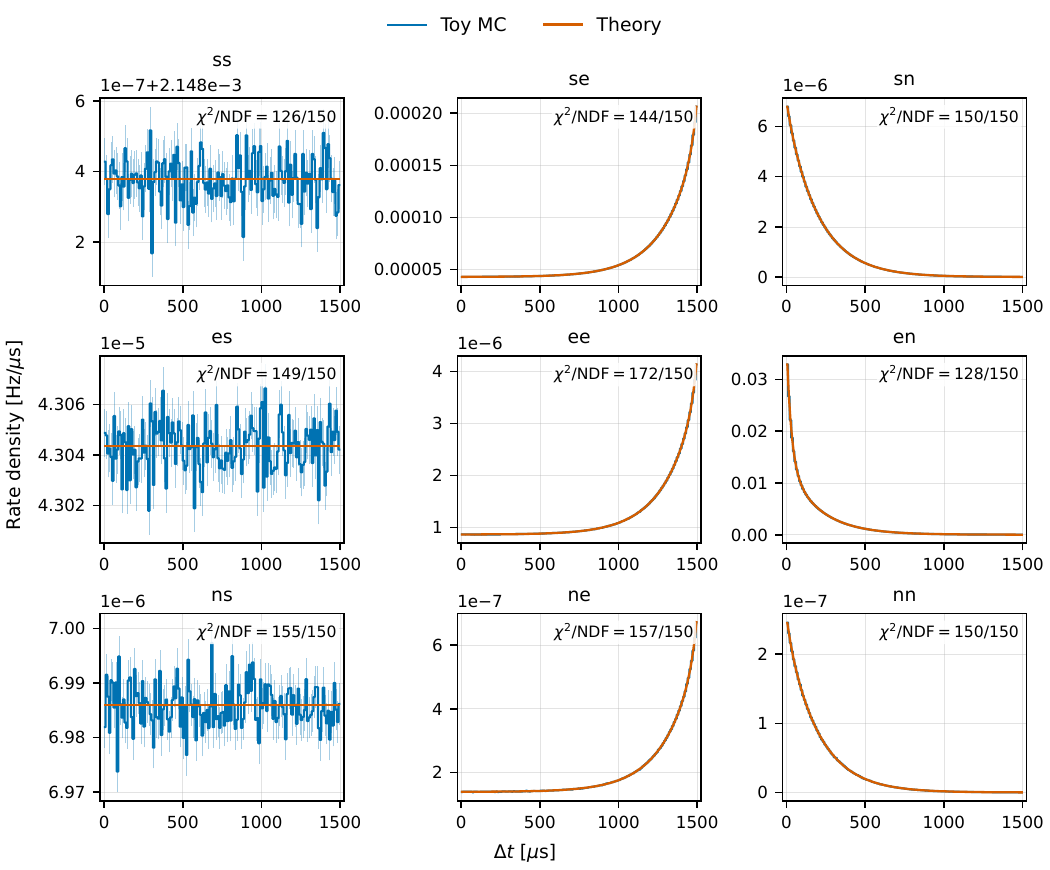}\hfill
  \includegraphics[width=0.49\textwidth]{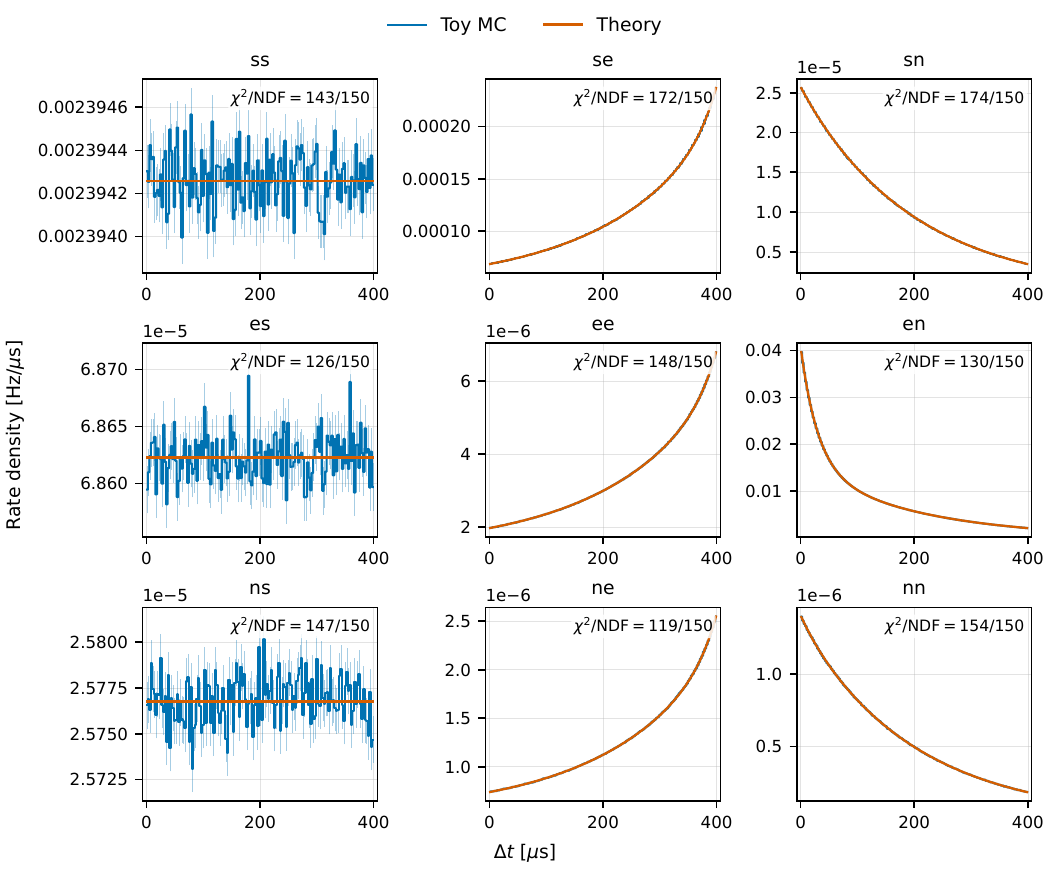}\\[1ex]
  \includegraphics[width=0.49\textwidth]{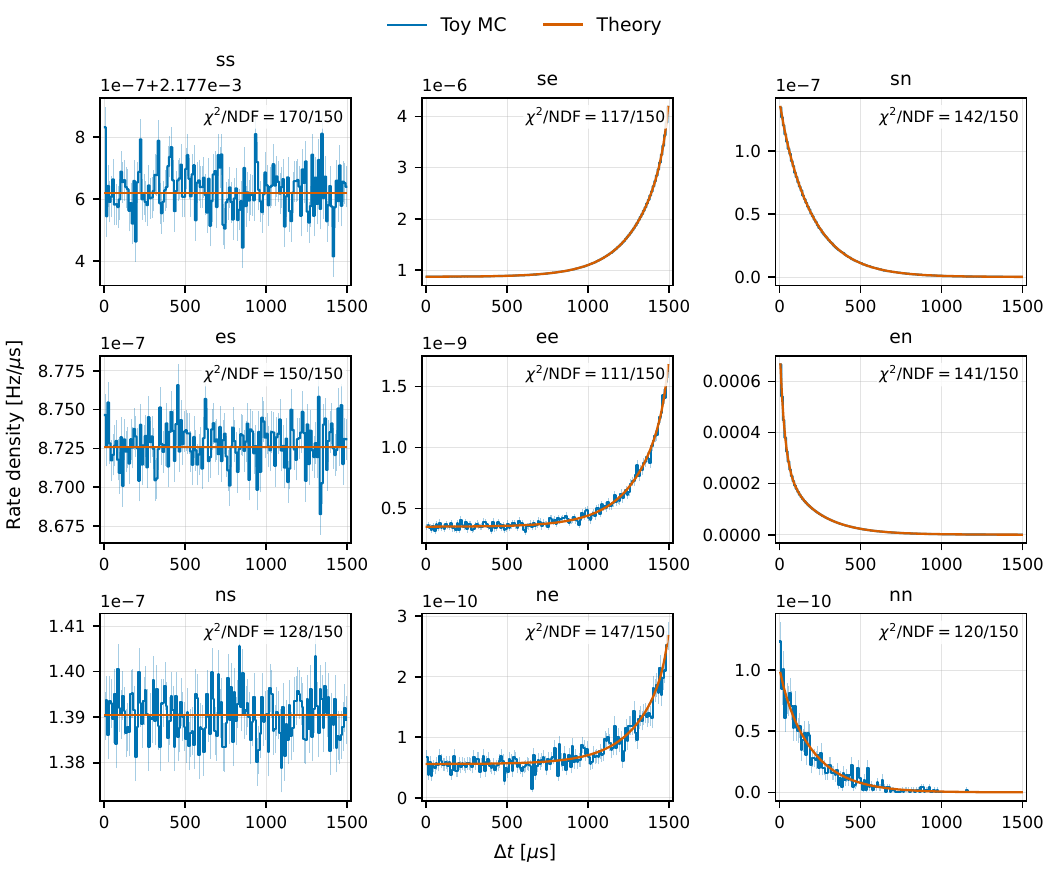}\hfill
  \includegraphics[width=0.49\textwidth]{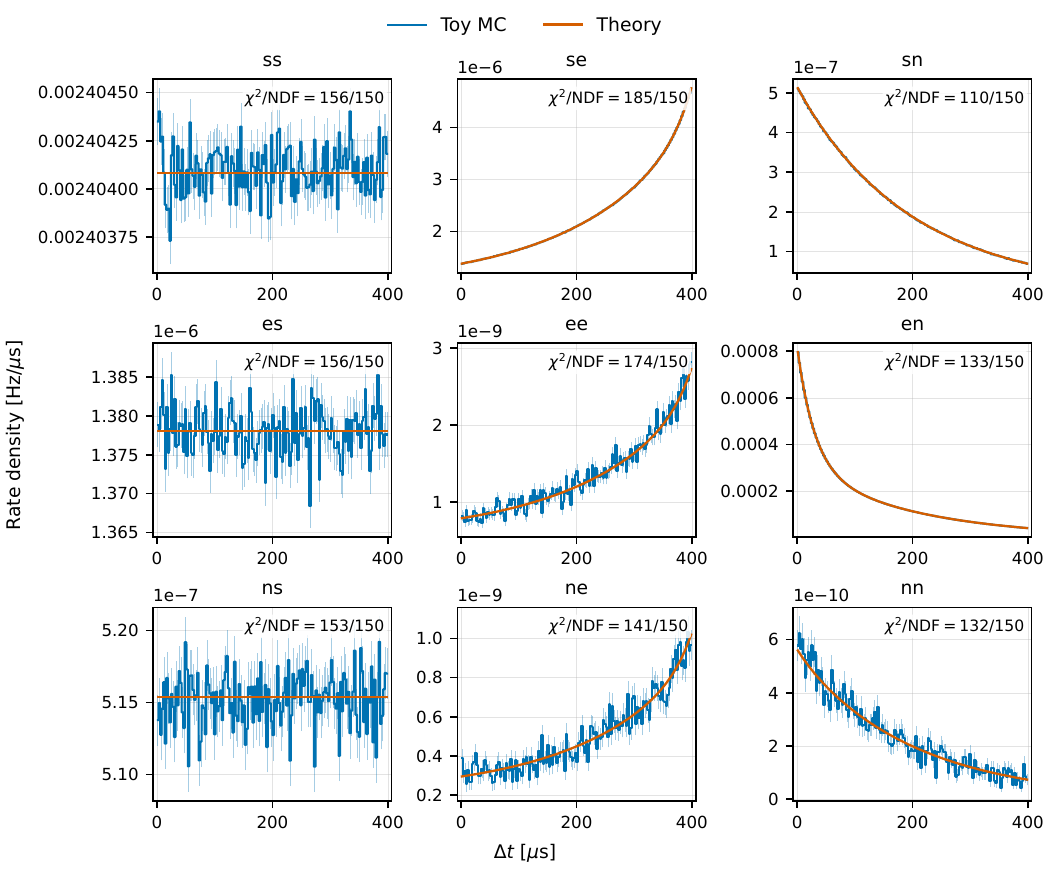}
  \caption{Forward time distributions of all nine two-fold sequences in the
  four validation configurations of the \(2\times10^{13}\)-event campaign
  at \(T_0=0\), where the three dead-time conventions coincide.  The
  upper row has
  \(R_{\rm corr}=\SI{5}{Hz}\), the lower row
  \(R_{\rm corr}=\SI{0.1}{Hz}\); within each row
  \(T_c=\SI{1500}{\micro s}\) and \(\SI{400}{\micro s}\), respectively.
  Step histograms show the independent toy-MC densities and smooth curves
  the exact ordered propagator \(G^{(2)}\), integrated over the same
  \(\Delta t\) bins.  Error bars are toy-MC counting uncertainties; no
  factorized density enters this comparison.  Each panel quotes
  \(\chi^2/{\rm NDF}\) of the toy-MC bin counts against the exact
  prediction over its 150 bins (Poisson variance of the predicted count,
  no fitted parameter); Supplemental Figs.~S4--S7 reproduce the four
  panels at full size.}
  \label{fig:prd-dt}
\end{figure*}

Figure~\ref{fig:prd-dt} directly tests the shape information in the exact
two-fold propagator \(G^{(2)}\): all nine \(\Delta t\) densities agree with
the independent event stream over the four configurations.  Each curve is
the exact differential \(G^{(2)}\) prediction integrated over the plotted
bin, so the comparison tests the within-window time dependence as well as
the already validated integral yield.  The per-panel \(\chi^2/{\rm NDF}\)
values lie between \(0.73\) and \(1.24\) for 150 degrees of freedom,
where the expected spread is \(\sqrt{2/150}=0.12\), so no panel shows a
shape distortion beyond its counting resolution.  The flat, rising, and
falling shape families anticipated in Table~\ref{tab:dt-shapes} are visible
across both source rates and both window lengths.

\subsection{Nonzero-dead-time conventions}
\label{subsec:prd-conventions}

The \(T_0=0\) grid cannot distinguish dead-time conventions.  A dedicated
\(R_{\rm corr}=\SI{0.2}{Hz}\), \(T_c=\SI{400}{\micro s}\) campaign therefore
compares window-close re-arming, global non-paralyzable, and global
paralyzable analyzers with their matched kernels.  The two non-paralyzable
rules use \(T_0=\SI{50}{\micro s}\) and \(T_c/2=\SI{200}{\micro s}\); the
paralyzable rule uses \(T_0=\SI{100}{\micro s}\) and
\(\SI{200}{\micro s}\).  Seeds depend only on \((T_0,\mathrm{shard})\), so
rules at a shared \(T_0\) analyze bit-identical \(10^{12}\)-event streams.
At \(T_0=T_c/2\), the active follower region collapses to one span and all
27 three-fold channels are kinematically zero in both theory and simulation.

Of the 78 rate rows at each convention, 51 are analytically nonzero.
Forty-eight have nonzero Monte Carlo counts; the other three per convention expect
at most 0.074 counts and observe zero.  Away from these unresolved rows, the
largest pulls are 2.36 for the non-paralyzable rules and 2.65 for the
paralyzable rule.  The statistically sharpest \(s\) rate has
\((R_{\rm model}-R_{\rm MC})/R_{\rm model}=-8.0\times10^{-7}\)
(\(-0.31\sigma\)) at \(T_0=\SI{50}{\micro s}\), \(+2.5\times10^{-6}\)
(\(+0.95\sigma\)) at \(\SI{100}{\micro s}\), and
\(+5.2\times10^{-6}\) (\(+2.0\sigma\)) at \(T_c/2\); rules sharing a
stream agree within \(0.01\sigma\).

The \(n\)-trigger family (\(n\), \(ne\), \(ns\)) provides the
discriminating test.  For its one-fold \(n\) channel under the global
non-paralyzable rule, the matched pulls are \(+0.71\) and \(+0.33\) at
\(T_0=\SI{50}{\micro s}\) and \(T_c/2\), whereas the same simulated
\(n\) rates are \(22\sigma\) and \(153\sigma\) from the window-close
prediction.  The
paralyzable rule shifts weight from fold two to fold one because extended
blindness absorbs followers; its matched kernel reproduces this reshuffling,
including ordinary \(nn\) pulls of \(-0.9\) and \(+1.2\) at
\(T_0=\SI{100}{\micro s}\) and \(T_c/2\).
Table~\ref{tab:deadtime-conventions} lists the exact relative shifts of
the two global conventions from window-close at
\(T_0=\SI{100}{\micro s}\) for the \(R_{\rm corr}=\SI{5}{Hz}\) grid
points.

\begin{table}[H]
  \centering
  \caption{Relative shifts \(R^{\rm conv}/R^{\rm WC}-1\) from window-close
  (WC) at \(T_0=\SI{100}{\micro s}\) and \(R_{\rm corr}=\SI{5}{Hz}\):
  global non-paralyzable (G) and global paralyzable (GP).  GP is
  evaluated at \(T_c=\SI{400}{\micro s}\) only: its mark sum
  (Sec.~\ref{sec:history-chain}) grows exponentially in \(T_c/T_0\),
  and the grid restricts this convention to \(T_c/T_0\le4\).}
  \label{tab:deadtime-conventions}
  \scriptsize
  \setlength{\tabcolsep}{3.2pt}
  \begin{tabular}{lrrr}
    \toprule
    & \multicolumn{1}{c}{G, \(T_c=\SI{1500}{\micro s}\)}
    & \multicolumn{1}{c}{G, \(T_c=\SI{400}{\micro s}\)}
    & \multicolumn{1}{c}{GP, \(T_c=\SI{400}{\micro s}\)}\\
    \midrule
    \(s\)  & \(-6.7\times10^{-6}\) & \(-2.4\times10^{-5}\) & \(+6.9\times10^{-5}\)\\
    \(e\)  & \(-6.7\times10^{-6}\) & \(-2.4\times10^{-5}\) & \(+2.0\times10^{-3}\)\\
    \(n\)  & \(-1.9\times10^{-2}\) & \(-8.0\times10^{-3}\) & \(-8.1\times10^{-3}\)\\
    \(ss\) & \(-6.7\times10^{-6}\) & \(-2.4\times10^{-5}\) & \(-8.3\times10^{-4}\)\\
    \(se\) & \(-6.7\times10^{-6}\) & \(-2.4\times10^{-5}\) & \(+1.5\times10^{-4}\)\\
    \(en\) & \(-7.3\times10^{-6}\) & \(-2.4\times10^{-5}\) & \(-1.6\times10^{-3}\)\\
    \(ne\) & \(-1.9\times10^{-2}\) & \(-8.0\times10^{-3}\) & \(-8.0\times10^{-3}\)\\
    \(nn\) & \(-1.9\times10^{-2}\) & \(-9.2\times10^{-2}\) & \(-2.6\times10^{-1}\)\\
    \bottomrule
  \end{tabular}
\end{table}

The mechanism is a blind tail beyond the window edge.  An \(e\) follower
recorded within \(T_0\) of the close keeps a global analyzer blind into the
gap, where its delayed capture can no longer open the next window.  At
\(T_0=T_c/2\), the global \(n\), \(ne\), and \(ns\) rates are depleted by
\(1.89\times10^{-2}\) and \(nn\), with two free neutrons, by \(0.13\);
\(sn\) carries a diluted
\(-1.4\times10^{-3}\) echo, while channels with no free neutron move by
about \(-4.7\times10^{-5}\).  At \(T_0=\SI{50}{\micro s}\), the
\(n\), \(ne\), and \(ns\) shift is \(-2.7\times10^{-3}\).  The ratio of these shifts is
7.0, below the factor 16 of a pure quadratic, and quantifies saturation
toward the stress point.  The difference vanishes quadratically for
\(T_0/T_c\to0\); at larger ratios the matching exact kernel should be used.

\subsection{Aggregate multiplicity \texorpdfstring{\(\ge4\)}{>=4}}
\label{subsec:prd-ge4}

Although individual ordered rates are displayed through multiplicity three,
the total opening rate fixes the aggregate higher-multiplicity channel:
\begin{equation}
 R_{\ge4}=R_{\rm open}-M_1-M_2-M_3,
 \label{eq:ge4-identity}
\end{equation}
where \(M_k\) is the sum of all \(3^k\) ordered fold-\(k\) rates.  The
event-level validation directly counts the same object as windows with
more than three recorded events.  Because the subtraction loses
\(\log_{10}(R_{\rm open}/R_{\ge4})\) digits to cancellation,
Table~\ref{tab:ge4} lists this cancellation factor beside each
comparison.  The direct count agrees with the subtraction on every
nonzero configuration, with
maximum absolute pulls 1.31 on the \(2\times10^{13}\)-event \(T_0=0\)
grid, 1.39 on its \(T_0=\SI{1}{\micro s}\) companion, and 1.34 in the
nonzero-dead-time campaign of Section~\ref{subsec:prd-conventions}.  At
\(T_0=T_c/2\), both return the exact kinematic zero.

\begin{table*}[t]
  \centering
  \caption{Aggregate multiplicity-\(\ge4\) rates: the exact subtraction
    of Eq.~(\ref{eq:ge4-identity}), the cancellation factor
    \(R_{\rm open}/R_{\ge4}\) it loses, the toy-MC overflow counter with
    its \(\sqrt{N}\) error, and the pull
    \((R_{\rm exact}-R_{\rm MC})/\sigma_{\rm MC}\).  Rows abbreviate
    \(R_{\rm corr}\) in Hz over \(T_c\) in \si{\micro s}; the last block is
    the \(R_{\rm corr}=\SI{0.2}{Hz}\), \(T_c=\SI{400}{\micro s}\)
    campaign of Section~\ref{subsec:prd-conventions}, each convention
    (WC window close, G global non-paralyzable, GP global paralyzable)
    against its matched kernel.  At \(T_0=T_c/2\) no window can hold
    four recorded events and both sides are exactly zero.}
  \label{tab:ge4}
  \footnotesize
  \setlength{\tabcolsep}{6pt}
  \begin{tabular}{lcccc}
    \toprule
    Setting & Exact (Hz) & \(R_{\rm open}/R_{\ge4}\) & Toy MC (Hz) & Pull \\
    \midrule
    \multicolumn{5}{l}{\(2\times10^{13}\) events, \(T_0=0\)} \\
    \(5/1500\) & \(0.0533101\) & \(9.66\times 10^{2}\) & \(0.0533115 \pm 0.0000011\) & \(-1.31\) \\
    \(0.1/1500\) & \(0.00369935\) & \(1.27\times 10^{4}\) & \(0.00369939 \pm 0.00000027\) & \(-0.113\) \\
    \(5/400\) & \(0.00552180\) & \(9.85\times 10^{3}\) & \(0.00552179 \pm 0.00000030\) & \(0.0410\) \\
    \(0.1/400\) & \(9.98752\times 10^{-5}\) & \(4.92\times 10^{5}\) & \((9.9898 \pm 0.0040)\times 10^{-5}\) & \(-0.557\) \\
    \multicolumn{5}{l}{\(2\times10^{13}\) events, \(T_0=\SI{1}{\micro s}\), WC} \\
    \(5/1500\) & \(0.0523782\) & \(9.83\times 10^{2}\) & \(0.0523796 \pm 0.0000010\) & \(-1.39\) \\
    \(0.1/1500\) & \(0.00367214\) & \(1.28\times 10^{4}\) & \(0.00367217 \pm 0.00000027\) & \(-0.119\) \\
    \(5/400\) & \(0.00531949\) & \(1.02\times 10^{4}\) & \(0.00531945 \pm 0.00000030\) & \(0.133\) \\
    \(0.1/400\) & \(9.73146\times 10^{-5}\) & \(5.05\times 10^{5}\) & \((9.7333 \pm 0.0040)\times 10^{-5}\) & \(-0.476\) \\
    \multicolumn{5}{l}{\(10^{12}\) events, \(0.2/400\), \(T_0\) and convention} \\
    \SI{50}{\micro s}, WC & \(2.86419\times 10^{-5}\) & \(1.72\times 10^{6}\) & \((2.8513 \pm 0.0096)\times 10^{-5}\) & \(1.34\) \\
    \SI{50}{\micro s}, G & \(2.86418\times 10^{-5}\) & \(1.72\times 10^{6}\) & \((2.8513 \pm 0.0096)\times 10^{-5}\) & \(1.34\) \\
    \SI{100}{\micro s}, GP & \(1.68418\times 10^{-6}\) & \(2.93\times 10^{7}\) & \((1.660 \pm 0.023)\times 10^{-6}\) & \(1.02\) \\
    \SI{200}{\micro s}, all & \(0\) & n/a & \(0\) & n/a \\
    \bottomrule
  \end{tabular}
\end{table*}

\subsection{Optional external spatial cross-check}
\label{sec:time-distance}
\label{subsec:distance-closure}

The temporal operator of this paper carries no vertex information, so
its accidental sum can also be confronted, in a real detector, with a
data-driven distance estimate: a mixed-event or off-time distance
template for the sum of accidental classes, normalized in a
signal-free long-distance sideband and transferred to the signal-distance
region under identical time, energy, and multiplicity selections.  Such a
cross-check is external to the present construction and has limits of its
own.  It needs reliable vertices for both members of the pair; its
sideband must be free of the displacement tail of genuine correlated
pairs, otherwise the normalization absorbs signal; and its template must
transfer unchanged from the sideband to the signal region, which position
dependence of the singles rate or of the reconstruction can spoil.  Where
these conditions hold it constrains the same accidental sum that
Eq.~(\ref{eq:exact-master}) predicts, with no detector-geometry model
beyond the empirical template; where they fail, its bias is not
controlled by the temporal calculation.  The present validation makes no
use of it.

\section{Physical regimes of the accidental \texorpdfstring{\(en\)}{en} rate}
\label{sec:regimes}

The False (accidental) component of the \(en\) rate is the contribution where
the delayed neutron was already pending from an earlier IBD event.  The
natural first estimate of this accidental rate is the Poisson product
\(R_e\cdot R_n\cdot T_c\), where \(R_e\) and \(R_n\) are the exact one-fold
prompt and neutron-capture rates.

\begin{figure}[t]
  \centering
  \includegraphics[width=\columnwidth]{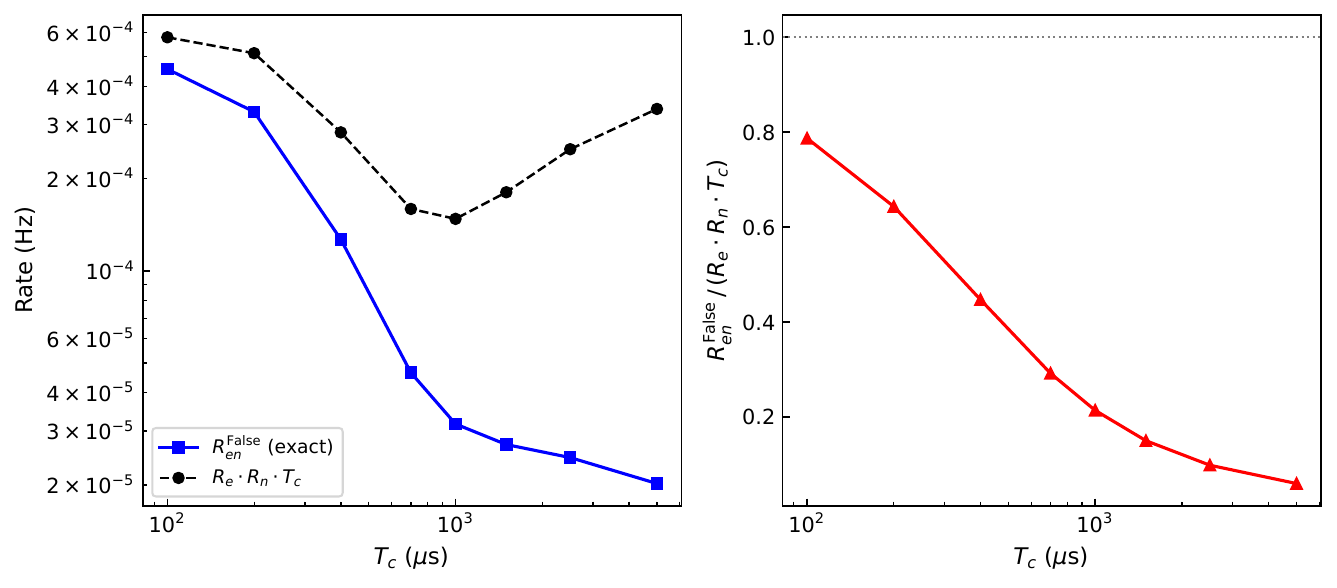}
  \caption{Exact accidental (False) \(en\) rate versus the Poisson product
  \(R_e\cdot R_n\cdot T_c\) as \(T_c\) varies from \SI{100}{\micro s} to
  \SI{5}{ms}.  Left panel: absolute rates; right panel: the suppression ratio
  \(R_{en}^{\rm False}/(R_e\cdot R_n\cdot T_c)\).  At short \(T_c\) the False
  rate and the Poisson product are comparable (ratio \(\approx 0.8\));
  over the scan the Poisson product passes through a minimum at
  \(T_c=\SI{1}{ms}\) and then rises, whereas the False rate keeps
  falling, to \(\simeq2\times10^{-5}\,\text{Hz}\), driving the ratio to
  0.060 at \(T_c=\SI{5}{ms}\).  Parameters: \(R_s=\SI{50}{Hz}\),
  \(R_{\rm corr}=\SI{5}{Hz}\), \(R_\mu=\SI{200}{Hz}\), \(\epsilon_n=0.8\),
  \(T_0=0\), and the capture-time mixture of Sec.~\ref{sec:model}.}
  \label{fig:prd-regime}
\end{figure}

Figure~\ref{fig:prd-regime} shows the comparison across a factor of 50 in
\(T_c\) at the high-IBD configuration (\(R_{\rm corr}=\SI{5}{Hz}\)).  At small
\(T_c\) (\(\leq\SI{200}{\micro s}\)), the False \(en\) rate and the Poisson
product \(R_e\cdot R_n\cdot T_c\) are within a factor of two (ratio
\(\approx 0.6\text{--}0.8\)); the correction from muon-veto suppression and
pending-neutron memory is modest.  As \(T_c\) grows toward the characteristic
reset spacing \((R_\mu)^{-1}\approx\SI{5}{ms}\), the False rate decreases
from \(4.6\times10^{-4}\,\mathrm{Hz}\) at
\(T_c=\SI{100}{\micro s}\) to \(2.0\times10^{-5}\,\mathrm{Hz}\) at
\(T_c=\SI{5}{ms}\).  The Poisson product is not monotonic, because
\(R_e\) and \(R_n\) themselves depend on \(T_c\): it falls from
\(5.8\times10^{-4}\,\mathrm{Hz}\) at \(\SI{100}{\micro s}\) to
\(1.5\times10^{-4}\,\mathrm{Hz}\) at \(\SI{1}{ms}\) and then rises to
\(3.4\times10^{-4}\,\mathrm{Hz}\) at \(\SI{5}{ms}\).  At
\(T_c=\SI{5}{ms}\) the suppression ratio is
0.060: long coincidence windows require the reset-boundary and
pending-neutron conditioning, not the naive Poisson product.  The
exact derivation captures the short-window and long-window limits within a
single closed-form expression.  The scan holds \(\epsilon_n=0.8\)
fixed.  At \(T_c=\SI{1500}{\micro s}\), repeating the calculation at
\(\epsilon_n=0.6\) leaves the suppression ratio unchanged to four
significant figures (0.1501 for both values), because numerator and
denominator carry the same leading \(\epsilon_n\) scaling.  The
absolute False component rises from \(2.71\times10^{-5}\) to
\(4.06\times10^{-5}\,\mathrm{Hz}\), tracking the corresponding increase
of the Poisson product \(R_e\cdot R_n\cdot T_c\).  Physically, fewer self
neutrons arrive to convert would-be \(en\) windows into three-fold
\(enn\)-type windows, leaving more windows able to close as a clean
prompt plus old neutron.

Finally, the validation grid deliberately uses IBD-rich rates
(\(R_{\rm corr}=5\) and \(\SI{0.1}{Hz}\)) so that the simulation
resolves the rare channels.  Nothing in the formulas is
perturbative in \(R_{\rm corr}\), and no hidden breakdown appears as
the rate decreases: evaluating the exact rates at
\(R_{\rm corr}=\SI{0.01}{Hz}\) with all other grid parameters fixed,
the pair efficiency is
constant along the decade
(\(R_{en}^{\rm True}/R_{\rm corr}=0.69653\) at \(\SI{0.1}{Hz}\)
against \(0.69670\) at \(\SI{0.01}{Hz}\) for
\(T_c=\SI{1500}{\micro s}\); \(0.68601\) against \(0.68607\) at
\(T_c=\SI{400}{\micro s}\)), with the residual drift the expected
second-order pileup, and the False component falls quadratically
(\(1.08\times10^{-8}\) to \(1.08\times10^{-10}\,\mathrm{Hz}\) over the
same decade at \(T_c=\SI{1500}{\micro s}\)), the independent-pairing
scaling of this section.  The IBD-rich grid is a statistics choice for
the validation, not a domain boundary of the formulas.

\section{Discussion}
\label{sec:discussion}

The production calculation uses history cap \(N_{\max}=4\) and current
headroom \(H=3\), hence current cap \(C=7\).  At \(T_0=0\), where the
three dead-time conventions coincide, a convergence study at
\(R_{\rm corr}=\SI{5}{Hz}\) and \(T_c=\SI{1500}{\micro s}\) (the
history-cap sweep behind Table~\ref{tab:cap-scan}) shows:

\begin{center}
\begin{tabular}{ccc}
\toprule
\(N_{\max}\) & \(R_{en}^{\rm Total}\) (Hz) & Rel.\ diff.\ vs.\ \(N_{\max}=4\) \\
\midrule
2 & 3.4359105376412 & \(5.3\times10^{-12}\) \\
3 & 3.4359105376230 & \(7.8\times10^{-16}\) \\
4 & 3.4359105376230 & 0 (reference) \\
5 & 3.4359105376230 & 0 \\
\bottomrule
\end{tabular}
\end{center}

The \(N_{\max}=2\) truncation differs by \(5\times10^{-12}\) relative;
\(N_{\max}=3\) is already below the numerical resolution of the reported
cap comparison; \(N_{\max}=4\) introduces no detectable error relative to
\(N_{\max}=5\).  This is the numerical
half of the exactness contract stated in the introduction: all rate
formulas are exact on the retained history and current-window spaces.
The history overflow is monitored through the mass that the
sub-stochastic kernel rows lose to the
\(\sum_i h_i > N_{\max}\) boundary
(Sec.~\ref{subsec:general-KN}); current-window overflow is controlled
separately at \(\sum_i(o_i+m_i)>C=N_{\max}+H\) by the error bound of
Sec.~\ref{subsec:truncation-contract}.  The convergence statement is a
one-point numerical scan, not a uniform bound over the parameter
domain; occupancy regimes far above the validation grid should rerun
the scan at their own parameters.

\section{Conclusion}
\label{sec:conclusion}

We have derived ordered coincidence rates for latent uncorrelated-single and
prompt--delayed source processes passed through a specified recording state
machine.  The resulting recorded stream is segmented by Poisson reset
boundaries, with pending
delayed events carried across boundaries as a Markov state and event dead
time treated under explicit conventions.  The construction yields all
one-, two-, and three-fold \(s/e/n\) rates, including the True/False \(en\)
split, multiplicity efficiencies, and two-fold time densities.  Every central
rate, efficiency, and time-density result reported here is an
ordered-propagator output; the factorized and earlier leading-order formulas
serve only as comparison calculations.  The finite-cap error bound is
complete for the window-close convention and conservative for the
strictly global one.

An independent streaming toy Monte Carlo validates the rates over
\(2\times10^{13}\) events per source-rate group: the 149 directly observable
channels on the \(T_0=0\) grid have an RMS pull of 1.00, with no coherent
residual.  That exposure is where the test was stopped, not where the
calculation ends: the rates are exact within the stated model, and a
larger sample would only sharpen the same comparison.  At
\(R_{\rm corr}=\SI{5}{Hz}\), \(T_c=\SI{400}{\micro s}\), and
\(\epsilon_n=0.8\), the exact \(en\) rate is \(\SI{3.416}{Hz}\),
\(14.6\%\) below the naive bound \(R_{\rm corr}\epsilon_n\).  This difference
shows why accepted-window, veto, and delayed-history conditioning must be
kept together in precision coincidence accounting.

\begin{acknowledgments}
This work was supported in part by the National Natural Science
Foundation of China (Grant No.~12305117), the Scientific Research
Startup Fund of Hunan University (Grant No.~531118011073), and the
Hunan Provincial Major Basic Research Project (Grant No.~2026JC0006).
I thank Zhe Wang and Shaomin Chen for helping me understand the earlier
approximate treatment of Ref.~\cite{YuWangChen2015}, and Jiajie Ling for
discussions of the different implementations of accidental-background
estimation.  The experience gained in the Daya Bay experiment was
invaluable to this work.
\end{acknowledgments}

\bibliography{coincidence_refs}

\end{document}